\pdfoutput=1
\documentclass[11pt]{article}
\usepackage{jheppub}

\usepackage{customprelude}

\newcommand{\Ytop}{\ensuremath{\bf 10\, \bf 10\, \bf 5}}

\newcommand{\be}{\begin{equation}}  
\newcommand{\ee}{\end{equation}}  
\newcommand{\bp}{\begin{pmatrix*}[r]}  
\newcommand{\ep}{\end{pmatrix*}}  
\newcommand{\bpp}{\begin{pmatrix}}  
\newcommand{\epp}{\end{pmatrix}}  

\usepackage{bbm}

\newcommand{\includegraphicstikz}[1]{\raisebox{-0.5\height}{\includegraphics{#1}}}

\def\P{\mathbb{P}}

\def\C{\mathbb{C}}
\def\cO{\mathcal{O}}

\def\1{\mathbbm{1}}

\title{\centering $E_6$ Yukawa couplings in F-theory\\as D-brane instanton effects}

\author[\spadesuit]{Andr\'es Collinucci}
\author[\clubsuit]{and I\~naki Garc\'ia-Etxebarria}

\affiliation[\spadesuit]{Physique Th\'eorique et Math\'ematique and International Solvay Institutes,\\
Universit\'e Libre de Bruxelles, C.P. 231, 1050 Bruxelles, Belgium}
\affiliation[\clubsuit]{Max Planck Institute for Physics,\\F\"ohringer Ring 6, 80805 Munich, Germany}

\emailAdd{collinucci.phys@gmail.com}
\emailAdd{inaki@mpp.mpg.de}

\abstract{In a weak coupling limit the neighborhood of $E_6$ Yukawa
  points in $SU(5)$ GUT \mbox{F-theory} models is described by a
  non-resolvable orientifold of the conifold. We explicitly show,
  first directly in IIB and then via a mirror symmetry argument, that
  in this limit the $E_6$ Yukawa coupling is better described as
  coming from the non-perturbative contribution of a
  euclidean D1-brane wrapping the non-resolvable cycle. We also
  discuss how the M-theory description interpolates between the weak
  and strong coupling viewpoints.}

\begin{document}

\makeatletter
\let\old@fpheader\@fpheader
\renewcommand{\@fpheader}{\old@fpheader\hfill
MPP-2016-349}
\makeatother

\maketitle

\newpage

\section{Introduction}

In recent years, starting with
\cite{Donagi:2008ca,Beasley:2008dc,Hayashi:2008ba,Beasley:2008kw,Donagi:2008kj},
F-theory \cite{Vafa:1996xn} has emerged as a rich and powerful
framework in which to do string model building, and more specifically
GUT model building. The main desirable feature that distinguishes
F-theory from ordinary type IIB model building at weak coupling is the
natural appearance of the $\Ytop$ Yukawa coupling in $SU(5)$ GUT
F-theory models, which in weakly coupled IIB models can only realized
via D-brane instanton effects, i.e. nonperturbatively, and is thus
expected to be fairly suppressed. Since phenomenological constraints
require the $\Ytop$ coupling associated to the top quark multiplet to
be of order one, a non-perturbative suppression factor makes realistic
model building in IIB a challenge. The situation in F-theory seems to
be better, with computations of the physical Yukawa couplings in toy
models yielding promising results
\cite{Font:2012wq,Font:2013ida,Marchesano:2015dfa,Carta:2015eoh}. To a
large extent this feature of F-theory justifies the considerable
effort spent in developing the theoretical tools behind
\mbox{F-theory} model building in the last few years, trying to
address the significantly increased technical difficulties involved in
computing important physical aspects of the backgrounds (when compared
to the weakly coupled IIB approach), such as quantum corrections to
K\"ahler potentials \cite{GarciaEtxebarria:2012zm,Minasian:2015bxa,
  Grimm:2013bha}, non-chiral spectra \cite{Bies:2014sra, Collinucci:2014taa}, D3/M5-instanton effects \cite{Anderson:2015yzz,Bianchi:2012kt,Bianchi:2012pn,Bianchi:2011qh,Blumenhagen:2010ja,Cvetic:2009ah,Cvetic:2009mt,Cvetic:2010rq,Cvetic:2012ts,Cvetic:2011gp,Donagi:2010pd,Grimm:2011sk,Grimm:2011dj,Heckman:2008es,Kerstan:2014dva,Kerstan:2012cy,Marsano:2008py,Marsano:2011nn,Martucci:2014ema}, or fluxes on 7-branes \cite{Bies:2014sra,Braun:2012nk,Braun:2011zm,Braun:2014pva,Braun:2014xka,Collinucci:2012as,Grimm:2011fx,Intriligator:2012ue,Jockers:2016bwi,Krause:2011xj,Krause:2012yh,Kuntzler:2012bu,Lin:2015qsa,Lin:2016vus,Lin:2016zha,Marsano:2010ix,Marsano:2011nn,Marsano:2012bf,Marsano:2011hv,Martucci:2015oaa,Martucci:2015dxa,Mayrhofer:2013ara,Palti:2012dd}

What we will show in this paper is that, despite superficial
appearances to the contrary, there is no qualitative distinction
between the F-theory and IIB approaches when it comes to the
generation of the $\Ytop$ coupling: when we take the F-theory models
generating the $\Ytop$ Yukawa to small coupling we reach a
complementary weakly coupled description in which the same $\Ytop$
coupling arises from a D-brane instanton effect.

We will provide strong evidence for this assertion from various dual
viewpoints via a careful technical analysis, but there are a priori
reasons to expect this connection to exist. The key observation is
that, as explained in \cite{Donagi:2009ra,Esole:2011sm}, in a specific
weakly coupled limit the neighborhood of the $E_6$ point\footnote{As
  is conventional we will often refer to the point where the $\Ytop$
  coupling is generated as the $E_6$ point. As discussed in detail in
  \cite{Esole:2011sm,Marsano:2011hv}, and reviewed in
  \S\ref{sec:strong-coupling-codim-3}, the terminology here is
  somewhat misleading, since the resolved fiber over the Yukawa point
  is not of $IV^*$ type, but we will stick to the usual nomenclature
  henceforth.\label{ft:E6-caveat}} in $SU(5)$ GUT models becomes an
orientifold of the conifold. This orientifold is peculiar in that it
projects out the small resolution mode of the conifold, so attempts to
study this configuration using ordinary singularity resolution
techniques in algebraic geometry are not applicable.

Nevertheless, the non-resolvability of the conifold singularity, in
itself, does not obstruct the existence of a perfectly sensible weakly
coupled description of the system. Holomorphicity of the Yukawa
couplings in the superpotential, together with the fact that the
weakly coupled limit is nothing but motion in complex structure moduli
space of the fourfold, suggests then that one should equally well be
able to compute the $\Ytop$ coupling at the $E_6$ point using purely
weakly coupled language. Since in the IIB description the $\Ytop$
coupling is forbidden in perturbation theory, it must be generated
non-perturbatively by D-brane instantons. These considerations lend
support to the idea that the $\Ytop$ coupling should be generated by
\mbox{D1-brane} instantons at weak coupling (as Donagi and Wijnholt
already suggested in \cite{Donagi:2009ra}). The goal of this paper
will be to show that this conclusion is indeed correct, and to
initiate the study of some of its implications.

\medskip

We have organized this work as follows. In \S\ref{sec:weak-coupling}
we review how to take the relevant weak coupling limit for a
neighborhood of the $E_6$ point. The core of our paper is
\S\ref{sec:IIB}, where we compute, in the weakly coupled IIB
description, the instanton contribution to the superpotential, showing
that as expected it generates a $\Ytop$ coupling. In particular, the
actual computation of the superpotential coupling by integration of
instanton zero modes is in \S\ref{sec:instanton-effect}. We then
rederive the same result in the mirror IIA description in
\S\ref{sec:mirror}. In \S\ref{sec:M-theory} we analyze the weak
coupling limit from the geometric M-theory viewpoint, and reproduce
some of the features of the weakly coupled analysis directly in this
language. Appendix~\ref{app:transport} contains a technical result we
use in the text, and appendix~\ref{app:globalres} reviews the M-theory
description of the $\Ytop$ coupling away from weak coupling.

\section{IIB description of the $E_6$ point}
\label{sec:weak-coupling}

In this section we briefly review the derivation in
\cite{Donagi:2009ra} for the convenience of the reader, and to set
notation. We will focus on a $SU(5)$ model with a $\bf 5$ and a
$\bf 10$ multiplet which couple via a $\Ytop$ coupling. Such couplings
can be naturally engineered in the context of local F-theory models
(in the small angle limit) by an unfolding of an exceptional
singularity. In particular, the $\bf 5$ and $\bf 10$ curves should
intersect over a point where the enhancement is of $E_6$
type.\footnote{In order to have a proper mass hierarchy coming from
  the $\Ytop$ coupling we want to have a single intersection of
  $\bf 10$ and $\bf 5$ curves, which requires the introduction of
  non-trivial T-brane data \cite{Cecotti:2010bp, Hayashi:2009ge}. This
  explains why the geometry of fiber is not exactly that of affine
  $E_6$ \cite{Esole:2011sm,Braun:2013cb}.} This can be easily achieved
by writing the local structure of the fibration in Tate form
\cite{Tate1975,Bershadsky:1996nh}
\begin{equation}
  y^2 + a_1 xyz + a_3 yz^3 = x^3 + a_2 x^2z^2 + a_4 xz^4 + a_6z^6\, .
\end{equation}
We choose to denote the transverse coordinate to the $SU(5)$ stack by
$\sigma$, so we impose
\begin{equation}
  \label{eq:Tate-SU(5)}
  a_1 = -b_5, \quad a_2 = \sigma b_4,\quad a_3 = -\sigma^2b_3, \quad
  a_4 = \sigma^3 b_2, \quad a_6 = \sigma^5 b_0
\end{equation}
with the $b_i$ generic polynomials in $\sigma$ nonvanishing at
$\sigma=0$. Following \cite{Tate1975} we introduce
\begin{align}
  \ssb_2 & = a_1^2 + 4a_2 & \ssb_8 & =
                                   \frac{1}{4}(\ssb_2\ssb_6-\ssb_4^2)\\
  \ssb_4 & = a_1a_3 + 2a_4 & \Delta & = -\ssb_2^2\ssb_8 - 8\ssb_4^3 -
                                      27\ssb_6^2 +
                                      9\ssb_2\ssb_4\ssb_6\\
  \ssb_6 & = a_3^2 + 4a_6
\end{align}
We are interested in taking this configuration to weak coupling. As
discussed in \cite{Donagi:2009ra} we can achieve this by replacing
\begin{equation}
  a_3\to \epsilon a_3, \quad a_4\to \epsilon a_4, a_6 \to \epsilon^2a_6
\end{equation}
and taking the $\epsilon\to 0$ limit. In this limit
\begin{equation}
  \Delta \sim -\frac{1}{4}\epsilon^2\ssb_2^2(\ssb_2\ssb_6 - \ssb_4^2)
  + \cO(\epsilon^3)
\end{equation}
and the string coupling goes to zero almost everywhere. The $\ssb_2=0$
component of the discriminant was identified in
\cite{Sen:1996vd,Sen:1997gv} as the location of the O7$^-$ plane at
weak coupling, and the $\ssb_2\ssb_6 - \ssb_4^2=0$ component as the
location of D7 branes. For the particular ansatz~\eqref{eq:Tate-SU(5)}
we have
\begin{align}
  \ssb_2 & = b_5^2 + 4\sigma b_4\, ,\\
  \ssb_2\ssb_6-\ssb_4^2 & = \sigma^5(4b_3^2b_4 - 4b_2b_3b_5 +
                          4b_0b_5^2 + \sigma(16 b_0b_4 - 4b_2^2))\, .
\end{align}
In this last line we identify the $\sigma^5$ component as the $SU(5)$
stack, and the rest of the expression as the flavor brane responsible
for the existence of the $\bf 5$ representation.

The Calabi-Yau threefold where the IIB theory is formulated is given
by the double cover
\begin{equation}
  \xi^2 = \ssb_2
\end{equation}
branched at the orientifold locus. We are particularly interested in
the neighborhood of the $E_6$ Yukawa coupling point, which is located
at \cite{Bershadsky:1996nh,Donagi:2009ra}
\begin{equation}
  P_{E_6} = \{b_5 = b_4 = 0\}\, .
\end{equation}
If we introduce the variables $u'=b_5$ and $w'=4b_4$ we arrive to the
conifold equation \cite{Donagi:2009ra,Esole:2011sm}
\begin{equation}
  \xi^2 = (u')^2 + \sigma w'\, .
\end{equation}
Although this describes a CY threefold with standard conifold singularities, any attempt at a small resolution will be projected out by the orientifold involution. Concretely, there would be two small resolutions, constructed by taking as an ambient space the product 
\begin{equation} \label{2smallres}
\C^4 \times \P^1: \quad \{(\xi, u', w', \sigma); [z_1: z_2] \}
\end{equation}
and as a subvariety either one of the following
\begin{equation}
\begin{pmatrix} u'-\xi & \sigma \\ -w' & u'+\xi \end{pmatrix} \cdot \begin{pmatrix} z_1\\z_2 \end{pmatrix}=0 \,, \quad {\rm or} \quad  \begin{pmatrix} u'+\xi & \sigma \\ -w' & u'-\xi  \end{pmatrix} \cdot \begin{pmatrix} z_1\\z_2 \end{pmatrix}=0\,.
\end{equation}
The involution $\xi \mapsto -\xi$ maps one small resolution into the
other. Therefore, neither one yields an orientifold-invariant smooth
threefold.

One way around this problem is to change the way one takes the weak
coupling limit in F-theory. This strategy was studied in generality in
\cite{Aluffi:2009tm}, and more specifically for this setup in
\cite{Esole:2012tf}. Although in this way one manages to have a model
with a smooth threefold, the nature of the model changes
significantly, letting the sought-for Yukawa coupling elude us.

Our strategy will be to deal with the singular space directly, using
the language of noncommutative crepant resolutions, which physically
entails studying the quiver gauge theory on D-branes probing this
conifold singularity.

For our purposes it will be convenient to define a slightly
different set of coordinates in which the flavor branes have a simpler
expression. We introduce
\begin{align}
  u & = u' + 2\frac{b_2}{b_3} \sigma\\
  w & = w' - 4\frac{b_2}{b_3} u' +4\frac{b_0}{b_3^2} (u')^2
      +\sigma\left[4\frac{b_0}{b_3^2}w' -
      4\left(\frac{b_2}{b_3}\right)^2\right]\, .
\end{align}
so that we have
\begin{equation}
  \label{eq:approximate-conifold}
  \xi^2 = u^2 + \sigma w + \ldots
\end{equation}
where we have ignored cubic and higher terms in $(u,\sigma,w)$, which
do not affect the singular behavior close to the singular point at
$\xi=u=\sigma=w=0$. The virtue of these coordinates is that they will
make manifest the algebraic structure of the $SU(5)$ and flavor stacks
close to the conifold singularity. In particular, notice that the
$SU(5)$ stack is at $\sigma=0$, while the flavor stack is at $w=0$.

The algebraic structure of these divisors is best understood in terms
of the GLSM for the conifold
\begin{equation}
\begin{tabular}{|cccc|}
$\alpha_1$ & $\alpha_2$ & $\beta_1$ & $\beta_2$\\
$1$ & $1$ & $-1$ & $-1$ 
\end{tabular}
\end{equation}
with the D-term constraint (in 2d GLSM language)
\begin{equation}
\sum_{i=1}^2 |\alpha_i|^2 - |\beta_i|^2 = \begin{cases}  t > 0, \,  \\  t < 0, \, \end{cases}\,.
\end{equation}
In the first case, the homogeneous coordinates $\alpha_i$ represent
the coordinates of the exceptional $\mathbb{P}^1$, which is located at
the locus $\beta_i=0$. This exceptional $\bP^1$ has volume given by $|t|$.

\begin{figure}
  \centering
  \includegraphics[height=4cm]{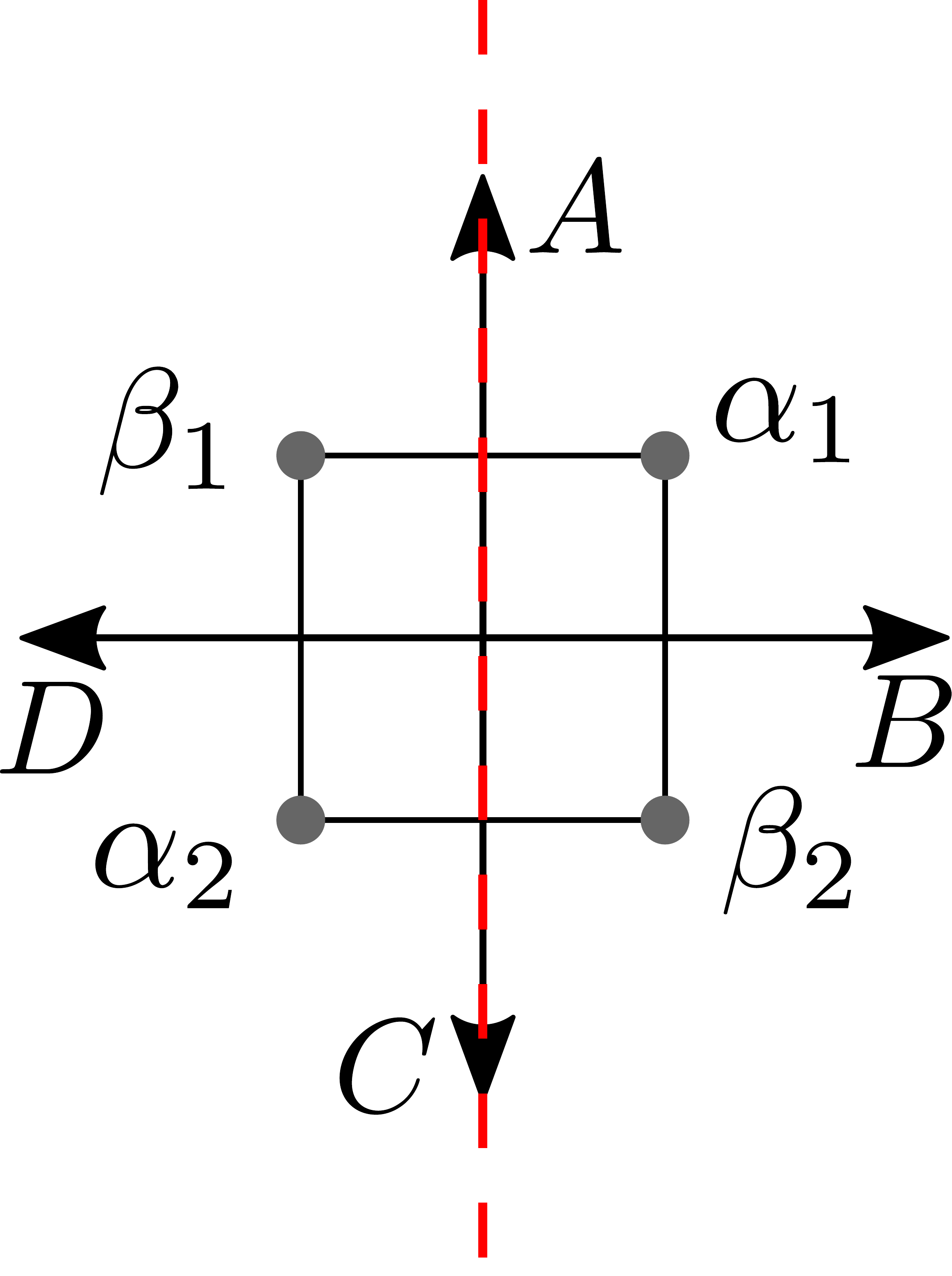}
  \caption{Toric and web diagram for the conifold. We have indicated
    the action of the orientifold involution studied in the text by
    the vertical red dashed line. The external legs in the web diagram
    are labeled for later convenience.}
  \label{fig:coniWeb}
\end{figure}

The correspondence between the toric and algebraic descriptions is explicit under the following map:
\begin{equation}
(\vec{\alpha}, \vec{\beta}) \mapsto (\xi, u, \sigma, w) = (\tfrac{1}{2}\,(\alpha_1 \beta_2-\alpha_2 \beta_1), \, \tfrac{1}{2}\,(\alpha_1 \beta_2+\alpha_2 \beta_1), \,-\alpha_1 \beta_1, \, \alpha_2 \beta_2)\,.
\end{equation}
In these variables the orientifold involution acts as follows:
\begin{equation}
  \label{eq:GLSM-orientifold}
  \alpha_i \leftrightarrow \beta_i\, .
\end{equation}
We summarize the toric data for the resulting geometry in
figure~\ref{fig:coniWeb}. Note, that this involution is not an
automorphism of the resolved conifold. Instead, it is a map from a
resolved conifold with K\"ahler modulus $t$ to a flopped conifold with
modulus $-t$.
Only for $t=0$ can we regard this as an
involution that maps the conifold into itself.

As a consequence, the orientifold involution projects out the
resolution mode of the conifold. On the other hand, the integral of
the $B$-field over the exceptional $\bP^1$ survives, since $B$ has an
intrinsic minus sign under the action of $(-1)^{F_L}\Omega$.

Coming back to the $SU(5)$ and flavor stack, notice that the
$\sigma=0$ divisor intersects the conifold at $\xi=\pm u$, which
factorizes into a stack and its image, in accordance with the fact
that the gauge symmetry is $SU(5)$. The $\{\xi=u\}\cap\{\sigma=0\}$
locus can be written in terms of GLSM coordinates as
$\{\alpha_2\beta_1=\alpha_1\beta_1=0\}$, so we associate it with the
Weil toric divisor $\beta_1=0$. Similarly, its $SU(5)$ image
$\{\xi=-u\}\cap\{\sigma=0\}$ can be written as
$\{\alpha_1\beta_2=\alpha_1\beta_1=0\}$, which is described by the
Weil divisor $\alpha_1=0$, in accordance with the orientifold
action~\eqref{eq:GLSM-orientifold}. A similar exercise for $w=0$ shows
that close to the conifold locus the flavor brane splits into a
brane-image brane pair, associated with the $\alpha_2=0$, $\beta_2=0$
pair of toric Weil divisors.

Note that the $SU(5)$ stack factorizes without any approximation
necessary, while factorization for the flavor stack at $w=0$ only
happens as we zoom into the conifold singularity, since
in~\eqref{eq:approximate-conifold} we dropped higher order terms. This
distinction is not particularly important for the analysis in the rest
of the paper. The subleading terms will affect the precise form of the
effective action, but not the existence of the instanton contribution
to the superpotential that we find, since the instanton lives at the
singular locus.

\section{IIB instanton computation via noncommutative crepant resolution}
\label{sec:IIB}

Our goal in this paper will be to understand the behavior of D-brane
instantons living at the singularity appearing at weak
coupling.\footnote{One may worry about the fact that $g_s$ formally
  diverges close to the O7 plane, which is precisely where we want to
  do our computation. This effect is not incompatible with the
  existence of a weakly coupled description of the system at any given
  energy scale. For instance, consider a D3 probe of a O7$^-$
  \cite{Banks:1996nj}. The divergence of $g_s$ on the O7 signals that
  the probe theory confines \cite{Seiberg:1994rs}, but the dynamical
  scale of this theory can be made arbitrarily small by tuning the
  ambient string coupling. The conifold can be obtained by partially
  smoothing an orbifold of this configuration, so it can also be made
  arbitrarily weakly coupled.} The main difficulty in doing this is
that, as we have just seen in the previous section, having F-theory
$SU(5)$ models with a $\Ytop$ coupling of order one means that, in the
weak coupling limit, one is forced to deal with a conifold singularity
whose resolution mode is projected out by the
orientifolding.\footnote{Note that the resolution mode \emph{had} to
  be projected out in order for the D1 instanton to have a chance of
  contributing to the superpotential. Otherwise, by resolving the
  conifold we could misalign the central charge of the D1 with respect
  to that of the $\cN=1$ background, and this would imply
  \cite{GarciaEtxebarria:2007zv,GarciaEtxebarria:2008pi} that we could
  at best generate a higher F-term
  \cite{Beasley:2004ys,Beasley:2005iu}, instead of a superpotential
  contribution.}

Indeed, this has been the main obstruction to studying these setups in
perturbative string theory, since elementary algebro-geometric methods
are not reliable on a singular space. Techniques for analyzing such
systems, based largely on mirror symmetry, are known
\cite{Hanany:2005ve,Franco:2005rj,Feng:2005gw,Franco:2006es,Franco:2007ii,Forcella:2008au},
and the result of applying such techniques will be briefly reviewed in
\S\ref{sec:mirror} below. Nevertheless the application of these
techniques involves a certain amount of heuristics (at least at the
level that they are currently developed), so in the interest of making
our derivation as assumption-free as possible, we have opted
to give a first principles derivation of the physics at the
singularity using the powerful technology of \emph{non-commutative
  crepant resolutions} \cite{Bergh:aa, Berenstein:aa} (\emph{NCCRs} in
what follows).
In practice, the NCCR approach gives us a concrete way of defining
D-branes on the singular space, without referring to a small
resolution. We will see that computing open string spectra is
surprisingly easy in this language.

Armed with the knowledge of the instanton zero-modes present in our
system, and how these couple to the background D7/D7-strings, we will
be able, in \S\ref{sec:instanton-effect}, to reproduce very
straightforwardly the $\Ytop$ interaction. We encourage the impatient
reader to skip ahead to the derivation of the $\Ytop$ coupling in
\S\ref{sec:instanton-effect}, and then return here for the systematic
justification of the basic ingredients going into the computation.

\subsection{NCCRs} \label{subsec:nccr}

Intuitively, the NCCR construction can be understood as a replacement
of the coordinate ring\footnote{In what follows we will use various
  elementary notions in category theory freely. For introductions for
  physicists see \cite{Aspinwall:2004jr,Herbst:2008jq}.}
$R = \C[u, \xi, \sigma, w]/(-\xi^2+u^2+\sigma w)$, describing the
singular space, with the ring $A$ of open string modes of probe
branes. This new ring $A$, which is also an algebra over $R$, can be
thought of as the path algebra of the quiver. It is non-commutative,
since paths cannot be composed in arbitrary order. The fact that the
ring $R$ is singular implies that one cannot describe fractional
branes easily, as these correspond to modules with infinitely long
resolutions. On the other hand, the noncommutative ring $A$ is such
that any module will admit a finite resolution. This is the essence of
the noncommutative resolution. One further demands that $A$ be
\emph{Cohen-Macaulay}. This is the ring-theoretic analog of requiring
a trivial canonical bundle, leading to an noncommutative
\emph{crepant} resolution (NCCR).

We define the conifold threefold algebraically by the equation
\begin{equation}
  -\xi^2 + u^2 + \sigma \,w \,.
\end{equation}
This variety has a coordinate ring $R = \mathbb{C}[\xi, u, \sigma, w]/(u^2-\xi^2+\sigma w)$, and admits a so-called \emph{matrix factorization}, a pair of square matrices\footnote{Technically, a matrix factorization is an ordered pair of matrices. Hence, given a pair $(\phi, \psi)$, we also have $(\psi, \phi)$ as another matrix factorization.} $(\phi, \psi)$ such that $\phi \cdot \psi =\psi \cdot \phi  = (u^2-\xi^2+\sigma w) \cdot \1$. From these two matrices, we can define two so-called  \emph{maximal Cohen-Macaulay} (MCM) modules over R. Essentially these are $R$-modules defined as the cokernels of the matrices
\begin{equation}
  \phi = \begin{pmatrix} u-\xi & \sigma \\ -w & u+\xi \end{pmatrix} \,, \quad {\rm and} \quad \psi = \begin{pmatrix} u+\xi & -\sigma \\ w & u-\xi  \end{pmatrix} \, .
  \end{equation}
We define
\begin{eqnarray}
M&\equiv&  {\rm coker}\big(\,R^{\oplus 2} \stackrel{\psi}\longrightarrow R^{\oplus 2} \, \big)\, ,\\
\tilde M&\equiv&  {\rm coker}\big(\,R^{\oplus 2} \stackrel{\phi}\longrightarrow R^{\oplus 2} \, \big)\, .
\end{eqnarray}
For the conifold, these are all the non-trivial irreducible MCM
modules up to isomorphism. They are akin to line-bundles
over the conifold\footnote{After choosing an appropriate small
  resolution, these pullback to $\cO(1)$ and $\cO(-1)$,
  respectively.}, except that they fail to be locally free at the
singularity. This can be understood by noticing that the matrices $\psi, \phi$ have rank one on the conifold, leaving a one-dimensional cokernel over every non-singular point. At the singularity, the matrices vanish, and the cokernels jump to dimension two. So we can think of these as Calabi-Yau filling branes with an added point-like brane at the origin. 

These modules $M$ and $\tilde M$ over $R$ can be fit into exact complexes as follows:
\begin{equation}
  \includegraphicstikz{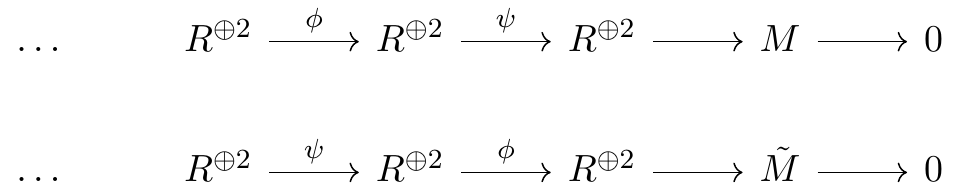}
\end{equation} 
We can think of $M$ and $\tilde M$ as the cokernels of the maps $\psi$ and $\phi$, respectively. Henceforth, we
will replace $M$ or $\tilde M$ by their resolution complex as follows:
\begin{equation}
  \includegraphicstikz{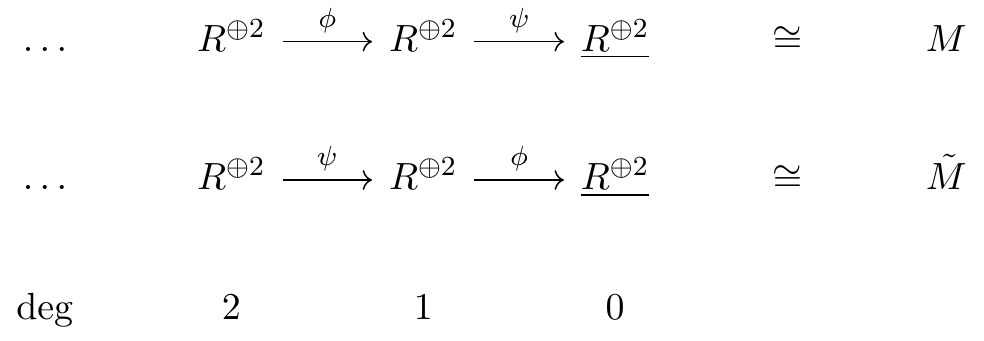}
\end{equation} 
For our complexes, we use cohomological degree staring at zero on the
right, and increasing as we move left. We will underline the zeroth
position for clarity. The isomorphisms here state that our modules are
simply the $H^0$ modules for these complexes.

Note, that these resolutions are semi-infinite, a hallmark of singular spaces. The goal of an NCCR is to replace $R$ with a ring such that all modules admit finite resolutions.

The NCCR for the singular ring $R$ is defined by picking one of these two modules, say $M$, and defining the endomorphism algebra $A = {\rm End}(R \oplus M)$. $A$ turns out to be a noncommutative ring that will serve as our NCCR. It can be decomposed into four pieces
\begin{equation}
A = \underbrace{{\rm Hom}(R, R)}_{\cong R} \quad \oplus \quad \underbrace{{\rm Hom}(M, M)}_{\cong R} \quad \oplus \quad \underbrace{ {\rm Hom}(R, M)}_{\cong M} \quad \oplus \quad \underbrace{{\rm Hom}(M, R)}_{\cong \tilde M}\, ,
\end{equation}
and can be encoded as the path algebra of the following quiver with relations:
\begin{center}
  \includegraphics{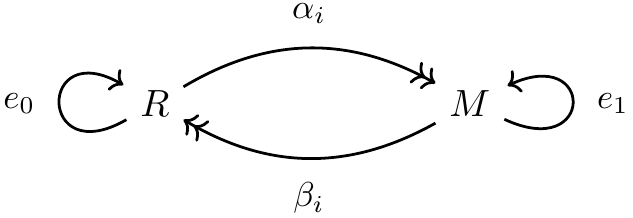}
\end{center}
Here, the arrows are morphisms that can be represented as cochain maps
between resolution complexes. Take two complexes
$K_\bullet, C_\bullet$. Concretely, an element in
Hom$(K_\bullet, C_\bullet)$ corresponds to a collection of vertical
maps $f_\bullet$

\begin{equation} \label{homotopydef}
  \includegraphicstikz{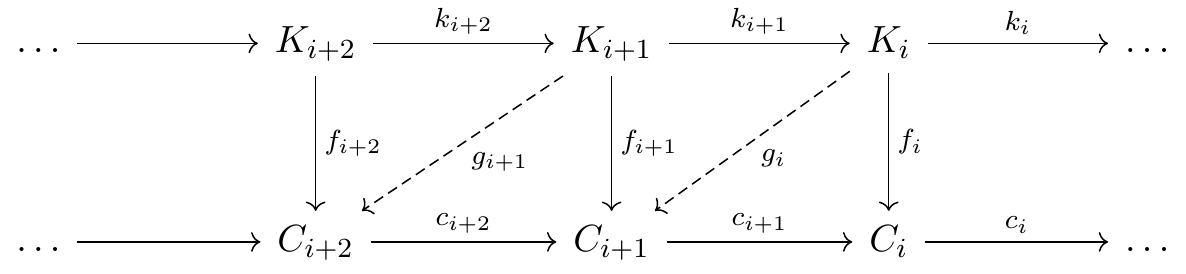}
\end{equation}
such that each square commutes, i.e.
$ f_i \circ k_{i+1} = c_{i+1}\circ f_{i+1}$, and modulo homotopies,
i.e. $f_{i+1} \sim f_{i+1}+ g_{i} \circ k_{i+1}-c_{i+2} \circ g_{i+1}$
for some collection of the diagonal dotted $g_i$. These notions are
reviewed in many physics papers, see \cite{Aspinwall:2004jr} for a
general account, and \cite{Collinucci:2014taa,Collinucci:2014qfa} for
concrete examples.

Back to the conifold, the $\alpha_{1, 2}: R \longrightarrow M$ can be written as follows
\begin{equation}
  \includegraphicstikz{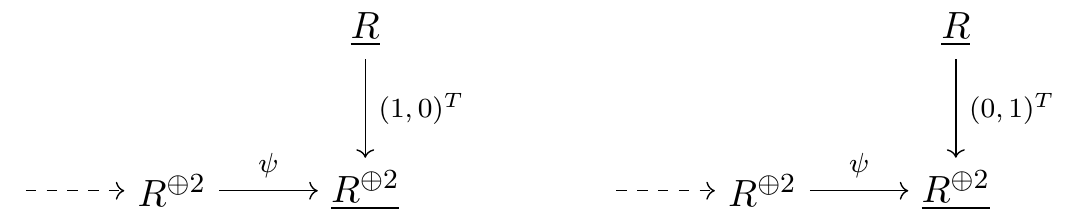}
\end{equation}
and for $\beta_{1, 2}: M \longrightarrow R$
\begin{equation}
  \includegraphicstikz{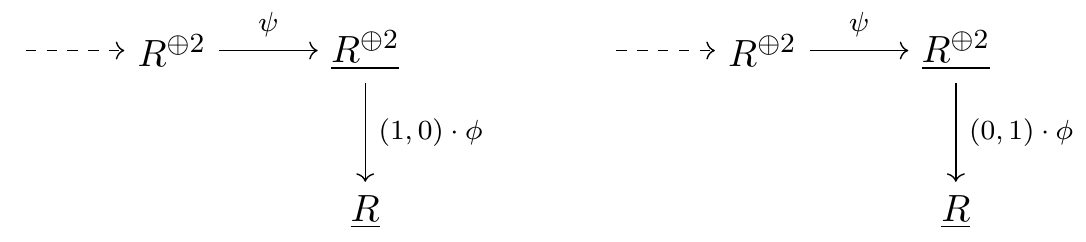}
\end{equation}
Finally, $e_0 \in {\rm Hom}(R, R) \cong R$ and $e_1 \in {\rm Hom}(M, M) \cong R$ are the multiplicative identities (actually idempotents) of the endomorphism ring of each node.
These morphisms satisfy the relations:
\begin{equation}
\alpha_1  \beta_i \alpha_2 = \alpha_2  \beta_i \alpha_1\,, \quad {\rm and} \quad \beta_1 \alpha_i \beta_2 = \beta_2 \alpha_i \beta_1\, \quad {\rm for} \quad i=1, 2\,,
\end{equation}
where composition is defined from right to left.
For instance, we see that
\begin{eqnarray}
\alpha_1 \beta_1 \alpha_2 &=& \begin{pmatrix} 1 \\ 0 \end{pmatrix} \cdot (1, \, 0) \cdot \phi \begin{pmatrix} 0 \\ 1 \end{pmatrix} = \sigma \begin{pmatrix} 1 \\ 0 \end{pmatrix}\\ 
\alpha_2 \beta_1 \alpha_1 &=& \begin{pmatrix} 0 \\ 1 \end{pmatrix} \cdot (1, \, 0) \cdot \psi \begin{pmatrix} 1 \\ 0 \end{pmatrix} = (u-\xi) \begin{pmatrix} 0 \\ 1 \end{pmatrix}
\end{eqnarray}
The two column vectors differ by an element of the image of $\psi$:
\begin{equation}
\alpha_2 \beta_1 \alpha_1 - \alpha_1 \beta_1 \alpha_2 = \begin{pmatrix} -\sigma \\ u-\xi \end{pmatrix} = \psi \begin{pmatrix} 0 \\ 1\end{pmatrix}\,.
\end{equation}
Such a morphism is discarded in the \emph{homotopy category} as being gauge-equivalent to zero, and it actually corresponds to the zero morphism at the level of the cohomology.

The relations can be encoded in the superpotential
$W = \alpha_1 \beta_1 \alpha_2 \beta_2 - \alpha_1 \beta_2 \alpha_2
\beta_1$ which we recognize as the superpotential of the
Klebanov-Witten theory \cite{Klebanov:1998hh}.

The noncommutative crepant resolution is then simply defined as the ring $A$, which can be identified with the path algebra of the quiver modulo the relations derived from the superpotential. The product of the ring is the concatenation of arrows, and the sum is simply taking complex linear combinations of arrows.

D-branes are described in this formulation as complexes of right
$A$-modules. More precisely, the bounded derived category
D$^b({{\rm mod-}A})$ comprises complexes of $A$-modules modulo certain
equivalences known as quasi-isomorphism. The interest in the NCCR
stems from the fact that this category D$^b({{\rm mod-}A})$, which is
defined entirely in the singular geometry, is known to be derived
equivalent to the bounded derived category of coherent sheaves of the
resolved conifold in both resolution phases. By studying D-branes
through the NCCR, we sit in the middle between the two flopped phases,
and are never forced to break our orientifold
involution.

\subsection{Quiver representations}

$A$-modules are equivalent to representations of the quiver. We will
be interested in two kinds of modules:
\begin{enumerate}
\item Modules that correspond to
D7-branes. These can be regarded as infinite-dimensional
representations.
\item Modules corresponding to fractional D$(-1)$-branes, which means
  D1 and anti-D1-branes wrapping the vanishing $\P^1$ of the
  conifold. These correspond to finite-dimensional representations.
\end{enumerate}
We now explain how to construct these modules.

\subsubsection{Recovering the conifold}

As a warmup, we will study the finite-dimensional representation
corresponding to a complete (non-fractional)
D$(-1)$-brane. Physically, we expect the moduli space of this brane to
be its transverse space, i.e. the conifold. By switching on FI terms
in the quiver gauge theory, we can reach both flopped phases.

A (finite-dimensional) quiver representation consists in assigning a vector space to each node of the quiver, and promoting the arrows to linear maps between the vector spaces it connects. The idempotents $e_0$ and $e_1$ are set to one and left out of the diagram. In our case, we are interested in the following representation:

\begin{center}
  \includegraphics{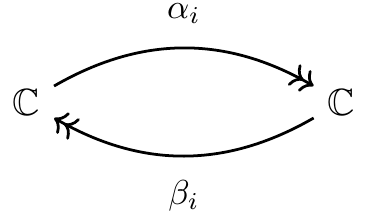}
\end{center}
Denoting a representation by its dimension vector $\vec{d}$, in this
case we have $\vec{d}=(1, 1)$. We can interpret this as a single
D$(-1)$-brane splitting up into one fractional brane on each node,
leading to an abelian quiver theory. Here, the $\alpha_i$ and
$\beta_i$ are simply complex numbers that transform in the
(anti-)bifundamental of the relative $U(1) \subset U(1)\times
U(1)$. In order to study the moduli space of $\vec{d}=(1,1)$
representations, we quotient out by the action of the relative
$U(1)$. This leads us to a toric variety
\begin{equation}
\arraycolsep=10pt
\begin{array}{|c|c|c|c|} 
\hline
\alpha_1 &\alpha_2 &\beta_1 &\beta_2\\ \hline
1 & 1 & -1 & -1 \\ \hline
\end{array} 
\end{equation}
We must also impose a D-term constraint
\begin{equation}
|\alpha_1|^2+|\alpha_2|^2-|\beta_1|^2-|\beta_2|^2 = t\,.
\end{equation}
The resolved phases correspond to $t >0$ and $t<0$. In our case, the
orientifold involution~\eqref{eq:GLSM-orientifold} acts as
$t \mapsto -t$, so it is not possible to respect it for non-zero $t$.

\subsubsection{Non-compact branes}

There are two basic infinite-dimensional representations $P_0$ and $P_1$, which are the projective right $A$-modules, defined as follows:
\begin{eqnarray}
P_0 &=& e_0 \cdot A = \{ \text{linear combinations of all paths ending on the lhs node `0'}\}\\ 
P_1 &=& e_1 \cdot A =  \{ \text{linear combinations of all paths ending on the rhs node `1'}\}
\end{eqnarray}

These can be thought of as CY-filling branes with line bundles on them.
From these two modules, we can build all branes of interest through complexes. Let us now define how the orientifold involution $\sigma$ acts on a complex. Take a complex $K_\bullet$ of the form
\begin{equation}
  \includegraphicstikz{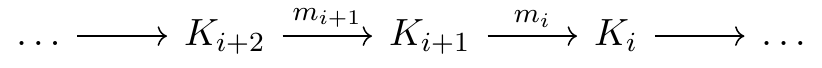}
\end{equation}
where the cohomological degree increases from left to right. Then, the orientifold image of this complex is another complex $\tilde K_\bullet$ defined through the mapping $\Sigma$ which maps to objects of a complex as follows:
\begin{equation} \label{sigmarules1}
\Sigma: \quad K_{i} \longrightarrow \tilde K_{i} \equiv \sigma^*(K_{1-i})\,,
\end{equation}
where depending on the choice of orientifold involution $\sigma^*(K_{1-i})$ may further act on the object $K_{1-i}$ (we will determine the proper action below), and it acts on the maps as follows:
\begin{equation}\label{sigmarules2}
\Sigma: \quad m_{i} \longrightarrow \tilde K_{i} \equiv \sigma^*(m_{i})^T\,.
\end{equation}

Concretely, on a complex $K_\bullet$, we have the following action:
\begin{equation} 
  \includegraphicstikz{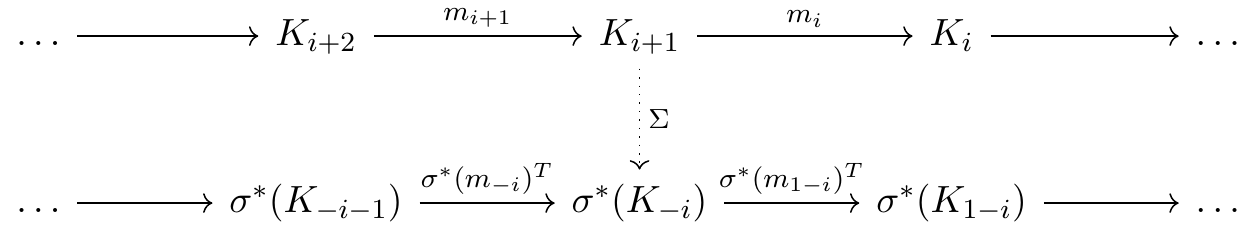}
\end{equation}

Let us now define the non-compact D7-branes of interest, which are supported over divisors. We will introduce a stack of brane/image-brane pairs called $G_0$ and $G_1$, where our $SU(5)$ theory will live:
\begin{equation}
G_0: \quad P_1 \stackrel{\beta_1}\longrightarrow \underline{P_0}\,,
\qquad G_1: \quad P_0 \stackrel{\alpha_1}\longrightarrow
\underline{P_1}\, .
\end{equation}
We will also need a `flavor' brane/image-brane pair:
\begin{equation}
F_0: \quad P_1 \stackrel{\beta_2}\longrightarrow \underline{P_0}\,,
\qquad F_1: \quad P_0 \stackrel{\alpha_2}\longrightarrow
\underline{P_1}\, .
\end{equation}

We can easily see that $\Sigma: G_0 \mapsto G_1$, and
$\Sigma: F_0 \mapsto F_1$ by applying the rules in \eqref{sigmarules1}
and \eqref{sigmarules2}, if we assume that
$\sigma^*(P_{0, 1}) = P_{0, 1}$, which is the choice compatible with
the geometric action of our orientifold (as will become clear
momentarily, when we map these complexes to sheaves).

How can we interpret these modules? One way is to study what happens when we promote the maps in the complexes to coordinates in the moduli space of the $\vec{d}=(1,1)$ representation. As we saw in the previous section, the $\alpha_i$ and $\beta_i$ become toric homogeneous coordinates for the conifold. Hence, we clearly see by inspection that, for instance, $G_0$ will correspond to a D-brane supported on the divisor $\beta_1=0$. The other piece of information we need is how to transport the projective modules to $P_0$ and $P_1$ to sheaves on the moduli space. In this case, one choice is to send
\begin{equation}
P_0 \mapsto \mathcal{O}\,, \qquad P_1 \mapsto \mathcal{O}(1)\,.
\end{equation}
Hence, we conclude that our modules are mapped to sheaves as follows:
\begin{eqnarray} \label{noncompmapping}
G_0:&& \quad \cO(1) \stackrel{\beta_1}\longrightarrow \underline{\cO}\,, \qquad G_1: \quad \cO \stackrel{\alpha_1}\longrightarrow \underline{\cO(1)}\\ \nonumber
F_0:&& \quad \cO(1) \stackrel{\beta_2}\longrightarrow \underline{\cO}\,, \qquad F_1: \quad \cO \stackrel{\alpha_2}\longrightarrow \underline{\cO(1)}
\end{eqnarray}

\subsubsection{Fractional branes from the $(1,0)$ and $(0,1)$}
The fractional branes, which in our case will be fractional D$(-1)$-instantons, correspond to the two \emph{simple} (i.e. not admitting subrepresentations), one-dimensional representations: $S_0$ with $\vec{d} = (1,0)$, and $S_1$ with $\vec{d} = (0,1)$. As $A$-modules, these are simply defined as the modules generated by $e_0$ and $e_1$, respectively:
\begin{eqnarray}
S_0 &=& \C\langle e_0 \rangle = \{\text{linear combinations of unique loop of length zero at node `0'} \}\\
S_1 &=& \C\langle e_1 \rangle = \{\text{linear combinations of  unique loop of length zero at node `1'} \}
\end{eqnarray}
As quiver representations, each consists of a node with a $\C$, and
the self-arrow $e_0 = 1$. Hence, these have no moduli. This is a
reflection of the fact that these are fractional branes wrapping a
rigid vanishing curve. These fractional branes admit the following
projective resolutions:
\begin{equation} \label{fractionalmappings}
  \includegraphicstikz{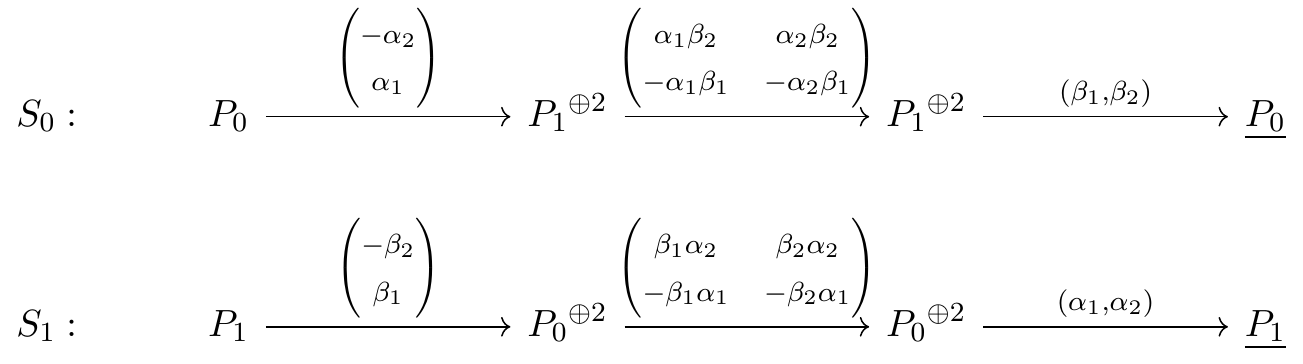}
\end{equation}
In appendix~\ref{app:transport} we show that these transport to D1 and
anti-D1 branes wrapping the vanishing $\P^1$.

We can compute the Ext$^1$'s between these complexes by brute force. However, there is a much quicker way to do this, which exploits the fact that these are finite-dimensional representations. Suppose we want to compute Ext$^1(S_1, S_0)$. This is equivalent to the so-called Yoneda-Ext group that parametrizes all possible non-trivial extensions of $S_0$ by $S_1$, i.e., exact sequences of the form
\begin{equation}
S_1 \stackrel{a}\longrightarrow E \stackrel{b}\longrightarrow S_0
\end{equation}
such that $E \not \cong S_0 \oplus S_1$. Clearly, the middle object $E$, which will depend on the choice of maps $(a, b)$, must always have dimensions $\vec{d} = (1, 1)$. Hence, we can draw an exact sequence of quiver representations as follows:
\begin{equation}
  \includegraphicstikz{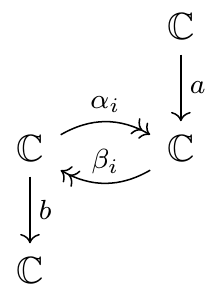}
\end{equation}
In this sequence, we have omitted the zeroes corresponding to empty quiver nodes. However, these are important, because arrows going to and from empty nodes are the zero map. A sequence of quiver representations means all squares must commute. In the upper $(0,1)$ quiver representation, the arrows going to the left are zero. Hence, we must impose that the composition $\beta_i \circ a =0$ for $i=1,2$, which implies $\beta_i=0$. 

For each choice of the pair $\alpha_i$ we have an exact sequence of
quiver representations. Hence,
Ext$^1(S_1, S_0) \cong \mathbb{C}^2 \cong (\alpha_1, \alpha_2)$. For
the zero element of this group, i.e. $\vec{\alpha}=0$, we see indeed
that the middle element is isomorphic to $S_0 \oplus S_1$. We can take
this one step further by modding out these maps by the $\mathbb{C}^*$
that acts homogeneously on these two maps, and excluding the trivial
map. This yields a $\mathbb{P}^1$. We can think of this as the
exceptional curve after resolving, since it corresponds to the moduli
space for the recombination tachyon between an D1 and and anti-D1 with
a net flux difference of $\mathcal{O}(1)$. In other words, we could
think of this $\mathbb{P}^1$ as the moduli space of the induced
D$(-1)$-brane trapped on this pair. Alternatively, in field theory
terms we have that the fact that we have $\beta_i=0$ implies that all
mesons vanish, so we are stuck at the exceptional locus of the
resolution.

Similarly, we deduce that Ext$^1(S_0, S_1) \cong
\mathbb{C}^2$.
Finally, we will define our fractional branes as follows:
\begin{equation}
I_0 \equiv S_0[-1]\,, \qquad I_1 \equiv S_1[-1]\,.
\end{equation}
Where we have shifted our complexes one position to the right (in the
conventions of \cite{Aspinwall:2004jr}) to take into account that the
D1 brane mutually supersymmetric with respect to a D7 brane wrapping
an arbitrary cycle is, from the perspective of sheaves, the anti-brane
of the skyscraper sheaf \cite{Bershadsky:1995qy}. The spectrum between
these two will not change under a simultaneous shift. However, the
spectrum with the non-compact branes will.

\subsection{Ext's}

Let us now compute the Ext groups involving the flavor branes. We have
already established in the previous section that:
\begin{equation}
{\rm Ext}^1(I_0, I_1) \equiv {\rm Ext}^1(I_1, I_0) \equiv \mathbb{C}^2\,.
\end{equation}
Let us now calculate the spectra involving the non-compact branes. We
will do a few sample calculations and then list the results for all
sectors. Let us start with the spectrum between $G_0$ and $I_0$. It is
given by the Ext$^i(G_0, I_0)$ groups which correspond to cochain
maps, whereby the degree $i$ instructs us to shift the second complex
to the right as follows:
\begin{equation}
{\rm Ext}^i(G_0, I_0) \equiv {\rm Hom}_{D^b({\rm mod-}A)}(G_0, I_0[i])\, .
\end{equation}
The fact that this is computed in the derived category of modules
means that we have to quotient out by quasi-isomorphisms. Fortunately,
it is known fact that if a ring has enough projectives, as it
is the case here, it is sufficient to consider projective resolutions
and quotient out by cochain homotopies (explained around formula \eqref{homotopydef}). 

Let us then compute Ext$^1(G_0, I_0)$. Since $I_0 = S_0[-1]$, we see that Ext$^1(G_0, I_0) \cong$ Ext$^0(G_0, S_0)$. So the task is to find the vertical maps $(f_0, f_1)$ in the following diagram:
\begin{equation}
  \includegraphicstikz{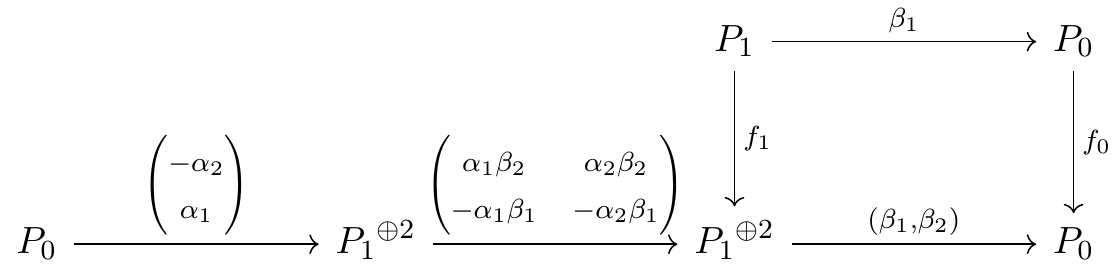}
\end{equation}
Let us start with $f_0$. For consistency, it is necessarily of the
form
\begin{equation}
  f_0 = z + (\ldots) \circ \beta_1+(\ldots) \circ \beta_2
\end{equation}
with $z$ a complex number. The elided parts can be arbitrarily
complex, but since they are composed with $\beta_i$ on the right they
always give contributions to $f_0$ which are homotopy equivalent to
zero, so we can simply forget about these and set $f_0 = z$.  Then, in
order for the square to commute, the only solution is
$f_1 = (z, 0)^T$, which is also non-trivial w.r.t. homotopy.  In
conclusion, Ext$^1(G_0, I_0) \cong \C$, where $\C$ is parametrized by
the number $z$. The rest of the D1/D7 Ext groups are computed in the
same manner.

We list the Ext groups between fractional and non-compact branes here:
\begin{eqnarray}
{\rm Ext}^i(G_0, I_0) &=& (0, \mathbb{C}, 0, 0)\,, \qquad {\rm Ext}^i(G_0, I_1) = (0,0, \mathbb{C}, 0)\,, \\
{\rm Ext}^i(G_1, I_0) &=& (0,0, \mathbb{C}, 0)\,, \qquad {\rm Ext}^i(G_1, I_1) = (0, \mathbb{C}, 0, 0)\,, \\
{\rm Ext}^i(F_0, I_0) &=& (0, \mathbb{C}, 0, 0)\,, \qquad {\rm Ext}^i(F_0, I_1) = (0,0, \mathbb{C}, 0)\,, \\
{\rm Ext}^i(F_1, I_0) &=& (0,0, \mathbb{C}, 0)\,, \qquad {\rm Ext}^i(F_1, I_1) = (0, \mathbb{C}, 0, 0)\,.
\end{eqnarray}

Since the spectrum between two non-compact branes is infinite-dimensional, the calculation is qualitatively different. Let us perform a sample calculation, say Ext$^1(G_0, G_1)$. It consists in the set of vertical maps $f_1 \sim f_1 +g_0\circ \beta_1-\alpha_1 \circ g_1$ modulo homotopies given by $(g_0, g_1)$
\begin{equation}
  \includegraphicstikz{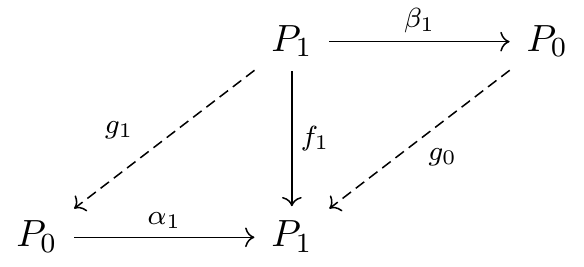}
\end{equation}
Here, we see that any loop from node `1' to itself not made of
$\beta_1$ at its start or $\alpha_1$ at its end is allowed. In other
words, we can take any linear combination of any power of
$\beta_2 \circ \alpha_2$. Hence, we conclude that our Ext group is a
polynomial ring Ext$^1(G_0, G_1)\cong \C[\beta_2 \alpha_2]$. This is
nothing other than the ring of polynomials on the complex plane. This
reflects the fact that our D7-branes intersect over a complex plane,
and the low energy bifundamental strings are simply fields defined on
it. In this non-compact setup, this spectrum is an
infinite-dimensional vector space. However, upon compactifying the
intersection curve of the two D7-branes, the spectrum will become
finite-dimensional, and upon switching on an appropriate flux, it will
be chiral. Note that this information is \emph{input} from the point
of view of the theory at the conifold singularity. Below we will
choose this input to be that appearing in F-theory GUT models.

\begin{figure}
  \centering
  \includegraphics[width=0.6\textwidth]{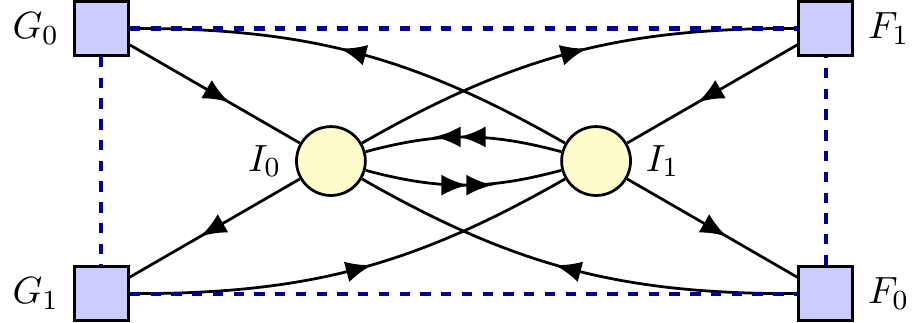}
  \caption{Quiver representation for the theory discussed in the text,
    before taking the orientifold involution (described by a
    reflection with respect to a horizontal line through the quiver
    nodes, as described in \S\ref{sec:orientifold-projection}). We
    have indicated the matter between the flavor branes by the blue
    dashed line, the actual matter content coupling to the quiver
    degrees of freedom will depend on choices of flux away from the
    singularity.}
  \label{fig:full-quiver}
\end{figure}

The other D7/D7 spectra are computed in the same manner, with the
results:
\begin{eqnarray}
{\rm Ext}^i(G_0, G_1) &=& (0, \mathbb{C}[\alpha_2 \beta_2], 0, 0)\,, \qquad {\rm Ext}^i(G_0, F_1) = (0, \mathbb{C}[\alpha_1 \beta_2], 0, 0)\,, \\
{\rm Ext}^i(G_1, G_0) &=& (0, \mathbb{C}[\beta_2 \alpha_2], 0, 0)\,, \qquad {\rm Ext}^i(G_1, F_0) = (0, \mathbb{C}[\beta_1 \alpha_2], 0, 0)\,, \\
{\rm Ext}^i(F_0, F_1) &=& (0, \mathbb{C}[\alpha_1 \beta_1], 0, 0)\,, \qquad {\rm Ext}^i(F_0, G_1) = (0, \mathbb{C}[\alpha_2 \beta_1], 0, 0)\,, \\
{\rm Ext}^i(F_1, F_0) &=& (0, \mathbb{C}[\beta_1 \alpha_1], 0 , 0)\,, \qquad {\rm Ext}^i(F_1, G_0) = (0, \mathbb{C}[\beta_2 \alpha_1], 0, 0)\,.
\end{eqnarray}
The resulting quiver is shown in figure~\ref{fig:full-quiver}.

\subsection{Orientifold projection}
\label{sec:orientifold-projection}

\begin{figure}
  \centering
  \begin{subfigure}[b]{0.4\textwidth}
    \centering
    \includegraphics[height=5cm]{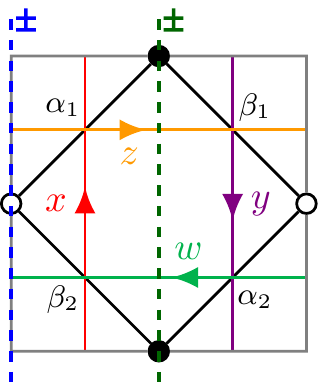}
    \caption{Action on the dimer model.}
    \label{sfig:orientifolded-conifold-dimer}
  \end{subfigure}
  \hspace{1cm}
  \begin{subfigure}[b]{0.4\textwidth}
    \centering
    \includegraphics[width=\textwidth]{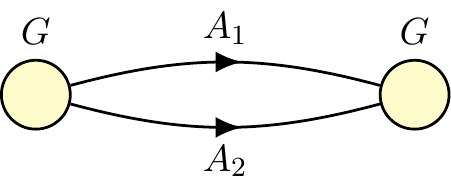}
    \caption{Resulting quiver.}
    \label{sfig:orientifolded-conifold-quiver}
  \end{subfigure}
  \caption{Action of the orientifold studied in the text on the quiver
    theory of the conifold. \subref{sfig:orientifolded-conifold-dimer}
    The action on the dimer model
    \cite{Hanany:2005ve,Franco:2005rj,Feng:2005gw} is given by a line
    orientifold, in the terminology of \cite{Franco:2007ii}. The
    choice of equal signs for the two fixed lines is associated with
    having an O7 plane instead of an O3
    plane. \subref{sfig:orientifolded-conifold-quiver} Resulting
    quiver gauge theory. We will determine in the text that
    $G=\Sp(2N)$.}
  \label{fig:orientifolded-conifold}
\end{figure}

We would now like to understand the nature of the orientifold
projection acting on the singular conifold. I.e. which projection
should we take on the quiver found in
figure~\ref{fig:full-quiver}. Recall from \S\ref{sec:weak-coupling}
that, for a conifold defined as
\begin{equation}
  \xi^2=u^2 + \sigma w
\end{equation}
the orientifold involution acted geometrically as $\xi\to-\xi$,
leaving the other coordinates invariant. If we rewrite this conifold
as $xy=zw$, with $(x,y,z,w)=(u+\xi, u-\xi,-\sigma,w)$, the
orientifold action is
\begin{equation}
  \label{eq:meson-action}
  x\leftrightarrow y \quad ; \quad z \leftrightarrow z \quad ; \quad w
  \leftrightarrow w\, .
\end{equation}
For the $(1,1)$ representation of the quiver, these geometric
coordinates are in one-to-one correspondence with vevs of the
elementary mesons of the quiver theory. The conifold theory admits
various involutions, but it is easy to see (from the classification in
\cite{Franco:2007ii}, for instance) that the only class of involutions
compatible with the action~\eqref{eq:meson-action} on the mesons is
the action shown in figure~\ref{sfig:orientifolded-conifold-dimer}. In
principle the projection associated to the two fixed lines can be
independently chosen, but as discussed in detail in
\cite{Garcia-Etxebarria:2015lif} the choice with opposite relative
signs corresponds to an O3 plane, instead of an O7 plane, so we need
to choose the two fixed lines to have the same sign. The resulting
theory is depicted in figure~\ref{fig:orientifolded-conifold}. We
still have the ambiguity of whether the gauge group on the nodes is of
type $SO(N)$ or $\Sp(2N)$.\footnote{We could choose the rank of the
  two nodes to be different, but this possibility plays no role in our
  analysis.}

One can easily see that the right choice is $G=\Sp(2N)$ by a probe
argument, as follows.  In our setup we have an O7$^-$ plane coming out
of the conifold point. We know that in flat space a single D3 brane on
top of an O7$^-$ gives rise to a four dimensional $\cN=2$ $\Sp(2)$
theory (it would be a $SO(2)$ algebra for a O7$^+$). This will also be
true at low energies for a D3 on the orientifolded conifold
background, as long as the D3 is on top of the O7$^-$ plane but away
from the singular point. This implies that there should be a
submanifold on the moduli space of the $G\times G$ theory for $N=1$
where at low energies one recovers the $\cN=2$ $\Sp(2)$ theory. A
single probe brane on the orientifolded conifold background will have
gauge group $SO(2)\times SO(2)$ or $\Sp(2)\times \Sp(2)$, since it
corresponds to two D3 branes on the covering space. Out of these two
choices, the only one that can Higgs to a $\Sp(2)$ subgroup anywhere
in the moduli space is $\Sp(2)\times \Sp(2)$, so we conclude that the
right projection is $\Sp(2N)\times \Sp(2N)$.

The fact that we obtain a $\Sp(2N)\times \Sp(2N)$ projection on the
quiver will be essential in order to be able to understand the $E_6$
Yukawa coupling in the superpotential as coming from an instanton: in
the F-theory background we have no gauge dynamics localized at the
Yukawa point, so we need to set $N=0$, and $\Sp(0)$ nodes in empty
quivers are precisely those which can give rise to stringy instanton
effects
\cite{Aganagic:2003xq,Intriligator:2003xs,Aganagic:2007py,GarciaEtxebarria:2008iw}. The
microscopic reason for this is that the projection on spacetime
filling branes and the corresponding instantonic branes is reversed
\cite{Argurio:2007qk,Argurio:2007vqa,Bianchi:2007wy,Ibanez:2007rs}
(see also \cite{Blumenhagen:2009qh} for a review), so a $\Sp$
projection on fractional D3 branes implies the existence of an
orthogonal projection on the fractional D$(-1)$ branes. This is
precisely the projection necessary for eliminating the neutral
$\ov\tau$ modes on D-brane instantons, and allowing for the existence
of a superpotential contribution.\footnote{It is also clear that the
  fractional D$(-1)$ has no deformation modes, so the only zero modes
  to worry about, in order to make sure that one has a bona fide
  superpotential contribution (as opposed to a higher F-term
  \cite{Beasley:2004ys,Beasley:2005iu}), are the charged zero
  modes. We will deal with these in the next section.}
\medskip

The previous argument is somewhat indirect, and it may be illuminating
to see explicitly the enhancement in moduli space. In the rest of this
section we do this exercise. (The reader not interested in this
derivation should feel free to skip ahead to the next section.)

In addition to the quiver in
figure~\ref{sfig:orientifolded-conifold-quiver}, there is a
superpotential of the form \cite{Imai:2001cq}
\begin{equation}
  \label{eq:G-superpotential}
  W = \varepsilon_{ij} \varepsilon_{lm} \Tr(\gamma_G A_i \gamma_G A_l^t
  \gamma_G A_j \gamma_G A_m^t)\, .
\end{equation}
If $G$ is $\Sp(2N)$ then $\gamma_g=\diag(\sigma_2,\ldots,\sigma_2)$
with $\sigma_2$ the Pauli matrix
$\left(\begin{smallmatrix}0&-i\\i&0\end{smallmatrix}\right)$, and for
$G=SO(M)$ we have $\gamma_G=\1$. Notice that for $G=\Sp(2)$ the
superpotential~\eqref{eq:G-superpotential} identically
vanishes,\footnote{This can be argued as follows. For any $2 \times 2$
  matrix $A_i$, we have that
  $\sigma_2 A_i^t \sigma_2 = \Tr(A_i) \cdot \mathbb{I}-A_i$. By the
  Cayley-Hamilton theorem this implies that
  $A_i \cdot \sigma_2 A_i^t \sigma_2 = \sigma_2 A_i^t \sigma_2 \cdot
  A_i = \det(A_i) \cdot \mathbb{I}$.  Direct substitution
  into~\eqref{eq:G-superpotential} then shows that $W$ vanishes.} so
we only need to worry about D-terms. Decompose
$A_1=a^{(1)}_\mu \sigma_\mu$, with $\sigma_\mu=(\1,i\sigma_i)$ and
$\sigma_i$ the Pauli matrices. Elements of $\Sp(2)\times \Sp(2)$ act
on $a^{(1)}_\mu$ as $SO(4)$ rotations. Note that these rotations act
independently on the real an imaginary parts of $a_\mu$. Let us go to
a gauge in which $\Re(a^{(1)}_\mu)=(a_0,0,0,a_3)$ with
$(a_0,a_3)\neq (0,0)$. (This will be the generic case for either the
real or imaginary part of one of the $A_i$ fields, we choose to talk
about the real part of $A_1$ for concreteness. It may be the case that
$a_0=0$ or $a_3=0$ for all $A_i$ at certain points in moduli space, we
will analyze these points momentarily.)

Non-abelian D-terms for this theory are given by
\begin{equation}
  \label{eq:non-abelian-D-terms}
  \sum_i \Tr(A_i^\dagger A_i \sigma_k) = \sum_i \Tr(A_i A_i^\dagger
  \sigma_k) = 0
\end{equation}
for any of the three Pauli matrices $\sigma_k$. With the gauge choice
above, these imply that we can write
\begin{equation}
  A_i = z_i^0 + iz_i^3\sigma_3\, .
\end{equation}
We see that the $\Sp(2)\times\Sp(2)$ gauge symmetry is broken down to
a $U(1)\times U(1)$
\begin{equation}
  \label{eq:Ai-U(1)}
  A_i \to e^{-i\alpha\sigma_3} A_i e^{i\beta\alpha_3}
\end{equation}
together with the $\bZ_2$ action
\begin{equation}
  \label{eq:Z2-field-theory}
  A_i \to (-i\sigma_1) A_i (i\sigma_1)\, .
\end{equation}
There is an overall $U(1)$, obtained when $\alpha=\beta$, which acts
trivially on $A_i$. The nontrivial $O(2)=U(1)\rtimes\bZ_2$ symmetry
can be understood in the representation of the $A_i$ as four-vectors
as the rotations and reflection on the 2-plane of non-vanishing
directions, associated to $\sigma_0$ and $\sigma_3$.  Introduce the
coordinates $u_i^{\pm}=z_i^{0}\pm i z_i^3$. It is easy to see
that~\eqref{eq:Ai-U(1)} acts on these coordinates as
$u_i^\pm \to e^{\pm i (\beta-\alpha)} u_i^\pm$. Furthermore, the
non-abelian D-terms~\eqref{eq:non-abelian-D-terms} become in these
coordinates the single equation
\begin{equation}
  \sum_i |u_i^+|^2 = \sum_i |u_i^-|^2\, .
\end{equation}
We recognize this moduli space as the usual GLSM construction of the
conifold given above with $u_i^+=\alpha_i$ and $u_i^-=\beta_i$,
together with a decoupled $U(1)$ factor, associated with the $U(1)$
symmetry of a D3 in flat space.

The geometric involution we are interested in appears here as the
remnant discrete $\bZ_2$ action~\eqref{eq:Z2-field-theory}, which acts
on the $u_i^\pm$ coordinates as $u_i^\pm \leftrightarrow u_i^\mp$. The
fixed point locus is obtained whenever $u_i^\pm=u_i^\mp$ (in some
gauge). This implies $z_i^3=0$ (in this gauge, or equivalently
$z_i^0=0$ in a different gauge), and assuming $z^0_1z_2^0\neq 0$ on
this locus we have an enhanced $\Sp(2)$ symmetry, in agreement with
the expected behavior of the D3 on the O7$^-$ plane.

\subsection{Instanton effects in the orientifolded background}
\label{sec:instanton-effect}

\begin{figure}
  \centering
  \includegraphics[height=6cm]{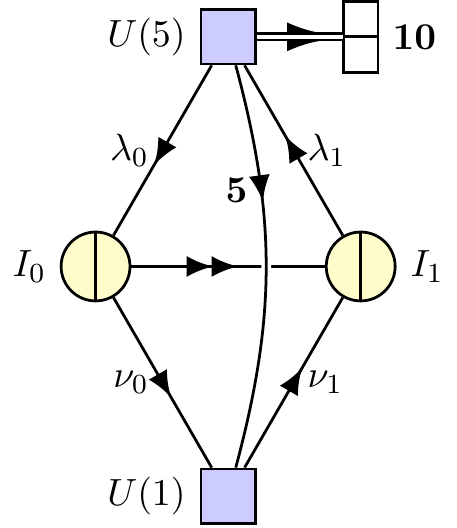}
  \caption{The theory obtained after orientifolding the theory in
    figure~\ref{fig:full-quiver}. The quiver nodes $I_i$ have
    symplectic gauge factors, formally given by $\Sp(0)$ in the
    absence of probe branes.}
  \label{fig:orientifolded-quiver}
\end{figure}

Let us now focus on the specific choice of ranks that will arise in
GUT models, as described in \S\ref{sec:weak-coupling}. We take in
particular multiplicity 5 for the $G_i$ stacks (so they will be
associated with the GUT stack) and multiplicity 1 for the $F_i$ stacks
(associated with the flavor brane). For simplicity, we choose the
spectrum between the $G_i$ branes to be given by a single chiral
bifundamental, which upon orientifolding will give rise to the {\bf
  10} representation. For phenomenological purposes it is often more
interesting to take three copies of the {\bf 10} representation, we
generalize the analysis below.  We also choose to have a single {\bf
  5} between the $G_i$ and $F_i$ stacks, with the same chirality as
the {\bf 10}. With these choices, the theory after orientifolding is
that described in figure~\ref{fig:orientifolded-quiver}.

The instanton effects come from euclidean D1 branes wrapping the
(empty) $I_i$ nodes. We will consider the case of a single fractional
instanton wrapping either $I_0$ or $I_1$. (Contributions due to
multi-instantons may be interesting, but their analysis is
significantly more subtle so we leave the analysis of these to the
interested reader.) Note that due to our choice of chirality for the
{\bf 10} and {\bf 5} representations the two nodes behave rather
differently. Let us start with $I_1$. In our conventions, the $I_1$
node admits terms in the instanton action coupling the zero modes
$\lambda_1,\nu_1$ to matter fields of the form
\begin{equation}
  \label{eq:inst-action}
  S_{\text{inst}}= c \, {\bf 10}_{[i j]} \, \lambda_{1}^{\bar i} \, \lambda_{1}^{\bar
  j} + d \, \lambda_{1}^{\bar i}\, {\bf 5}^i \, {\nu_1}
\end{equation}
for some $c,d\in\bC$. We have indicated the $U(5)$ indices
explicitly. This instanton action, upon integration of the fermionic
zero modes $\lambda_1,\nu_1$, will generate an effective
$\Ytop$ coupling in the low-energy
theory,\footnote{This configuration provides in fact a realization of
  the mechanism proposed in \cite{Blumenhagen:2007zk}.} in agreement
with the F-theory expectation.

On the other hand, the euclidean D-brane on $I_0$ behaves very
differently. Gauge invariance forbids any couplings analogous
to~\eqref{eq:inst-action}, so due to the unsaturated zero modes the
instanton will generate a higher F-term, instead of a superpotential
contribution \cite{Beasley:2004ys,Beasley:2005iu}.

Let us now argue that the couplings in~\eqref{eq:inst-action} do in
fact appear in the instanton action. In both terms, the matter fields
correspond to recombination modes of the flavor stacks. In the first
case, $G_0$ and $G_1$ are Weyl divisors that cannot leave the
singularity. Indeed, if the $\bf 10$ field is turned off, the charged
zero-modes are always massless. Recombination of the $G_0$ and $G_1$
divisors in order to obtain a Cartier divisor not intersecting the
singularity will manifest itself, from the point of view of the
instanton, in giving masses to the $\lambda_{1}^{\bar i}$ modes. The
mass term transforms in the $\bf 10$ representation of $SU(5)$, and
can indeed be seen as the expectation value of the corresponding GUT
matter field (which is, from the instanton worldvolume point of view,
a background field). A similar argument holds for the second term
in~\eqref{eq:inst-action}.

\subsection{Yukawa rank in the multiple family case}

For simplicity we have considered the case in which there is a single
${\bf 10}$ representation involved in the Yukawa coupling of
interest. In realistic models one would like to have three
generations, often all living on the same curve. The most general
instanton action is now given by
\begin{equation}
  S_{\text{inst}}= c_I \, {\bf 10}^I_{[i j]} \, \lambda_{1}^{\bar i} \, \lambda_{1}^{\bar
  j} + d \, \lambda_{1}^{\bar i}\, {\bf 5}^i \, {\nu_1}
\end{equation}
with $I$ the family index. The resulting Yukawa matrix is proportional
to
\begin{equation}
  \begin{pmatrix}
    c_1^2 & c_1 c_2 & c_1c_3 \\
    c_1c_2 & c_2^2 & c_2 c_3 \\
    c_1c_3 & c_2c_3 & c_3^2
  \end{pmatrix}
\end{equation}
which, as already remarked in \cite{Blumenhagen:2007zk}, is of rank
one. This is in agreement with the general expectation from the
F-theory analysis \cite{Cecotti:2009zf}.

\section{Mirror description}
\label{sec:mirror}

The previous results can also be justified, to some extent, from the
point of view of the IIA mirror description of the conifold using the
techniques in \cite{Hori:2000ck,Feng:2005gw}. In particular, the
structure of the mirror manifold in the presence of the orientifold
involution of interest to us has been discussed in
\cite{Franco:2007ii}. One additional ingredient with respect to the
discussion in \cite{Franco:2007ii} is the presence of flavor
branes. How to include these in the present context was discussed in
\cite{Franco:2006es,Forcella:2008au}. Note that the discussion in
these works involves a nontrivial amount of guesswork, which is why we
have chosen to give a more principled analysis above. Nevertheless the
mirror picture is technically simpler to analyze, so we will briefly
describe it here in order to provide some additional intuition for the
reader.

\begin{figure}
  \centering
  \includegraphics[width=0.6\textwidth]{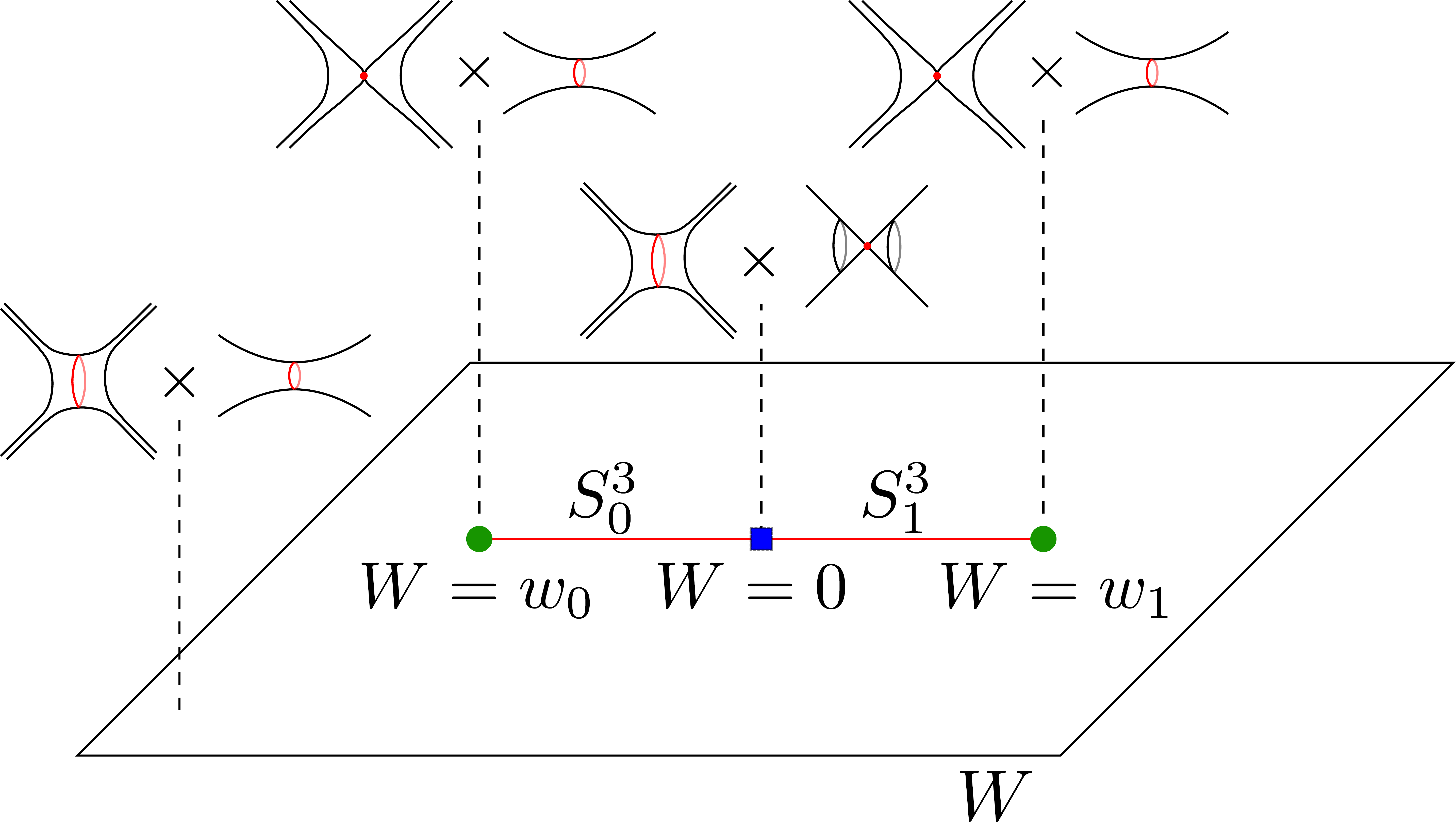}

  \caption{Mirror for the conifold, as a fibration over $\bC$,
    parameterized by a coordinate $W$. The generic fiber is $\bC^*$
    times $\bP^1$ punctured at four points. The $\bC^*$ fiber
    degenerates at $W=0$, and the punctured $\bP^1$ at
    $W\in\{w_0,w_1\}$. The mirror of the fractional branes on the
    conifold wrap the three-spheres $S^3_0$ and $S^3_1$, obtained by
    fibering the product of the degenerating 1-cycles in the fiber
    over the segment in the base.}
  \label{fig:mirror}
\end{figure}

As described in \cite{Hori:2000ck,Feng:2005gw}, for the purposes of
computing chiral information one can take the mirror of the conifold
to be described by a $\bC^*\times \Sigma$ fibration over $\bC$ given
by the equations
\begin{align}
  uv & = W \\
  P(x,y) & = W
\end{align}
with $u,v,W\in \bC$, $x,y\in \bC^*$. The complex plane is
parameterized by $W$. The first equation gives a $\bC^*$ fiber over
the $W$ plane, with a degeneration at $W=0$. The second equation
defines a Riemann surface fiber with four punctures. We pick
coordinates and a framing \cite{Bouchard:2011ya} such that
\begin{equation}
  P(x,y) = q + y + xy - xy^2\, .
\end{equation}
Here $q$ encodes the complex structure of the mirror, or equivalently
the complexified K\"ahler modulus of the original conifold. The
resulting Riemann surface is singular for $W=q$ and $W=q+1$; we denote
these points $w_0,w_1$ respectively. Consider the $T^2$ over any point
in the $W$ plane formed by the $S^1\subset\bC^*$ collapsing at $W=0$
times the one-cycle in $\Sigma$ associated with the degenerations at
$w_i$. The total space of this $T^2$ over a segment in the $W$ plane
connecting $W=0$ and $w_i$ has the topology of $S^3$, and the two
fractional branes of the conifold are obtained by wrapping D6 branes
on these cycles. The resulting geometry is depicted in
figure~\ref{fig:mirror}.

\begin{figure}
  \centering
  \includegraphics[width=0.4\textwidth]{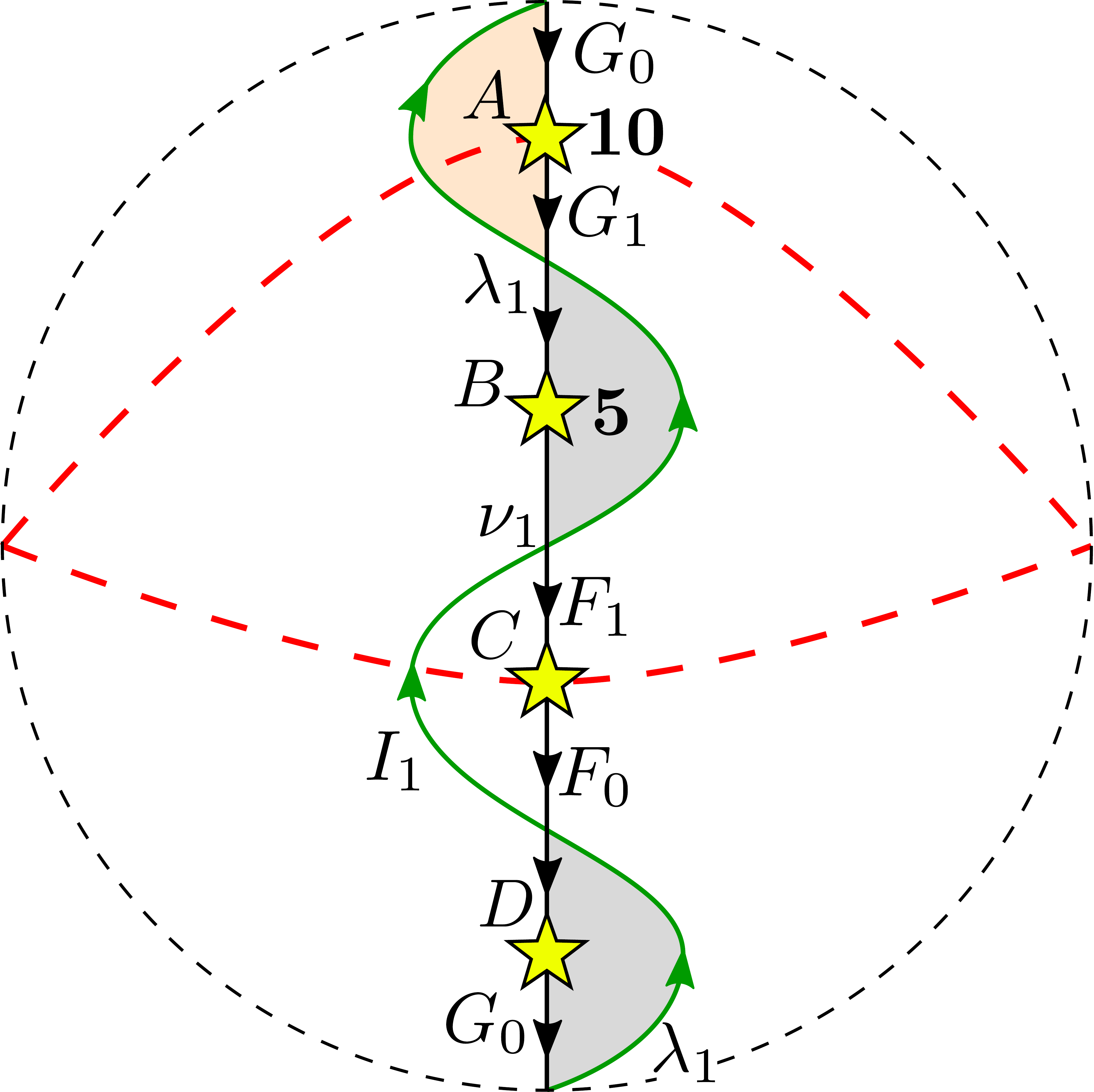}
  \caption{Schematic structure of the fiber of the mirror above the
    $W=0$ point. The fiber is topologically a disk with four punctures
    (denoted by stars), mapped here to a disk in which the dashed
    boundary should be identified to a point. The fixed locus of the
    orientifold involution is given by the red dashed line. The green
    line denotes the intersection of the $I_1$ euclidean D2 brane with
    the Riemann surface, and the black segments between puntures
    indicate the location of the flavor branes. We have also indicated
    by the shaded disks the worldsheet instantons giving rise to
    \eqref{eq:inst-action}.}
  \label{fig:coniRiemann}
\end{figure}

These two fractional branes intersect over $W=0$. We will focus on the
structure of the $\Sigma$ fiber over this point, which is described by
$\{q+y+xy-xy^2\}\subset (\bC^*)^2$. The structure of the fiber at this
point is summarized in figure~\ref{fig:coniRiemann}. As we see from
this figure, in this mirror description various nontrivial aspects of
the IIB analysis are manifest. The structure of zero modes and chiral
multiplets arises in a simple way from brane intersections (perhaps at
infinity along a puncture, as for the {\bf 5} and {\bf 10} matter
fields), and the couplings in the instanton effective
action~\eqref{eq:inst-action} arise naturally from worldsheet
instanton effects.

\section{M-theory description}
\label{sec:M-theory}

In this section we would like to reproduce some of the effects we have
observed in the weakly coupled IIB and IIA descriptions directly in
the more usual (in the F-theory model building literature) geometric
M-theory language. We will focus on two interrelated effects. The
first effect we would like to understand is simply how the usual
computation of Yukawa couplings in M-theory language, that is, in
terms of M2 branes wrapping appropriate chains with boundaries on the
matter states \cite{Marsano:2011hv,Martucci:2015dxa,Martucci:2015oaa},
connects to the instantonic picture.

At a heuristic level it is not hard to argue that there is a
connection, as follows. The IIA configuration T-dual to the fractional
D$(-1)$ on the conifold is given by D2 brane wrapping the T-duality
cycle, and the exceptional $\bP^1$ of the conifold. At the edges of
the T-duality interval we have O6 planes, and the $\bP^1$ of the
conifold necessarily collapses there. On a generic point along the
interval the $\bP^1$ is not necessarily of zero size. The D2 instanton
wraps the total space of this $\bP^1$ fibration over the interval,
which defines a non-trivial three-cycle $\Sigma_3$.

\begin{figure}
  \centering
  \begin{subfigure}{0.4\textwidth}
    \centering
    \includegraphics[height=4cm]{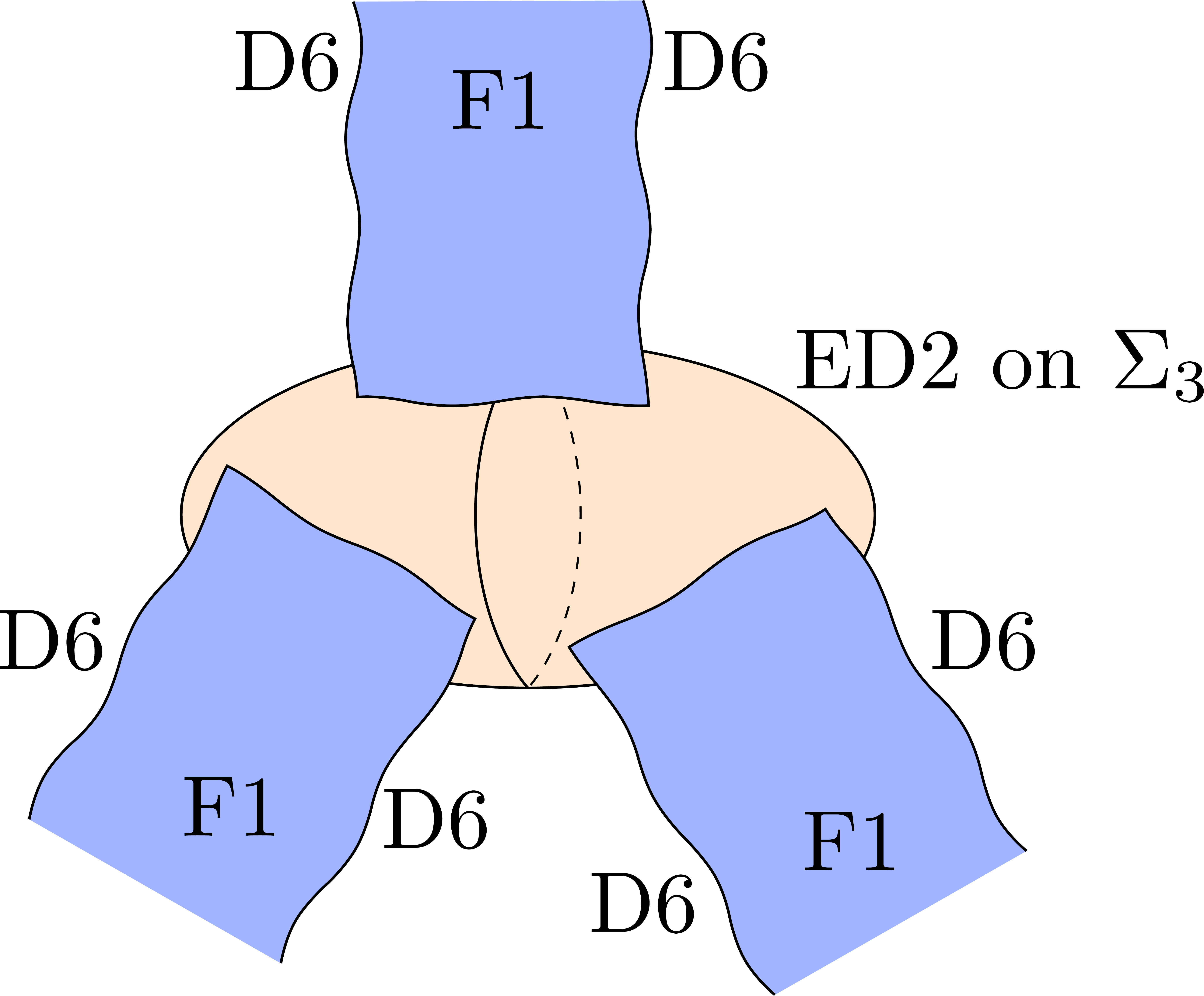}
    \caption{Instanton mediated coupling.}
    \label{sfig:D2-worldsheets}
  \end{subfigure}
  \hspace{1cm}
  \begin{subfigure}{0.4\textwidth}
    \centering
    \includegraphics[height=4cm]{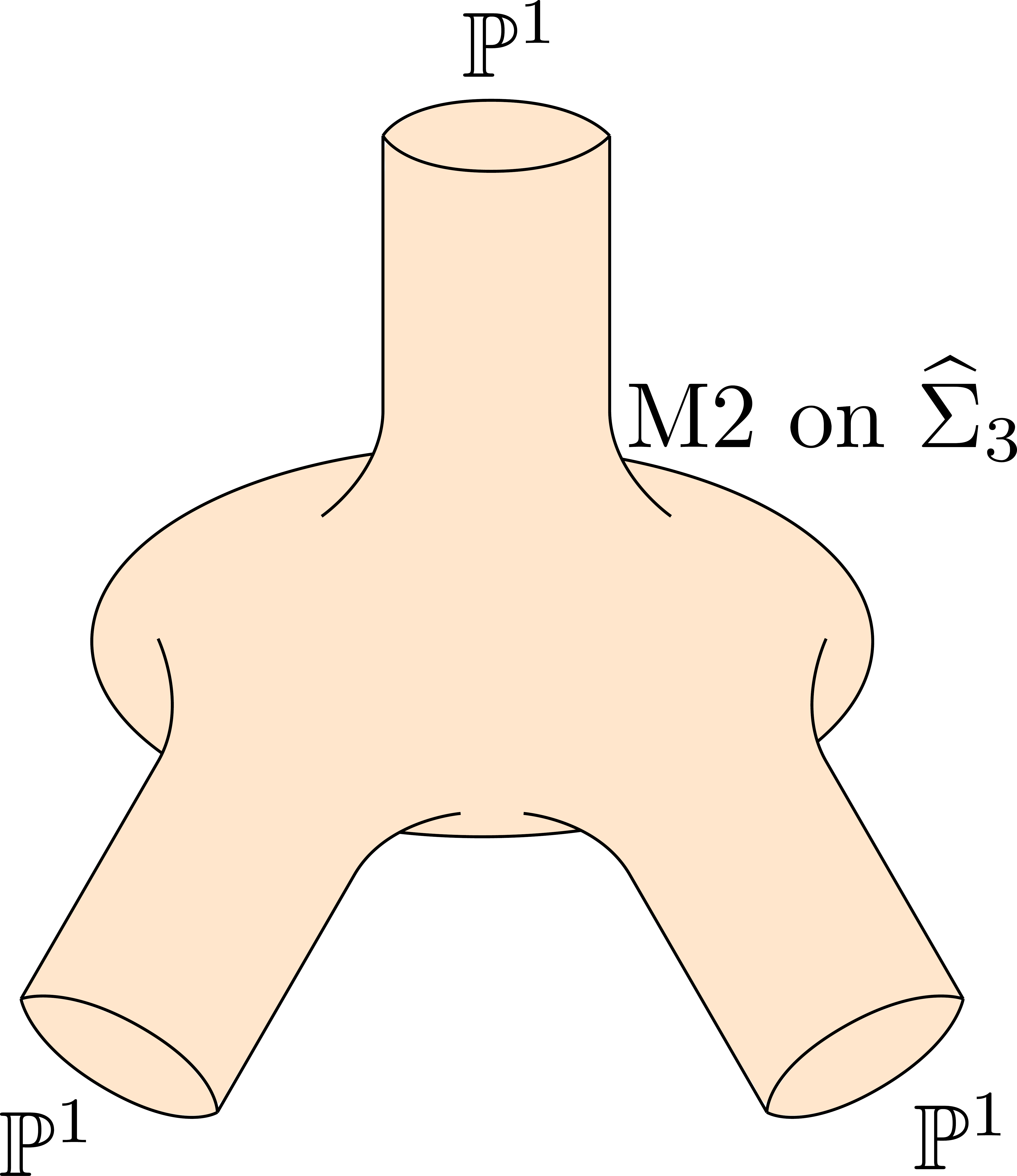}
    \caption{M-theory uplift.}
    \label{sfig:M2-recombined}
  \end{subfigure}
  \caption{\subref{sfig:D2-worldsheets} Schematic viewpoint of a
    instanton mediated coupling. Massless open strings between
    background D6 branes meet at the location of an euclidean D2
    brane, which (assuming the right zero mode structure) can mediate
    an effective coupling. \subref{sfig:M2-recombined} Its M-theory
    uplift, where the whole configuration has recombined into a
    non-compact three-cycle with three $\bP^1$ boundaries.}
  \label{fig:M2-heuristic}
\end{figure}

Consider now in the IIA picture for a coupling mediated by D2-brane
instantons. This is depicted in figure~\ref{sfig:D2-worldsheets}. We
have that various open string states meeting at a given spacetime
point, where the euclidean D2-brane lives, giving rise to the
effective vertex in the low energy action. In the M-theory picture
both the D2 and the F1 lift to M2 branes, so the lift of the IIA
configuration will be given by a recombined smooth M2, shown in
figure~\ref{sfig:M2-recombined}. This M2 wraps the three-chain
$\widehat{\Sigma}_3$, with boundaries given by $\bP^1$ cycles
associated with the states appearing in the coupling. This is
precisely the usual description for how perturbative couplings (such
as the $\Ytop$ coupling of interest to us) appear in the M-theory
description \cite{Marsano:2011hv,Martucci:2015dxa,Martucci:2015oaa}.

\begin{figure}
  \centering
  \begin{subfigure}{0.4\textwidth}
    \centering
    \includegraphics[height=3cm]{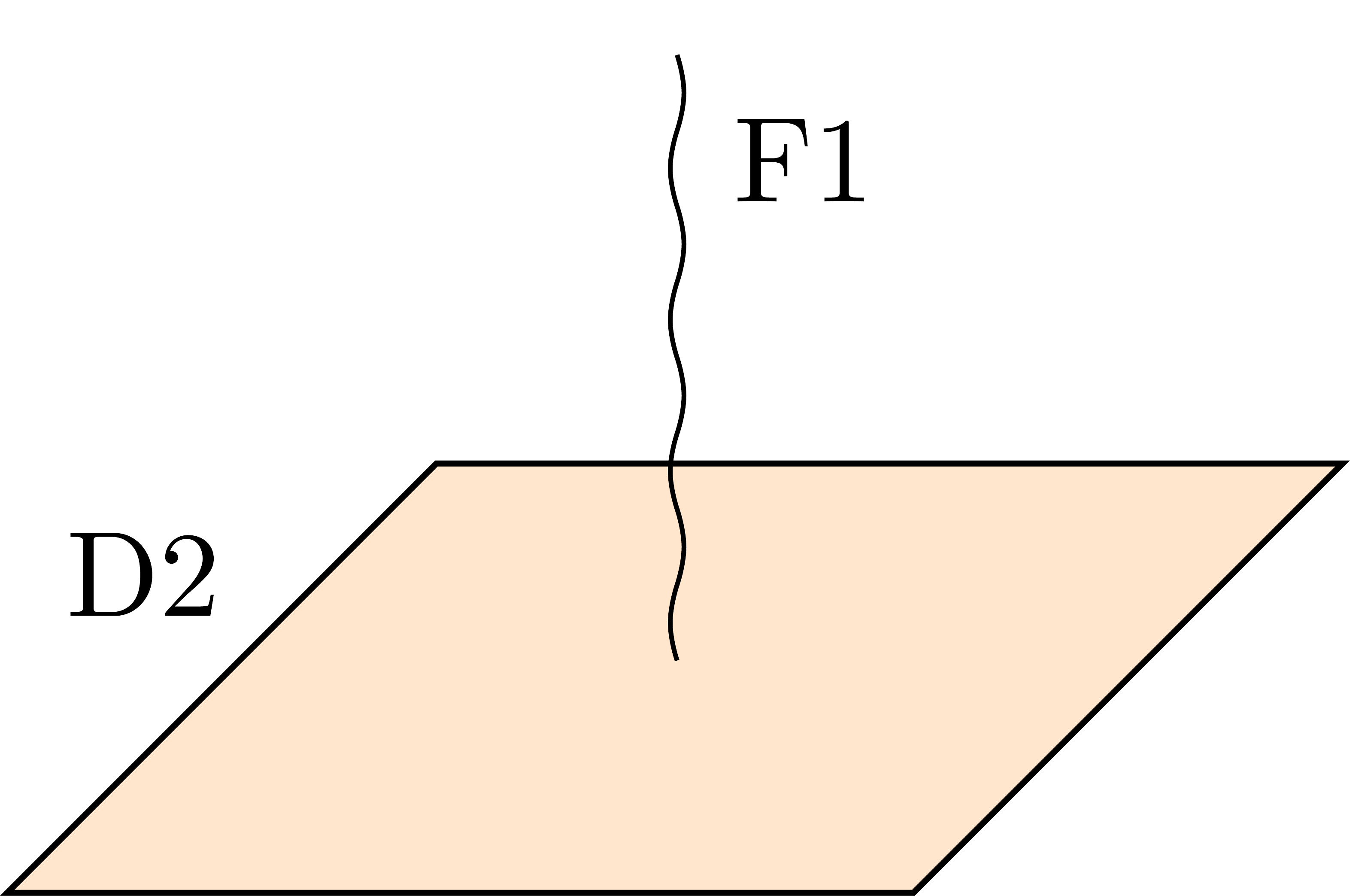}
    \caption{Weakly coupled viewpoint.}
  \end{subfigure}
  \hspace{1cm}
  \begin{subfigure}{0.4\textwidth}
    \centering
    \includegraphics[height=3cm]{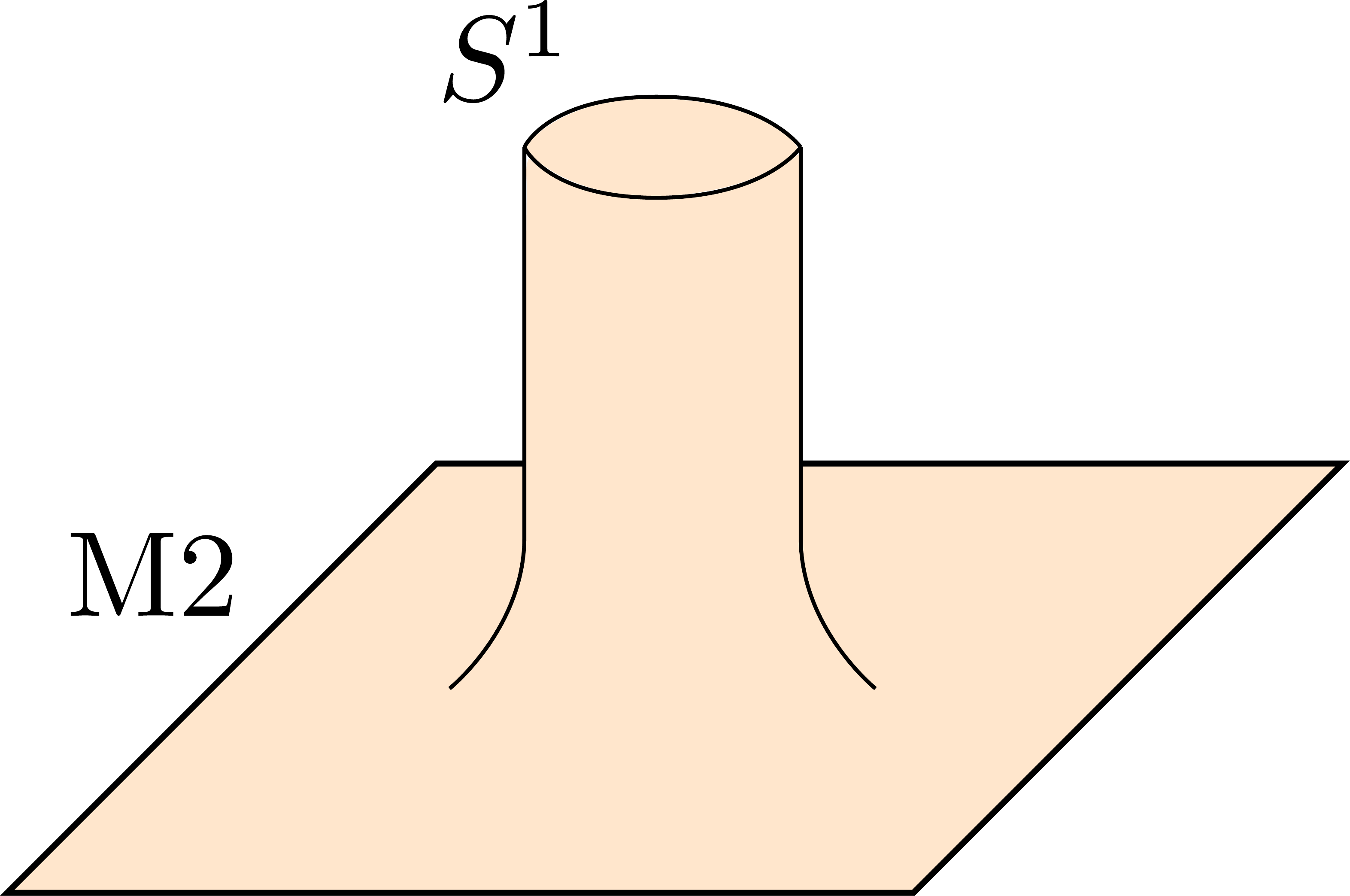}
    \caption{M-theory uplift.}
  \end{subfigure}
  \caption{A fundamental string ending on a D2 brane lifts to a smooth
    funnel-like solution for the M2 brane in M-theory.}
  \label{fig:D2-F1-recombination}
\end{figure}

We can understand the transition from figure~\ref{sfig:D2-worldsheets}
to~\ref{sfig:M2-recombined} as follows. Focus on a junction where an
open F1-worldsheet stretched between two D6-branes ends on an interval
on the Euclidean D2-brane, say the upper junction
in~\ref{sfig:D2-worldsheets}. Now take Euclidean $\tau$ time to run
horizontally from left to right in the diagrams. Then, at each slice
with constant $\tau$ we have a semi-infinite line from the F1 ending
on the D2. From the point of view of the D2-worldvolume, the F1 is an
electric source, and a backreacted solution creates a funnel-type
geometry, whereby the F1 is replaced by a smooth spike made entirely
of the D2-brane \cite{Callan:1997kz}. At any horizontal slice, the
boundary of this surface is a circle, whose radius grows with
$g_s$. This system lifts straightforwardly to a smooth M2 brane
configuration in M-theory. This recombination process is shown in
figure~\ref{fig:D2-F1-recombination}.

Now take a family of these funnel geometries parametrized by $\tau$,
covering the whole interval. At the two extremities $g_s$ vanishes,
since there are D6-branes there. Hence, the boundary circles of the M2
funnels shrink to zero size at the extremities of the interval. In the
end, the full boundary of this family is a circle fibration over the
interval collapsing over two points, i.e. an $S^2$.  Now we have
replaced the F1/D2-junction with a smooth D2 with boundaries, which
trivially uplifts to M-theory to an M2 with boundaries, as depicted
in~\ref{sfig:M2-recombined}.

In what follows we will present evidence supporting this heuristic
picture, by explicitly tracking the M-theory three-chain
$\widehat{\Sigma}_3$ to the weak coupling limit, and seeing that it
survives the limit, getting localized at the $E_6$ point. A full proof
of the connection requires an explicit identification of
$\widehat{\Sigma}_3$ with the uplift of the D2 brane. It would be very
interesting to work this out in detail, but we will not do this here.

Along the way we will find that in the strict weak coupling limit
($g_s=0$), the geometry develops a non-flat fiber at the $E_6$
point. This effect only appears at $g_s=0$, and disappears as long as
$g_s$ is finite, no matter how small. We conjecture that the physical
origin for this effect has to do with the light strings appearing for
vanishing $B$-field on the collapsed cycle.

This conjecture is based on the following observations. Notice first
that in the weakly coupled description this $B$-field period can be
arbitrarily tuned while staying on the quiver locus (contrary to what
is stated in \cite{Donagi:2009ra}), so it is possible to enhance the
contribution of the instanton by tuning the $B$-field. More in detail,
the instanton contribution to the superpotential goes as
\begin{equation}
  \Delta_{np} W = (\ldots) e^{-\frac{1}{g_s}\int_{\bP^1} B_2 + \int_{\bP^1} C_2} \, \Ytop
\end{equation}
where we have omitted the (unknown) dependence on the complex
structure moduli, and schematically indicated by $\int_{\bP^1}B_2$ and
$\int_{\bP^1}C_2$ the periods of the $B_2$ and $C_2$ fields on the
non-resolvable $\bP^1$. (This is schematic since the $\bP^1$ is never
present in the geometry as a proper algebraic cycle, but the
corresponding periods are still well defined.) We see that taking
$\int_{\bP^1}B_2\to 0$ we enhance the magnitude of the
non-perturbative effect. Nevertheless, we cannot simply take
$\int_{\bP^1}B_2$ to vanish. This is because in order to have a
standard field theory description of the conifold background one needs
the $B$ field on the collapsing $\bP^1$ not to vanish
\cite{Strominger:1995cz,Aspinwall:1995zi}. Otherwise one has massless
strings coming from D3 branes wrapping the collapsing $\bP^1$. The
mass of these states is, in fact, given by the same expression as the
instanton action. Our conjecture is that these D3 branes dualize to M5
branes wrapping the non-flat fiber. Assuming that this identification
is correct, the IIB analogue of the disappearance of the non-flat
fiber as we take $g_s$ nonzero is presumably related to some dynamical
effect on the D3 making it massive. Also, the D3 branes on the type
IIB side get a mass by turning on a $B$-field background, which should
imply that the M5 branes on the non-flat fiber get a mass upon turning
on an appropriate $C_3$ background. It would be very interesting to
verify (or refute) our conjectured identification, and to understand
the mechanisms behind both mass mechanisms, but we will not attempt to
do so here.

\subsection{The weak coupling limit}

We will start by reviewing the techniques in
\cite{Donagi:2012ts,Clingher:2012rg}, which provide an efficient way
of taking a weak coupling limit in F-theory such that one can more
easily recognize the IIB brane system encoded by an elliptic
fibration, elaborating on Sen's original proposal
\cite{Sen:1996vd,Sen:1997gv}. We will first describe the procedure in
\cite{Donagi:2012ts,Clingher:2012rg} in general, and then apply it to
the $SU(5)$ model.

Let us start with the following particular form for an elliptic fibration:
\be
W: \quad -y^2+x^3+b_2 x^2 z^2+2\,b_4 x z^4+b_6 z^6
\ee
Following Sen \cite{Sen:1996vd,Sen:1997gv}, we introduce a small parameter $\epsilon$ and rescale $(b_4, b_6) \mapsto (\epsilon b_4, \epsilon ^2 b_6)$, then we find that the $j$-function of this model blows up everywhere except at $b_2=0$. In this way, the discriminant has the following leading term:
\be
\Delta =\epsilon^2 b_2^2 (b_4^2-b_2 b_6) + \cO(\epsilon^3)\,.
\ee
As explained in \cite{Sen:1996vd,Sen:1997gv}, there is an O7-plane at $b_2=0$ and a D7-brane at $b_4^2-b_2 b_6=0$. The problem with this approach to the weak coupling limit stems from
the fact that taking $\epsilon \rightarrow 0$ does not commute with
computing the discriminant of the elliptic fibration. In order to see
Sen's result, one must first keep $\epsilon$ arbitrary, compute the
discriminant, and only then expand. If, on the other hand one first
expands the Tate equation in $\epsilon$ and then takes the limit, then
the model washes away all the D7-brane data, giving too crude a
rendition of the situation.

One can improve on this by adopting the philosophy of
\cite{Donagi:2012ts,Clingher:2012rg}, and promoting the parameter
$\epsilon$ to a coordinate. So we study the fivefold $X_5$, which is a
family of CY fourfolds, of which the limiting hypersurface
$\epsilon=0$ would be the weakly coupled F-theory model. $X_5$ is then
given by the hypersurface equation:
\begin{equation}
X_5: \quad  y^2=x^3+b_2 x^2 z^2+2\,\epsilon b_4 x z^4+\epsilon b_6 z^6\, .
\end{equation}
One first notices that this fivefold is singular at the ideal
$(y, x, \epsilon)$. Then, one proceeds to blow-up the fivefold at this
ideal. The result is the following ambient space
\[\arraycolsep=10pt
\begin{array}{|c|c|c|c|c|} \hline
\hat x & \hat y & z &\hat  t & v \\ \hline \hline
2 & 3 & 1 & 0 & 0 \\ \hline
1 & 1 & 0 & 1 & -1 \\ \hline
\end{array} \]
with the following irrelevant ideals \[ (\hat x,\hat  y, z),\, (\hat  x,\hat  y, \hat t),\, (z, v),\] and the following ideal describing the proper transform $\hat X_5$ of the fivefold:
\begin{equation} \label{hatfive}
\hat{X_5}: \quad  \hat y^2= \hat x^3 v+b_2 \, \hat x^2 z^2+2\,b_4\,  \hat x\hat  t z^4+b_6\, \hat  t^2 z^6\, .
\end{equation}
The blow-down map is
\be
(\hat x, \hat y, t, v) \mapsto (x = \hat x v,\, y = \hat y v,\, \epsilon = \hat t v)
\ee
From now on, we will drop the `hats', hoping not to cause confusion.

The fourfold $Y_0$ over the central fiber $\epsilon= t v=0$ breaks into two components $Y_0 = Y_{\rm pert} \cap Y_{\rm rem}$. The first component
\be
Y_{\rm pert}: \quad \left(v,\quad -y^2+b_2 x^2+2 b_4  x t+b_6 t^2 \right)
\ee
inside the ambient space
\[\arraycolsep=10pt
\begin{array}{|c|c|c|} \hline
x & y & t \\ \hline \hline
1 & 1 & 1 \\ \hline
\end{array} \]
with irrelevant ideal $(x, y, t)$, reveals the perturbative IIB data. For instance, its discriminant is simply
\be
\Delta_{\rm pert} = b_4^2-b_2 b_6\,,
\ee
i.e., the locus of the perturbative D7-brane. It is a (quadric) $\P^1$-fibration over the base $B_3$.

The second `remaining' component, $Y_{\rm rem}$, is given by
\be
Y_{\rm rem}: \left( t, \quad y^2-x^3 v+b_2 x^2 z^2 \right)
\ee
inside the ambient space
\[\arraycolsep=10pt
\begin{array}{|c|c|c|c|c|} \hline
x & y & z & v \\ \hline \hline
2 & 3 & 1 & 0 \\ \hline
1 & 1 & 0 & -1 \\ \hline
\end{array} \]
Since $x=0 \Rightarrow y=0$, which would be forbidden, we can fix $x \mapsto 1$. Now we end up with a linear equation in $v$, allowing us to eliminate it. Therefore, we are left with a purely toric space given by
\[\arraycolsep=10pt
\begin{array}{|c|c|} \hline
y & z  \\ \hline \hline
 1 & 1  \\ \hline
 \end{array} \]
which is a constant $\P^1$-bundle over $B_3$.

The two sphere fibrations meet at the following ideal:
\be
Y_{\rm pert} \cap Y_{\rm rem}: \quad \left(t, v, -y^2+b_2 \right) 
\ee
in the ambient space define by the complex $y$-plane. This is a double cover of $B_3$ branched over the locus $b_2=0$, which we recognize as the O7-plane. Put differently $Y_{\rm pert} \cap Y_{\rm rem}$ is isomorphic to the perturbative IIB CY threefold target space.

\subsection{Constructing the $SU(5)$ model from the bottom up}

Armed with this technology, we can easily construct the F-theory
fourfold $Y_4$ corresponding to any given perturbative IIB setup;
we simply need to run this machinery backwards. We start with a CY threefold $X_3$ that admits a description of the form 
\be
X_3: \quad \xi^2 = b_2
\ee
with orientifold involution $\xi \mapsto -\xi$. We give as input an D7-brane hypersurface
\be
\Delta_{\rm pert}  \equiv b_4^2 - b_2 b_6\,.
\ee
In \cite{Collinucci:2008pf}, it was shown that this is the most general admissible form for a D7-brane consistent with an O7-involution, even if the brane is reducible and non-reduced. 

In our case, we would like to have
\be
\Delta_{\rm pert}  = \sigma^5 (w + \sigma P) \approx \sigma^5 w \quad \text{near the GUT brane}\,,
\ee
where $\sigma$ and $w$ are coordinates in $B_3$, and $P$ is some polynomial. This ensures a stack of $5$ branes at $\sigma=0$, and that the intersection between that stack and the remainder is described by a simple equation, $(\sigma, w)$. 

We also know that, in order to have an $SU(5)$ gauge group as opposed to $Sp(5)$, we need the divisor $\sigma$ to be reducible into a brane/image-brane pair. This is achieved by requiring $X_3$ to have the following conifold structure
\be
X_3: \quad \xi^2 = u^2+\sigma w\,,
\ee
where $u$ is another base coordinate. In other words, we are defining $b_2 \equiv u^2+\sigma w$. Now the $SU(5)$ stack and its image are given by the ideals
\be
(\sigma, \, \xi+u) \leftrightarrow (\sigma,\, -\xi+u)
\ee
which are swapped by the involution $\xi \rightarrow -\xi$.
Now all we need to do is fix the forms of $b_4$ and $b_6$ by requiring
\be
\Delta_{\rm pert} = -\sigma^5 w = b_4^2-b_2 b_6 = b_4^2-(u^2+\sigma w) b_6 \, .
\ee
We choose
\be
b_4 = u \sigma^2, \quad b_6 = \sigma^4\, .
\ee
This leads us to the following ansatz for the Tate equation of the CY fourfold:
\be
W: \quad -y^2+x^3 v+(u^2+\sigma w) x^2 z^2+2\,u \sigma^2 x t z^4+\sigma^4 t^2 z^6\, .
\ee
The full ideal for $Y_\epsilon$ can therefore be written as follows:
\be \label{nonresgenric}
\left( -y^2+x^3 v+(u x z+\sigma^2 t z^3)^2+ \sigma w x^2 z^2, \quad t v = \epsilon \right)\, .
\ee

\subsection{Yukawa interactions at weak coupling}
Having reviewed what creates the $10\, 10\, 5$ Yukawa coupling in
F-theory, we now want to see this process through the weak coupling
limit. Since the point of this paper is that such a coupling exists in
perturbative IIB, then we should be able to see it as $Y_\epsilon$
degenerates into $Y_0 = Y_{\rm rem} \cup Y_{\rm pert}$. (The analysis
in the $\epsilon\neq 0$ case is well understood by now
\cite{Krause:2011xj,Martucci:2015dxa}, we review it in
appendix~\ref{app:globalres}).

Let us take our central fiber, i.e. $\epsilon=0$ in \eqref{nonresgenric}, and focus on the `perturbative' $Y_{\rm pert}$ branch defined by $v=0$. Explicitly, we take
\be
Y_{\rm pert}: \quad y^2 =(u x+\sigma^2 t)^2+ \sigma w x^2
\ee
inside the following toric ambient space:
\[\arraycolsep=10pt
\begin{array}{|c|c|c|c|c|c|} \hline
x & y & t & \sigma & u & w \\ \hline \hline
1 & 1 & 1 & 0 & 0 & 0 \\ \hline
\end{array} \]
with irrelevant ideal $(x, y, t)$. This space is singular at the ideal
$(x, y, \sigma)$. Since the physics that interests us is happening at
the singularity, we will focus on the patch $t \neq 0$ and gauge fix
that coordinate to one.
In this way we have reduced the problem to studying the hypersurface 
\be y^2 =(u x+\sigma^2)^2+ \sigma w x^2 \subset \C^{5}\,. \ee
To gain some intuition, notice that, away from the orientifold locus given by $b_2 \cong u^2+\sigma w=0$, we can rewrite this as
\be
y^2 = (u^2+\sigma w) (\tilde x)^2+\frac{w \sigma^5}{u^2+\sigma w}
\quad {\rm with} \quad \tilde x = x+\frac{u \sigma^2}{u^2+\sigma w}\, .
\ee
So, in a neighborhood where $b_2$ is constant, we can interpret this equation as a $\C^*$ fibration that degenerates over $\Delta_{\rm pert} = \sigma^5 w=0$, which matches our expectation about our perturbative D7-brane setup.

Before we can start performing resolutions, we will make the
convenient redefinition $Y \cong y+u x+\sigma^2$, such that our
fourfold is now given by
\begin{equation}
Y (-Y+2 u x+2 \sigma^2)+\sigma w x^2\, .
\end{equation}
Now we can resolve the singularity at $(Y, x, \sigma)$. It turns out that we will need two blow-ups and two further small resolutions. There are several possible resolution phases for this setup, as explored in \cite{Esole:2011sm}. We will pick one, leading to the following toric ambient space
in the following ambient toric fivefold:
\[\arraycolsep=10pt
\begin{array}{|c|c|c|c|c|c|c||c|c|} \hline
x & Y  & \sigma &v_1 & v_2 & v_3 & v_4 & u & w \\ \hline \hline
1 & 1 & 1&-1& 0 & 0 & 0& 0 & 0\\ \hline
1 & 1 & 0&1& -1 & 0 & 0& 0 & 0\\ \hline
0 & 1 & 0&0& 1 & -1& 0& 0 & 0 \\ \hline
0 & 1 & 0&1& 0 & 0 & -1& 0 & 0\\ \hline
\end{array} \]
with irrelevant ideals: 
\be
(x, Y, \sigma),\,    (v_1, Y),\,  (v_2, Y),\,(v_2, \sigma), \, (v_3,
\sigma),\,   (v_1, v_3),\, (v_4, x) \, .
\ee
The fully resolved fourfold is described by the hypersurface:
\be \label{Yperthat}
\hat Y_{\rm pert}: \quad Y \left(-Y v_3 v_4+2 u x+2 v_1 v_4 \sigma^2 \right)  + v_1 v_2 \sigma w x^2=0
\ee
and blow-down map
\be
(x ,\, Y,\, \sigma ,\, v_1 ,\, v_2 ,\, v_3,\, v_4 ) \mapsto (\underline{x} = x v_1 v_2^2 v_3^2 v_4,\quad \underline{Y} = Y v_1 v_2^2 v_3^3 v_4^2,\quad\underline{\sigma} = \sigma v_1 v_2 v_3 v_4 )
\ee
where the underlined coordinates are coordinates of the blown-down singular space.

Let us now study the fiber structure of this resolved space, and the various degenerations it undergoes as we restrict to special loci of increasing codimension.
\subsubsection{Codimension one}
We begin by studying the fiber over $\underline{\sigma}\equiv \sigma v_1 v_2 v_3 v_4=0$. Intersecting the various factors with $Y_{\rm pert}$, we obtain a pattern of curves. 

Two words on our notation: 
\begin{enumerate}
\item When we write a sum of ideals, we really mean the homological sum of the curves associated to the ideals, e.g. 
\be
({\rm eq}_1) = ({\rm eq}_2) +({\rm eq}_3) 
\ee
really means that the lhs corresponds to a curve that decomposes into the two curves on the rhs. Technically we should say that it is the intersection of the two ideals.
\item We will gauge-fix coordinates whenever possible without explicitly saying so. For instance, in the ideal $(v_1, -v_4+2 u x)$, $x$ cannot vanish. For this would require $v_4$ to vanish, even though $(x, v_4)$ is ruled out. So we can use a projective action to fix $x \rightarrow 1$ and simply write the ideal as $(v_1, -v_4+2 u)$.
\end{enumerate}

Let us now proceed:
\begin{align}
&{C}_\sigma:  &(\sigma) \cap Y_{\rm pert}  &\quad \cong \quad C_R=\big(\sigma,\, Y\big) \quad + \quad  C_L=\big(\sigma,\, -Y v_4+2 u x) \big)\\
&{A_4}^{(1)}: & (v_1) \cap Y_{\rm pert} &\quad \cong \quad  (v_1,\, -v_4+2 u x)\\
&{A_4}^{(2)}:  &(v_2) \cap Y_{\rm pert} &\quad \cong \quad  (v_2,\, -v_3 v_4+2 u x+2 v_1 v_4)\\
&{A_4}^{(3)}:  &(v_3) \cap Y_{\rm pert} &\quad \cong \quad  (v_3,\, Y (2 u x+2 v_4)+ v_2 w x^2)\\
&{A_4}^{(4)}:  &(v_4) \cap Y_{\rm pert} &\quad \cong \quad  \big(v_4,\, 2 Y u+v_1 v_2 \sigma w\big)
\end{align}
The intersection pattern of the full fiber has the following graph:

\begin{center}
  \includegraphics{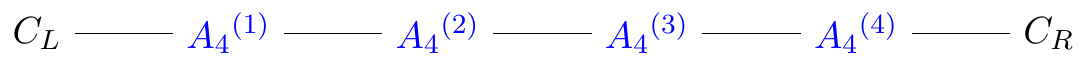}
\end{center} 
The blue part showcases a non-affine $A_4$ diagram. The extremities, $C_{L, R}$, are non-compact curves. We can think of this as a five-centered Taub-NUT space, such that, upon projecting onto a $C$-plane, we have a $C^*$-fibration that collapses over the D7-branes. The two extremal pieces are just the fiber expanding as we move away from the branes.

\subsubsection{Codimension two}
Over the loci $\Sigma_5 = (\underline{\sigma}, w)$ and $\Sigma_{10} = (\underline{\sigma}, u)$, we expect to see enhancements to $A_5$ and $D_5$, respectively. Now let us examine the fiber over these loci.

\begin{paragraph}{$\Sigma_{5}$ matter curve}
\begin{align}
&C_R &&\longrightarrow &C_R&=\big(\sigma,\, Y\big)\\
&C_L &&\longrightarrow &C_L&=\big(\sigma,\, -Y v_4+2 u x)\big)\\
&{A_4}^{(1)}&&\longrightarrow &{A_5}^{(1)}&=(v_1,\, -v_4+2 u x)\\
&{A_4}^{(2)}&&\longrightarrow &{A_5}^{(2)}&= (v_2,\, -v_3 v_4+2 u x+2 v_1 v_4)\\
&{A_4}^{(3)}&&\longrightarrow &{A_5}^{(3)}&=(v_3,\, 2 u x+2 v_4) \quad + \quad {A_5}^{(4)}=(v_3,\, Y) \\
&{A_4}^{(4)}&&\longrightarrow &{A_5}^{(5)}&=(v_4,\, Y)
\end{align}

\begin{center}
  \includegraphics{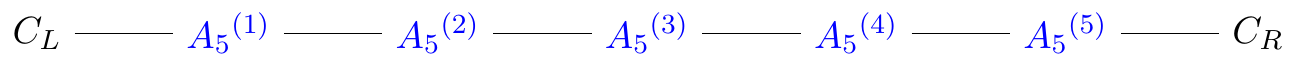}
\end{center} 
We clearly recognize an $A_5$ Dynkin diagram, sandwiched between the two complex planes, as expected.

\end{paragraph}

\begin{paragraph}{$\Sigma_{10}$ matter curve}
\begin{align}
&C_R &&\longrightarrow &\Gamma&=\big(\sigma,\, Y\big)\\
&C_L &&\longrightarrow &\Gamma&=\big(\sigma,\, Y\big)\quad + \quad {D_5}^{(1)}=\big(\sigma,\, v_4 \big)\\
&{A_4}^{(1)}&& \longrightarrow &{D_5}^{(2)}&=(v_1,\, v_4)\\
&{A_4}^{(2)}&& \longrightarrow &{D_5}^{(3)}&=(v_2,\, v_4) \quad + \quad {D_5}^{(4)}=(v_2,\, 2 v_1-v_3)\\
&{A_4}^{(3)}&& \longrightarrow &{D_5}^{(5)}&=(v_3,\, 2 Y v_4+ v_2 w x^2)\\
&{A_4}^{(4)}&& \longrightarrow &{D_5}^{(2)}&=(v_4,\, v_1) \quad + \quad {D_5}^{(3)} =(v_4,\, v_2) \quad + \quad {D_5}^{(1)} = (\sigma, v_4)
\end{align}

The arrangement has the following shape:
\begin{center}
  \includegraphics{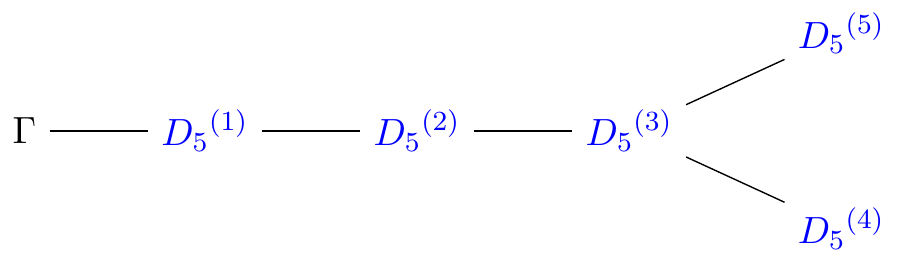}
\end{center} 
The blue part is a $D_5$ Dynkin diagram. This can be understood as an orbifold of an $A_5$ Dynkin diagram: The two external pairs plus the $C_L$ and $C_R$ get identified. The middle node gets itself orbifolded with two fixed points, which lead to two singularities, which after resolution give the ${D_5}^{(4)}$ and ${D_5}^{(5)}$ nodes.

\end{paragraph}

\subsubsection{Codimension three}
We will now study the fate of the fiber at the Yukawa
`$E_6$'-point. This is where things differ drastically from the more
conventional $\epsilon\neq 0$ case reviewed in
appendix~\ref{app:globalres}. What we are about to see is that our
fibration is non-flat, meaning that the fiber dimension will jump. In
this case, the fiber will decompose into three curves plus a surface!

\begin{paragraph}{Tensionless strings}

By taking \eqref{Yperthat} and setting $(\underline{\sigma}, u, w)= \vec{0}$, we get the following degenerate ideal:
\begin{align}
\hat Y_{\rm pert}|_{(u, w)}&: \quad \big(Y \left(-Y v_3+2 v_1 \sigma^2 \right) v_4,\, \sigma v_1 v_2 v_3 v_4 \big)\\
&=(Y^2,\, \sigma\big) +(-v_3+2 v_1,\, v_2\big)+(Y,\,v_3\big)+ (v_4)
\end{align}
Clearly, this fiber has a component given by $v_4=0$, which is a toric
divisor described by the following data
\[\label{toricdivisor}\arraycolsep=10pt
\begin{array}{|c|c|c|c|c|} \hline
 Y    & \sigma &v_1 & v_2 & v_3  \\ \hline \hline
 0   & 1&-2& 1 & 0 \\ \hline
 1   & 0&0& 1 & -1 \\ \hline
 1   & 0&1& 0 & 0 \\ \hline
\end{array} \]
with irrelevant ideals: 
\begin{eqnarray}
&& (v_1, Y),\, (v_2, Y),\,(v_2, \sigma), \, (v_3, \sigma),\,   (v_1, v_3)\,.
\end{eqnarray}
An M5-brane wrapping this divisor will become an effective tensionless
string upon blowing-down, unless (as discussed in the introduction to
the section) the $C_3$ field gives it a non-zero mass.

The configuration of the full fiber has the following shape:
\begin{center}
  \includegraphics{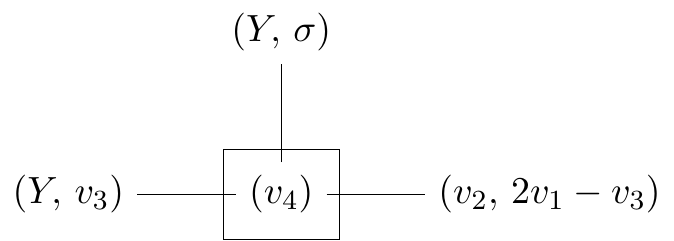}
\end{center} 
As one might expect, the weak coupling limit avoids creating the
non-affine $E_6$-diagram that develops in the strongly coupled
case. It is interesting to see that the way this is avoided is by
growing a vanishing four-cycle.

\end{paragraph}

\begin{paragraph}{``$E_6$'' enhancement: $\Sigma_{\bf 5} \rightarrow {\rm Yuk}_{E_6}$}
Now let us approach the Yukawa point $(\underline{\sigma}, u, w)$,
where all D-brane stacks meet each other and the O7-plane, coming from
the $\Sigma_{\bf 5}$-curve. We find the following splittings
\begin{align}
&C_R &&\longrightarrow &C_R&=\big(\sigma,\, Y\big)\\
&C_L &&\longrightarrow &C_R&=\big(\sigma,\, Y \big) \quad + \quad C =\big(\sigma,\, v_4 \big)\\
&{A_5}^{(1)}&&\longrightarrow &&(v_1,\, v_4)\\
&{A_5}^{(2)}&&\longrightarrow &&(v_2, v_4) \quad + \quad(v_2,\, 2 v_1-v_3)\\
&{A_5}^{(3)}&&\longrightarrow && (v_3, v_4)\\
&{A_5}^{(4)}&&\longrightarrow && (v_3, Y)\\
&{A_5}^{(5)}&&\longrightarrow &&(v_4,\, Y)
\end{align}

\end{paragraph}

\begin{paragraph}{Yukawa $\Sigma_{10}\rightarrow E_6$}
Now let us approach the Yukawa point from the ${\bf 10}$-matter curve:
\begin{align}
&\Gamma &&\longrightarrow  &\Gamma&=(\sigma,\, Y) \\
&{D_5}^{(1)}&& \longrightarrow &&=(v_4, \sigma)\\
&{D_5}^{(2)}&& \longrightarrow &&=(v_1,\, v_4)\\
&{D_5}^{(3)}&& \longrightarrow &&=(v_2,\, v_4)\\
&{D_5}^{(4)}&& \longrightarrow &&=(v_2,\, 2 v_1-v_3)\\
&{D_5}^{(5)}&& \longrightarrow &&=(v_3,\, Y) \quad + \quad (v_3,\, v_4)
\end{align}

\end{paragraph}

The aim now is to demonstrate that there is a homological relation of curves:
\be
\underbrace{{D_5}^{(1)}}_{(v_4, \sigma)} = \underbrace{{D_5}^{(3)}}_{(v_4, v_2)}+\underbrace{{A_5}^{(3)}}_{(v_4, v_3)}
\ee
just as in the non-perturbative situation. Here, we can easily see the relation as follows. By consulting the table \eqref{toricdivisor} which describes the toric divisor $v_4=0$, we see that the curves correspond to homogeneous coordinates that have the following weights under the three $\C^*$ actions:
\be
(\sigma) \leftrightarrow \begin{pmatrix} 1\\0\\0 \end{pmatrix} \quad (v_2) \leftrightarrow \begin{pmatrix} 1\\1\\0 \end{pmatrix}  \quad (v_3) \leftrightarrow \begin{pmatrix} 0\\-1\\0 \end{pmatrix}\,.
\ee
This implies that the homology classes add up as expected, thereby confirming the existence of a 3-chain connecting the three curves as in the non-perturbative case. In this situation, one can construct the following 3-chain $\Sigma_3$
\be 
\Sigma_3: (1-t) \sigma+t v_2 v_3 \qquad t \in [0, 1]\,.
\ee
We speculate that the D1-instanton discussed in \S\ref{sec:IIB} uplifts to an open M2-brane instanton wrapped on this 3-chain.

Note, that this 3-chain is entirely contained in the vanishing four-cycle given by $v_4=0$.

\subsection{Equivalence of M2-instanton effects}
In the previous two sections, we saw that there is a 3-chain mediating the transition
\be
{D_5}^{(1)} \rightarrow {D_5}^{(3)}+{A_5}^{(3)}
\ee
whereby an M2-state in the ${\bf 10}$ representation transitions two a sum of another state in the ${\bf 10}$ and one in the ${\bf 5}$ representations. A euclidean M2-brane that wraps such a 3-chain will indeed mediate such a transition.

What we showed in the previous sections and in
appendix~\ref{app:globalres} is that such a 3-chain exists both in
$Y_\epsilon$ with $\epsilon \neq 0$, and for $Y_{\rm pert}$. However,
the possibility still exists that these 3-chains are inequivalent.
Let us now show that this is not the case, by constructing a homotopy
relating the 3-chains as $\epsilon$ is taken to zero.

Let us first understand the 3-chain on the non-perturbative side.
The curves of interest are
\be 
{D_5}^{(1)}=(v_4, v v_2 v_3+ \sigma w),\quad {D_5}^{(3)}= (v_4, v_2),
\quad {A_5}^{(3)}=(v_3,\, u x+v_4)\, .
\ee
Let us restrict to the $\Sigma_{10}$-curve by setting $u=0$. Now we are looking at three curves inside the divisor $v_4=0$. So, our space is now
\[\arraycolsep=10pt
\begin{array}{|c|c|c|c|c|c|c|} \hline
 Y   & \sigma &v_1 & v_2 & v_3 & \epsilon & w \\ \hline \hline
 0  & 1&-2& 1 & 0 & 0 & 0\\ \hline
 1  & 0&0& 1 & -1 & 0 & 0\\ \hline
 1  & 0&1& 0 & 0 & 0 & 0 \\ \hline
\end{array} \]
with irrelevant ideals: 
$
      (v_1, Y),\, (v_2, Y),\,(v_2, \sigma), \, (v_3, \sigma),\,   (v_1, v_3)
$
and hypersurface equation
\be
v_1 v_2 \left(\epsilon v_2 v_3+\sigma w\right) =0\, .
\ee
The three curves are now
\be 
{D_5}^{(1)}=(\epsilon v_2 v_3+ \sigma w),\quad {D_5}^{(3)}= (v_2),
\quad {A_5}^{(3)}=(v_3)\, .
\ee
As we saw before, as $w \rightarrow 0$, we see that ${D_5}^{(1)} \rightarrow {D_5}^{(3)}+{A_5}^{(3)}$. Since these three curves are co-bordant, there is a 3-chain connecting them. Let us define \emph{for each fixed $\epsilon$} the 3-chain $\Sigma_\epsilon$ via the following ideal:
\be \label{3chain}
\Sigma_{\epsilon}: \quad \begin{pmatrix} \sigma & v_2 v_3 \\ -\epsilon & w \end{pmatrix} \cdot \begin{pmatrix} \sin \theta \\ \cos \theta \end{pmatrix} = 0\,, \qquad \theta \in [0, \frac{\pi}{2}]
\ee
such that
\be
{\Sigma_{\epsilon}}|_{\theta=\pi/2}: \quad (\sigma) \cap \, w\rightarrow \infty = {{D_5}^{(1)}}|_{w=\infty} \qquad {\rm and} \qquad {\Sigma_{\epsilon}}|_{\theta = 0}: \quad (v_2 v_3, w) = {{D_5}^{(3)}}+{{A_5}^{(3)}}|_{w=0}\,.
\ee
If we now take $\epsilon \rightarrow 0$, the equation \eqref{3chain} defining the 3-chain becomes
\be
\big(w \cos \theta ,\, \sigma \sin \theta+v_2 v_3 \cos\theta \big) = \big(\cos \theta,\, \sigma \big)+\big(w,\, \sigma \sin \theta+v_2 v_3 \cos\theta \big) \,.
\ee
This 3-chain can be understood as a piecewise construction: First there is a portion lying outside the non-flat fiber at $\sigma=0$. Once that chain touches the Yukawa region and enters the non-flat fiber, there is a chain connecting $\sigma=0 \rightarrow v_2 v_3=0$.

\section{Conclusions}

We have shown that the $\Ytop$ Yukawa coupling in $SU(5)$ GUT F-theory models is generated by a
D1-instanton effect in the weak coupling description of the
system. We have directly argued for this result from a couple of
different viewpoints, namely weakly coupled IIB string theory and its
weakly coupled IIA mirror. We have also presented evidence from the
geometric M-theory viewpoint that further supports our conclusions.

We find this result interesting in that it demystifies somewhat the
nature of the coupling, and brings F-theory closer to the much better
understood weakly coupled IIB model building. A particularly
interesting take-away from our result is that there is a second way of
generating $\Ytop$ couplings in weakly coupled IIB models, in addition
to the known process mediated by euclidean D3 branes (see for instance
\cite{Blumenhagen:2008zz} for concrete examples). The euclidean D1
contribution analyzed in this paper is an independent effect, which
may also be a useful ingredient in the IIB model builder's toolbox:
the instanton contribution requires the mild condition of the
Calabi-Yau background having a conifold singularity admitting an
appropriate involution, which is a condition that is not too
constraining, and can be imposed early when constructing the model
(along the lines of \cite{Garcia-Etxebarria:2015lif}, for instance).

Our observation also raises some interesting questions in itself,
particularly in connection to the usual M-theory description of the
$E_6$ coupling.

As explained in \S\ref{sec:M-theory}, at weak coupling there is
tension between having a significant contribution to the
superpotential and keeping perturbative control of the theory, due to
the presence of light strings. We expect this tension to relax as we
go away from the weak coupling limit, but it would sill be interesting
to follow the fate of the light string states at $\int_{\bP^1}B_2=0$
as we go away from weak coupling. We have started this analysis in
\S\ref{sec:M-theory}, but more work is needed to properly understand
the effect of these states in ordinary F-theory compactifications away
from weak coupling.

A second point concerns the transmutation of instantons into classical
couplings as we go towards strong coupling. As we argued in
\S\ref{sec:M-theory}, the fact that D1-instantons can describe some
classical couplings in F-theory is not too surprising, since both
effects lift in M-theory to M2-branes wrapping appropriate chains
\cite{Marsano:2011hv,Martucci:2015dxa,Martucci:2015oaa}. But it would
clearly be interesting to elucidate this point further via an explicit
dualization of the D1 instantons into M2 branes, and an explicit
matching of moduli in both pictures.

Finally, it would be interesting to understand whether we can describe
what happens at a Yukawa $E_6$ point in F-theory without resorting to fourfold
resolutions, such as those performed in \S\ref{sec:M-theory}. The NCCR approach we have used here to treat the conifold
singularity in perturbative IIB string theory begs for a counterpart
on the F-theory fourfold. Perhaps an approach along the lines of
\cite{Collinucci:2014taa} might be fruitful.

\acknowledgments

We thank Thomas Grimm, Fernando Marchesano, Luca Martucci, Raffaele
Savelli and Roberto Valandro for illuminating discussions, Diego
Regalado and Timo Weigand for comments on the draft, and each other's
institutions for generous hospitality while this work was being
completed.

This work was partially supported by FNRS - Belgium (convention 4.4503.15). A.C. is a Research Associate of the Fonds de la Recherche Scientifique F.N.R.S. (Belgium).

\appendix
\section{Transport from NCCR to resolved space}\label{app:transport}
In this section, we will explain how to transport objects of the
bounded derived category D$^b$(mod-$A$) of modules over the
noncommutative ring $A$ to objects of the bounded derived category of
coherent sheaves of the resolved space D$^b(X_{\rm smooth}$).
Intuitively, we will see how to transport a brane on the singular
space to a brane on the resolved space. We will be concise and show
the machinery directly in the cases of interest.

Let us repeat some definitions already explained in \S\ref{subsec:nccr} for convenience. On the singular side, we have a hypersurface ring
\begin{equation}
R = \C[\xi, u, \sigma, w]/(-\xi^2+u^2+\sigma w)
\end{equation}
admitting two \emph{matrix factorizations} $(\phi, \psi)$ and $(\psi, \phi)$, with
\begin{equation}
  \phi = \begin{pmatrix} u-\xi & \sigma \\ -w & u+\xi \end{pmatrix} \,, \quad {\rm and} \quad \psi = \begin{pmatrix} u+\xi & -\sigma \\ w & u-\xi  \end{pmatrix} 
  \end{equation}
such that $\phi \cdot \psi = \psi \cdot \phi =  (-\xi^2+u^2+\sigma w) \cdot \1$. From these matrices, two modules can be constructed as cokernels:
\begin{eqnarray}
M&\equiv&  {\rm coker}\big(\,R^{\oplus 2} \stackrel{\psi}\longrightarrow R^{\oplus 2} \, \big)\, ,\\
\tilde M&\equiv&  {\rm coker}\big(\,R^{\oplus 2} \stackrel{\phi}\longrightarrow R^{\oplus 2} \, \big)\, .
\end{eqnarray}
These are what are known as irreducible \emph{maximal Cohen-Macaulay} modules. In this case, the conifold admits only these two up to isomorphism. 
In order to perform a \emph{noncommutative crepant resolution} (NCCR), we are instructed to pick one of these two, say $M$, and construct the endomorphism algebra 
\begin{equation}
A \equiv {\rm End}(R \oplus M)
\end{equation}
This quiver has a path algebra encoded by the quiver
\begin{center}
  \includegraphics{conifold-path-algebra}
\end{center}
with superpotential $W = \alpha_1 \beta_1 \alpha_2 \beta_2 - \alpha_1 \beta_2 \alpha_2 \beta_1$
We define two projective right $A$-modules 
\begin{equation}
P_0 = e_0 \cdot A\,, \quad P_1 = e_1 \cdot A\,,
\end{equation}
of paths ending in the node in the label. 

In order to obtain a resolution, we construct the following representation of the quiver
\begin{center}
  \includegraphics{probe-representation}
\end{center}
Each node has a $\C^*$-action that redefines the basis of each $\C$
vector space. The arrows are complex numbers that transform under the
relative $\C^*$ action as the toric coordinates of the conifold:
\begin{equation}
\arraycolsep=10pt
\begin{array}{|c|c|c|c|} 
\hline
\alpha_1 &\alpha_2 &\beta_1 &\beta_2\\ \hline
1 & 1 & -1 & -1 \\ \hline
\end{array} 
\end{equation}
We must also impose a D-term constraint
\begin{equation}
|\alpha_1|^2+|\alpha_2|^2-|\beta_1|^2-|\beta_2|^2 = t\,.
\end{equation}
The two resolved phases $\tilde X_\pm$ correspond to $t >0$ and
$t < 0$. It is known \cite{Bondal:aa, Bergh:aa} that there is a
derived correspondence D$^b(X_+) \cong$ D$^b({\rm mod}-A) \cong$
D$^b(X_-)$. The correspondence in the case $\xi>0$ is
\begin{equation}
P_0 \mapsto \cO\,, \quad P_1 \mapsto \cO(1)\,.
\end{equation}
Indeed, we see heuristically that\footnote{Notice that we are abusing
  notation slightly here, by viewing $\alpha_i$ and $\beta_i$ both as
  paths in the quiver, and coordinates in the resolved space.}
\begin{eqnarray}
{\rm Hom}(P_0, P_1) &\cong& {\rm Hom}(\cO, \cO(1))  = \langle \alpha_1, \alpha_2 \rangle\, ,\\
{\rm Hom}(P_1, P_0) &\cong& {\rm Hom}(\cO(1), \cO)  = \langle \beta_1, \beta_2 \rangle\, .
\end{eqnarray}

Now we can study how our various branes are mapped from the singular
to the $X_+$ phase. For the non-compact branes, we can now readily
confirm the mappings in \eqref{noncompmapping}.  The fractional
branes, on the other hand, require more work. Let us define the brane
$S_0$ through its projective resolution \eqref{fractionalmappings}
\begin{equation}
  \includegraphicstikz{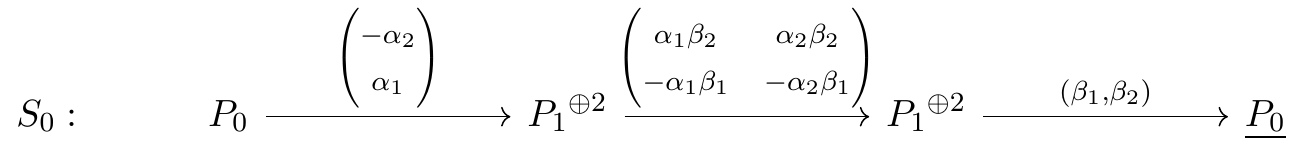}
\end{equation}
This is lifted to the following complex in the resolved $X_+$
\begin{equation}
  \includegraphicstikz{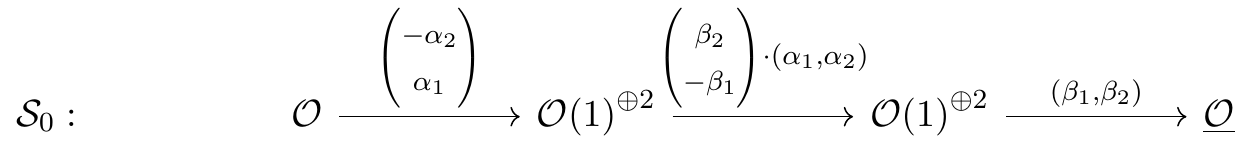}
\end{equation}
This object $\mathcal{S}_0 \in $ D$^b(X_+)$ can be rewritten after a basis transformation as follows:
\begin{equation}
  \includegraphicstikz{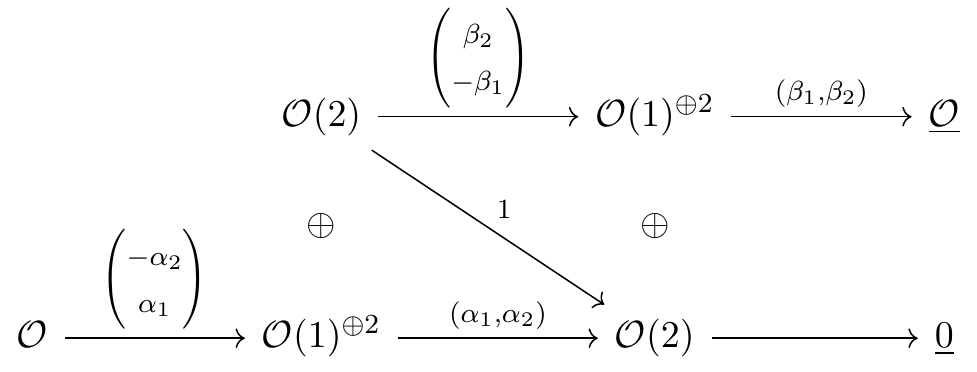}
\end{equation}
We have underlined the zero on the rhs to indicate that that is the
starting position, i.e. the degree zero object in the
complex.

This can be understood as a so-called \emph{mapping cone} between two
complexes. This essentially means that we can regard this object as a
bound state between the lower complex and the upper complex via
tachyon condensation. We will not go into this here, but refer instead
to \cite{Aspinwall:2009isa} for a general introduction to these
notions, and to \cite{Collinucci:2014qfa} for a concise introduction
in the string theory context. Suffice it to say that in this complex,
the lower part has trivial cohomology at every position. In fact, it
corresponds to the skyscraper sheaf over the deleted point
$(\alpha_1, \alpha_2) = (0,0)$.

By taking the cohomology of the full complex, we notice that the
portion containing $\cO(2) \stackrel{1}\rightarrow \cO(2)$ drops out
entirely. There exists a so-called \emph{quasi-isomorphism} that
simply maps it to the upper complex:
\begin{equation}
  \includegraphicstikz{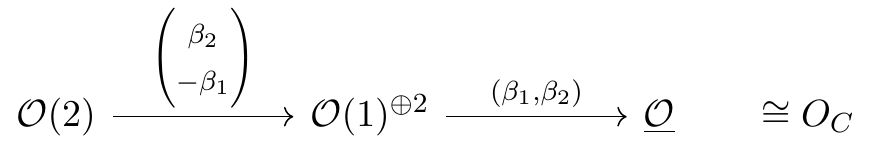}
\end{equation}
where $C$ is the resolution $\P^1$. 
To summarize, we can say that
\begin{equation}
S_0 \cong \cO_C
\end{equation}
corresponds to a D-brane wrapping the resolution curve. In our case, this will be a Euclidean D1.

Now let us run through the calculation for the other simple representation $S_1$:
\begin{equation}
  \includegraphicstikz{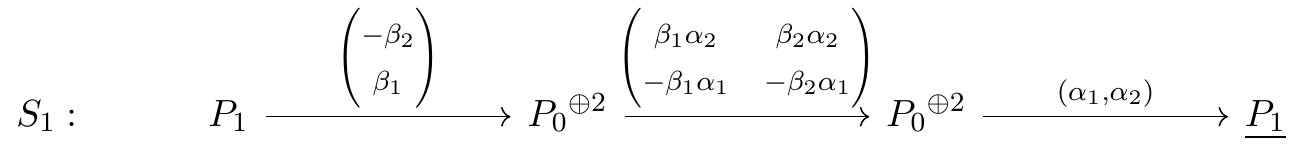}
\end{equation}
It transports in the resolved phase to the following complex of sheaves:
\begin{equation}
  \includegraphicstikz{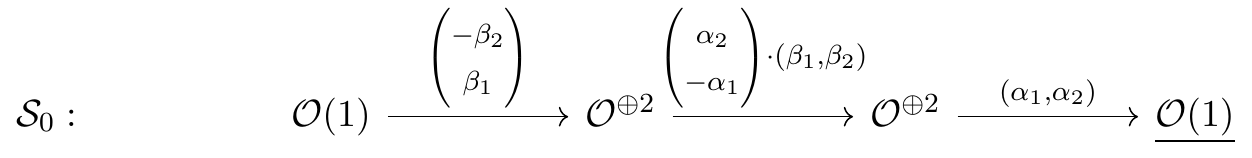}
\end{equation}
which can be rewritten as
\begin{equation}
  \includegraphicstikz{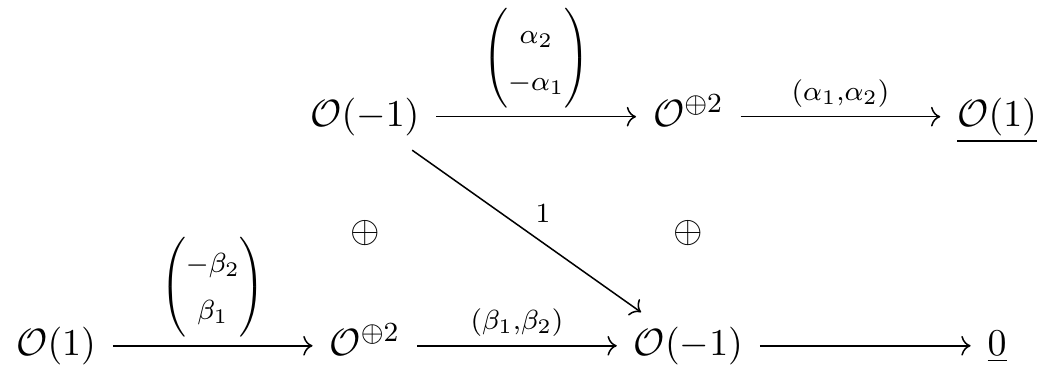}
\end{equation}
Now the upper half corresponds to a trivial object (i.e. one that has no cohomology). After a quasi-isomorphism to eliminate it, we are left with the following complex:
\begin{equation}
  \includegraphicstikz{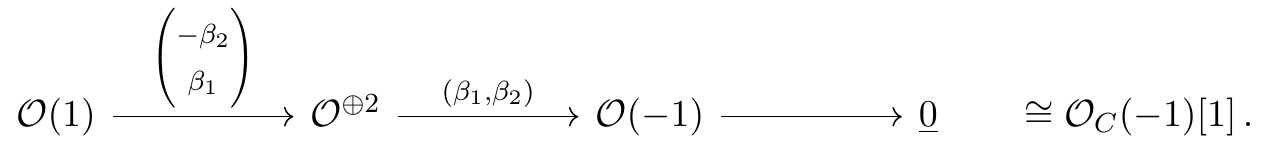}
\end{equation}
This shifted object corresponds to an \emph{anti}-D1-brane with flux of Chern number minus one wrapped on the resolution $\P^1$:
\begin{equation}
S_1 \cong \cO_C(-1)[1]\,.
\end{equation}
Indeed, now we see that these two fractional branes have net D1-charge zero, and net D$(-1)$-charge one, since the negative flux on an anti-D1-brane induces positive D$(-1)$-charge.

\section{Resolving the generic fourfold} \label{app:globalres}

We will now study the resolution of the model written in the previous
section. We will do it for a generic fiber $Y_\epsilon$ of the family
of fourfolds, with $\epsilon\neq 0$. We start with the full ideal for $Y_\epsilon$:
\be
\left( -y^2+x^3 v+(u x z+\sigma^2 t z^3)^2+ \sigma w x^2 z^2, \quad t
  v = \epsilon \right)\, .
\ee
In order to perform the resolution as economically as possible, we introduce an auxiliary coordinate $Y$, with the following relation
\be
Y \equiv y+u x z+v_1 \sigma^2 t z^3\, .
\ee
Now, our model is written as the following ideal:
\be
\big( Y \left(-Y+2 u x z+2 v_1 \sigma^2 t z^3\right)  + v_1 v_2
\left(x^3 v v_2+\sigma w x^2 z^2\right), \quad tv - \epsilon \big)\, .
\ee

The fully resolved fourfold $\tilde Y_\epsilon$ is then given by the
ideal
\be \label{Yeps}
\big( Y \left(-Y v_3 v_4+2 u x z+2 v_1 v_4 \sigma^2 t z^3\right)  + v_1 v_2 \left(x^3 v v_2 v_3+\sigma w x^2 z^2\right), \quad tv - \epsilon \big)
\ee
in the following ambient space
\[\arraycolsep=10pt
\begin{array}{|c|c|c|c|c|c|c|c|c|c|c|c|} \hline
x & Y &  z & t & v & \sigma &v_1 & v_2 & v_3 & v_4 & u & w \\ \hline \hline
2 & 3 &  1 & 0 & 0& 0&0& 0 & 0 & 0 & 0 & 0\\ \hline
1 & 1 & 0 & 1 & -1& 0&0& 0 & 0 & 0 & 0 & 0\\ \hline
1 & 1 & 0 & 0 & 0& 1&-1& 0 & 0 & 0 & 0 & 0\\ \hline
1 & 1 & 0 & 0 & 0& 0&1& -1 & 0 & 0 & 0 & 0\\ \hline
0 & 1 & 0 & 0 & 0& 0&0& 1 & -1& 0 & 0 & 0 \\ \hline
0 & 1 & 0 & 0 & 0& 0&1& 0 & 0 & -1 & 0 & 0\\ \hline
\end{array} \]
with irrelevant ideals: 
\begin{eqnarray}
&& (x, Y, z),\, (x, Y, t),\, (z, v),\, (x, Y, \sigma),\,  (t, v_i),\,   (z, v_i), \,  \\ \nonumber
&& (v_1, Y),\, (v_2, Y),\,(v_2, \sigma), \, (v_3, \sigma),\,   (v_1, v_3),\, (v_4, x) 
\end{eqnarray}
The blow-down map for this resolved space is the following
\begin{eqnarray} \label{blowdown}
&&(x,\, Y ,\,  z ,\, t ,\, v ,\, \sigma ,\, v_1 ,\, v_2 ,\, v_3 ,\, v_4) \\
&&\mapsto (\underline{x}=x v_1 v_2^2 v_3^2 v_4,\quad \underline{y}=y v_1 v_2^2 v_3^3 v_4^2, \quad \underline{z}=z,\quad \underline{t}=t, \quad \underline{v}=v, \quad \underline{\sigma}=\sigma v_1 v_2 v_3 v_4) \nonumber
\end{eqnarray}
where the underlined coordinates are those on the blown-down space.

\subsection{Yukawa interactions at strong coupling}

We will now study how representations and Yukawa couplings appear in
F-theory, following the logic of \cite{Krause:2011xj} and
\cite{Martucci:2015dxa}.

\subsubsection{Codimension one}

Let us study first the fiber over a generic point\footnote{Here, by
  `generic' we mean a point such that $u, w \neq 0$.} in
$\underline{\sigma} \equiv \sigma v_1 v_2 v_3 v_4 = 0$ for a generic
$Y_\epsilon$. In this case the fiber is comprised of the following
divisors:
\begin{align}
&{C}:  &(\sigma) \cap Y_\epsilon  &\quad \cong \quad \big(\sigma,\, Y (-Y v_4+2 u x z)+v_1 x^3 v) \big)\\
&{A_4}^{(1)}: & (v_1) \cap Y_\epsilon &\quad \cong \quad  (v_1,\, -v_4+2 u x)\\
&{A_4}^{(2)}:  &(v_2) \cap Y_\epsilon &\quad \cong \quad  (v_2,\, -v_3 v_4+2 u x+2 v_1 v_4)\\
&{A_4}^{(3)}:  &(v_3) \cap Y_\epsilon &\quad \cong \quad  (v_3,\, Y (2 u x+2 v_4)+ v_2 w x^2)\\
&{A_4}^{(4)}:  &(v_4) \cap Y_\epsilon &\quad \cong \quad  \big(v_4,\, 2 Y u+v_1 v_2 (v v_2 v_3+\sigma w)\big)
\end{align}
We have used the irrelevant ideals to `gauge-fix' coordinates to `1' whenever they are forbidden from vanishing. It can be shown that all of these divisors are $\P^1$'s, and that they have the intersection pattern of the affine Dynkin diagram for $A_4$, as expected:
\begin{center}
  \includegraphics{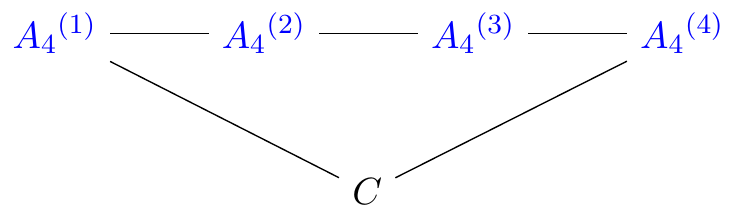}
\end{center} 
The blue part corresponds to the non-affine Dynkin diagram.

\subsubsection{Codimension two}

\begin{paragraph}{$\Sigma_5$ matter curve:}
The locus $(\underline{\sigma}, w)$ is the curve where the $SU(5)$ stack meets the extra `flavor' brane. Hence, we expect an $A_5$ enhancement. Indeed, the resolution shows that the ${A_4}^3$ curve splits, leading to the following map:

\begin{align}
&C &&\longrightarrow &C&=\big(\sigma,\, Y (-Y v_4+2 u x z)+v_1 x^3 v)\big)\\
&{A_4}^{(1)}&&\longrightarrow &{A_5}^{(1)}&=(v_1,\, -v_4+2 u x)\\
&{A_4}^{(2)}&&\longrightarrow &{A_5}^{(2)}&= (v_2,\, -v_3 v_4+2 u x+2 v_1 v_4)\\
&{A_4}^{(3)}&&\longrightarrow &{A_5}^{(3)}&=(v_3,\, 2 u x+2 v_4) \quad + \quad {A_5}^{(4)}=(v_3,\, Y) \\
&{A_4}^{(4)}&&\longrightarrow &{A_5}^{(5)}&=(v_4,\, 2 Y u+v v_1 v_2^2 v_3)
\end{align}

The arrangement takes the following form
\begin{center}
  \includegraphics{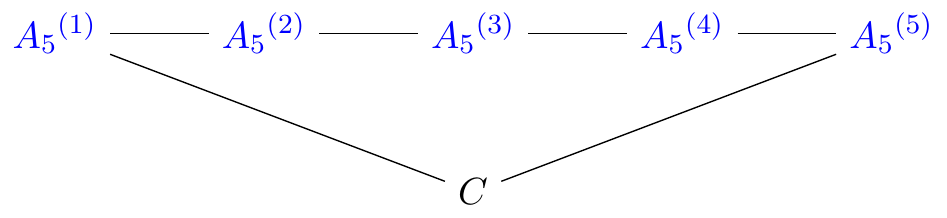}
\end{center}

\end{paragraph}
\begin{paragraph}{$\Sigma_{\bf 10}$ matter curve:}
Let us proceed to the $D_5$ enhancement at $(\underline{\sigma}, u)$. This is the locus where the GUT stack meets its image.

\begin{align}
&C &&\longrightarrow  &C&=(\sigma,\, -Y^2 v_4+v_1 x^3 v) \\
&{A_4}^{(1)}&& \longrightarrow &{D_5}^{(2)}&=(v_1,\, v_4)\\
&{A_4}^{(2)}&& \longrightarrow &{D_5}^{(3)}&=(v_2,\, v_4) \quad + \quad {D_5}^{(4)}=(v_2,\, 2 v_1-v_3)\\
&{A_4}^{(3)}&& \longrightarrow &{D_5}^{(5)}&=(v_3,\, 2 Y v_4+ v_2 w x^2)\\
&{A_4}^{(4)}&& \longrightarrow &{D_5}^{(2)}&=(v_4,\, v_1) \quad + \quad {D_5}^{(3)} =(v_4,\, v_2) \quad + \quad {D_5}^{(1)} = (v_4, v v_2 v_3+\sigma w)
\end{align}
The arrangement takes the following shape:
\begin{center}
  \includegraphics{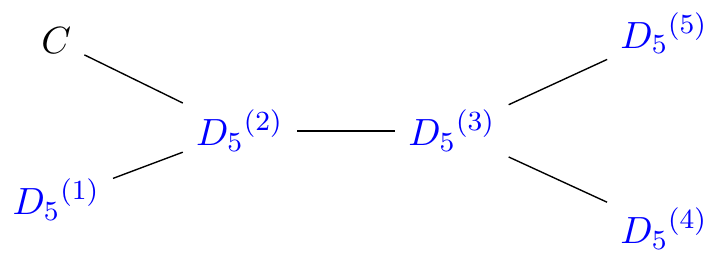}
\end{center}

\subsubsection{Codimension three}
\label{sec:strong-coupling-codim-3}

\end{paragraph}
\begin{paragraph}{``$E_6$'' enhancement: $\Sigma_{\bf 5} \rightarrow {\rm Yuk}_{E_6}$}
Now let us approach the Yukawa point $(\underline{\sigma}, u, w)$, where all D-brane stacks meet each other and the O7-plane, coming from the $\Sigma_{\bf 5}$-curve, we find the following splittings

\begin{align}
&C &&\longrightarrow &C&=\big(\sigma,\, -Y^2 v_4+v_1 x^3 v \big)\\
&{A_5}^{(1)}&&\longrightarrow &{E_6}^{(2)}&=(v_1,\, v_4)\\
&{A_5}^{(2)}&&\longrightarrow &{E_6}^{(3)}&=  (v_2, v_4) \quad + \quad {E_6}^{(6)}=(v_2,\, 2 v_1-v_3)\\
&{A_5}^{(3)}&&\longrightarrow &{E_6}^{(4)}&= (v_3, v_4) \label{hom2}\\
&{A_5}^{(4)}&&\longrightarrow &{E_6}^{(5)}&= (v_3, Y)\\
&{A_5}^{(5)}&&\longrightarrow &{E_6}^{(2)}&=(v_4,\, v_1) \quad + \quad 2 \times {E_6}^{(3)}=(v_4,\, v_2^2) \quad + \quad {E_6}^{(4)}= (v_4,\, v_3)
\end{align}
with the following shape of a non-extended $E_6$ Dynkin graph:
\begin{center}
  \includegraphics{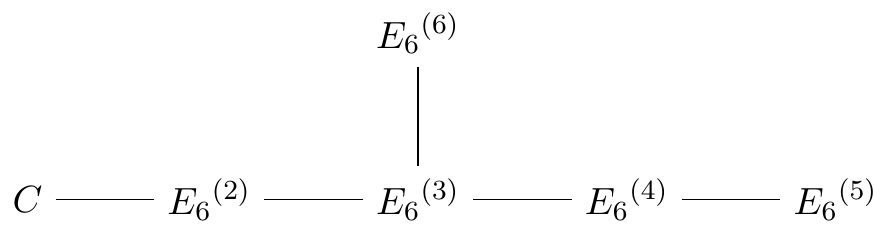}
\end{center} 
\end{paragraph}

Here, implicitly $C = {E_6}^{(1)}$. We stress that this is the
non-affine Dynkin diagram of $E_6$. This peculiarity was discovered in
\cite{Esole:2011sm}. Common lore would have us expect an affine-Dynkin
diagram. This paradox was solved in \cite{Braun:2013cb}, where the
authors showed that the standard Weierstrass/Tate model for $SU(5)$
actually has a T-brane built into it. In other words, a
non-diagonalizable vev of the $E_6$ Higgs field taking values in the
$SU(2) \times U(1)$, breaking it to $SU(5)$ is responsible for this
strange fiber geometry. By switching off this vev and following the
consequences for the elliptic fibration, one recovers the expected
affine Dynkin fiber. As explained in \cite{Cecotti:2010bp,
  Hayashi:2009ge} introducing this T-brane vev is crucial in order to
have the top quark much heavier than the other two generations.

\begin{paragraph}{``$E_6$'' enhancement: $\Sigma_{\bf 10} \rightarrow {\rm Yuk}_{E_6}$}
Let us now approach the Yukawa point from the $\Sigma_{\bf 10}$-curve.

\begin{align}
&C &&\longrightarrow  &C&=(\sigma,\, -Y^2 v_4+v_1 x^3 v) \\
&{D_5}^{(1)}&& \longrightarrow &{E_6}^{(3)}&=(v_4, v_2) \quad + \quad {E_6}^{(4)}= (v_4, v_3) \label{hom1}\\
&{D_5}^{(2)}&& \longrightarrow &{E_6}^{(2)}&=(v_1,\, v_4)\\
&{D_5}^{(3)}&& \longrightarrow &{E_6}^{(3)}&=(v_2,\, v_4) \label{hom3}\\
&{D_5}^{(4)}&& \longrightarrow &{E_6}^{(6)}&=(v_2,\, 2 v_1-v_3)\\
&{D_5}^{(5)}&& \longrightarrow &{E_6}^{(5)}&=(v_3,\, Y) \quad + \quad {E_6}^{(4)}=(v_3,\, v_4)
\end{align}
with the same arrangement, since the curves are the same as before.

\end{paragraph}

\subsubsection{Yukawa interactions}
Let us now study a possible Yukawa interaction. The various elements of the ${\bf 5}$ and ${\bf 10}$ representations of $SU(5)$ can be formal linear combinations of the $\P^1$'s at the respective enhancement loci. However, some elements are straightforward curves. We will focus on the following examples, 
\begin{itemize}
\item ${\bf 5}: \quad {A_5}^{(3)}$
\item ${\bf 10}: \quad {D_5}^{(1)}\, \quad {D_5}^{(3)}$
\end{itemize}
and exploit the relations \eqref{hom1}, \eqref{hom2}, \eqref{hom3}:
\begin{align}
{D_5}^{(1)} = {E_6}^{(3)}+{E_6}^{(4)}\qquad
{A_5}^{(3)} &= {E_6}^{(4)}\qquad
{D_5}^{(3)} =  {E_6}^{(3)}\\
\Longrightarrow \qquad {D_5}^{(1)} &= {D_5}^{(3)}+{A_5}^{(3)}
\end{align}
This equation couples two elements of the ${\bf 10}$ with one of the ${\bf 5}$.

Let us calculate the weights of the various membranes by intersecting with the vector of curves $\vec{v}\equiv ({A_4}^{(1)}, \ldots, {A_4}^{(4)})$:
\begin{align}
{\bf 5}&: \quad {A_5}^{(3)} \cdot \vec{v} = (0, 1, -1, 0) \leftrightarrow -C_{{\bf 5}_{3}}\\
{\bf 10}&: \quad {D_5}^{(3)} \cdot \vec{v} = (1, -1, 1, -1) \leftrightarrow -C_{{\bf 10}_{5}}\,,  \quad {D_5}^{(1)}\cdot \vec{v} = (1, 0,0,-1) \leftrightarrow C_{{\bf 10}_{6}}
\end{align}
Here, we have labeled curves following the conventions of \cite{Krause:2011xj}, by the order of appearance of an element in the construction of a representation, starting from the highest weight at number one. This confirms our expectation that these curves sit in the aforementioned representations. Therefore, the relation translates to
\begin{equation}
C_{{\bf 10}_{6}}+C_{{\bf 10}_{5}}+C_{{\bf 5}_{3}} = 0\,.
\end{equation}
This relation of homology classes implies the existence of a 3-chain
with these three curves as boundaries. If this 3-chain is wrapped by
an open M2-brane, it mediates a
$C_{{\bf 10}_{6}} \rightarrow -C_{{\bf 10}_{5}}-C_{{\bf 5}_{3}}$
transition, which implies the existence of the $\Ytop$ coupling.

\bibliographystyle{JHEP}
\bibliography{refs}

\providecommand{\href}[2]{#2}\begingroup\raggedright\begin{thebibliography}{100}

\bibitem{Donagi:2008ca}
R.~Donagi and M.~Wijnholt, \emph{{Model Building with F-Theory}},
  \href{http://dx.doi.org/10.4310/ATMP.2011.v15.n5.a2}{\emph{Adv. Theor. Math.
  Phys.} {\bf 15} (2011) 1237--1317},
  [\href{http://arxiv.org/abs/0802.2969}{{\tt 0802.2969}}].

\bibitem{Beasley:2008dc}
C.~Beasley, J.~J. Heckman and C.~Vafa, \emph{{GUTs and Exceptional Branes in
  F-theory - I}},
  \href{http://dx.doi.org/10.1088/1126-6708/2009/01/058}{\emph{JHEP} {\bf 01}
  (2009) 058}, [\href{http://arxiv.org/abs/0802.3391}{{\tt 0802.3391}}].

\bibitem{Hayashi:2008ba}
H.~Hayashi, R.~Tatar, Y.~Toda, T.~Watari and M.~Yamazaki, \emph{{New Aspects of
  Heterotic--F Theory Duality}},
  \href{http://dx.doi.org/10.1016/j.nuclphysb.2008.07.031}{\emph{Nucl. Phys.}
  {\bf B806} (2009) 224--299}, [\href{http://arxiv.org/abs/0805.1057}{{\tt
  0805.1057}}].

\bibitem{Beasley:2008kw}
C.~Beasley, J.~J. Heckman and C.~Vafa, \emph{{GUTs and Exceptional Branes in
  F-theory - II: Experimental Predictions}},
  \href{http://dx.doi.org/10.1088/1126-6708/2009/01/059}{\emph{JHEP} {\bf 01}
  (2009) 059}, [\href{http://arxiv.org/abs/0806.0102}{{\tt 0806.0102}}].

\bibitem{Donagi:2008kj}
R.~Donagi and M.~Wijnholt, \emph{{Breaking GUT Groups in F-Theory}},
  \href{http://dx.doi.org/10.4310/ATMP.2011.v15.n6.a1}{\emph{Adv. Theor. Math.
  Phys.} {\bf 15} (2011) 1523--1603},
  [\href{http://arxiv.org/abs/0808.2223}{{\tt 0808.2223}}].

\bibitem{Vafa:1996xn}
C.~Vafa, \emph{{Evidence for F theory}},
  \href{http://dx.doi.org/10.1016/0550-3213(96)00172-1}{\emph{Nucl. Phys.} {\bf
  B469} (1996) 403--418}, [\href{http://arxiv.org/abs/hep-th/9602022}{{\tt
  hep-th/9602022}}].

\bibitem{Font:2012wq}
A.~Font, L.~E. Ibanez, F.~Marchesano and D.~Regalado, \emph{{Non-perturbative
  effects and Yukawa hierarchies in F-theory SU(5) Unification}},
  \href{http://dx.doi.org/10.1007/JHEP03(2013)140,
  10.1007/JHEP07(2013)036}{\emph{JHEP} {\bf 03} (2013) 140},
  [\href{http://arxiv.org/abs/1211.6529}{{\tt 1211.6529}}].

\bibitem{Font:2013ida}
A.~Font, F.~Marchesano, D.~Regalado and G.~Zoccarato, \emph{{Up-type quark
  masses in SU(5) F-theory models}},
  \href{http://dx.doi.org/10.1007/JHEP11(2013)125}{\emph{JHEP} {\bf 11} (2013)
  125}, [\href{http://arxiv.org/abs/1307.8089}{{\tt 1307.8089}}].

\bibitem{Marchesano:2015dfa}
F.~Marchesano, D.~Regalado and G.~Zoccarato, \emph{{Yukawa hierarchies at the
  point of E$_{8}$ in F-theory}},
  \href{http://dx.doi.org/10.1007/JHEP04(2015)179}{\emph{JHEP} {\bf 04} (2015)
  179}, [\href{http://arxiv.org/abs/1503.02683}{{\tt 1503.02683}}].

\bibitem{Carta:2015eoh}
F.~Carta, F.~Marchesano and G.~Zoccarato, \emph{{Fitting fermion masses and
  mixings in F-theory GUTs}},
  \href{http://dx.doi.org/10.1007/JHEP03(2016)126}{\emph{JHEP} {\bf 03} (2016)
  126}, [\href{http://arxiv.org/abs/1512.04846}{{\tt 1512.04846}}].

\bibitem{GarciaEtxebarria:2012zm}
I.~Garcia-Etxebarria, H.~Hayashi, R.~Savelli and G.~Shiu, \emph{{On quantum
  corrected Kahler potentials in F-theory}},
  \href{http://dx.doi.org/10.1007/JHEP03(2013)005}{\emph{JHEP} {\bf 03} (2013)
  005}, [\href{http://arxiv.org/abs/1212.4831}{{\tt 1212.4831}}].

\bibitem{Minasian:2015bxa}
R.~Minasian, T.~G. Pugh and R.~Savelli, \emph{{F-theory at order $\alpha'^3$}},
  \href{http://dx.doi.org/10.1007/JHEP10(2015)050}{\emph{JHEP} {\bf 10} (2015)
  050}, [\href{http://arxiv.org/abs/1506.06756}{{\tt 1506.06756}}].

\bibitem{Grimm:2013bha}
T.~W. Grimm, J.~Keitel, R.~Savelli and M.~Weissenbacher, \emph{{From M-theory
  higher curvature terms to $\alpha'$ corrections in F-theory}},
  \href{http://dx.doi.org/10.1016/j.nuclphysb.2015.12.011}{\emph{Nucl. Phys.}
  {\bf B903} (2016) 325--359}, [\href{http://arxiv.org/abs/1312.1376}{{\tt
  1312.1376}}].

\bibitem{Bies:2014sra}
M.~Bies, C.~Mayrhofer, C.~Pehle and T.~Weigand, \emph{{Chow groups, Deligne
  cohomology and massless matter in F-theory}},
  \href{http://arxiv.org/abs/1402.5144}{{\tt 1402.5144}}.

\bibitem{Collinucci:2014taa}
A.~Collinucci and R.~Savelli, \emph{{F-theory on singular spaces}},
  \href{http://dx.doi.org/10.1007/JHEP09(2015)100}{\emph{JHEP} {\bf 09} (2015)
  100}, [\href{http://arxiv.org/abs/1410.4867}{{\tt 1410.4867}}].

\bibitem{Anderson:2015yzz}
L.~B. Anderson, F.~Apruzzi, X.~Gao, J.~Gray and S.-J. Lee, \emph{{Instanton
  superpotentials, Calabi-Yau geometry, and fibrations}},
  \href{http://dx.doi.org/10.1103/PhysRevD.93.086001}{\emph{Phys. Rev.} {\bf
  D93} (2016) 086001}, [\href{http://arxiv.org/abs/1511.05188}{{\tt
  1511.05188}}].

\bibitem{Bianchi:2012kt}
M.~Bianchi, G.~Inverso and L.~Martucci, \emph{{Brane instantons and fluxes in
  F-theory}}, \href{http://dx.doi.org/10.1007/JHEP07(2013)037}{\emph{JHEP} {\bf
  07} (2013) 037}, [\href{http://arxiv.org/abs/1212.0024}{{\tt 1212.0024}}].

\bibitem{Bianchi:2012pn}
M.~Bianchi, A.~Collinucci and L.~Martucci, \emph{{Freezing E3-brane instantons
  with fluxes}}, \href{http://dx.doi.org/10.1002/prop.201200030}{\emph{Fortsch.
  Phys.} {\bf 60} (2012) 914--920}, [\href{http://arxiv.org/abs/1202.5045}{{\tt
  1202.5045}}].

\bibitem{Bianchi:2011qh}
M.~Bianchi, A.~Collinucci and L.~Martucci, \emph{{Magnetized E3-brane
  instantons in F-theory}},
  \href{http://dx.doi.org/10.1007/JHEP12(2011)045}{\emph{JHEP} {\bf 12} (2011)
  045}, [\href{http://arxiv.org/abs/1107.3732}{{\tt 1107.3732}}].

\bibitem{Blumenhagen:2010ja}
R.~Blumenhagen, A.~Collinucci and B.~Jurke, \emph{{On Instanton Effects in
  F-theory}}, \href{http://dx.doi.org/10.1007/JHEP08(2010)079}{\emph{JHEP} {\bf
  08} (2010) 079}, [\href{http://arxiv.org/abs/1002.1894}{{\tt 1002.1894}}].

\bibitem{Cvetic:2009ah}
M.~Cvetic, I.~Garcia-Etxebarria and R.~Richter, \emph{{Branes and instantons at
  angles and the F-theory lift of O(1) instantons}},
  \href{http://dx.doi.org/10.1063/1.3327564}{\emph{AIP Conf. Proc.} {\bf 1200}
  (2010) 246--260}, [\href{http://arxiv.org/abs/0911.0012}{{\tt 0911.0012}}].

\bibitem{Cvetic:2009mt}
M.~Cvetic, I.~Garcia-Etxebarria and R.~Richter, \emph{{Branes and instantons
  intersecting at angles}},
  \href{http://dx.doi.org/10.1007/JHEP01(2010)005}{\emph{JHEP} {\bf 01} (2010)
  005}, [\href{http://arxiv.org/abs/0905.1694}{{\tt 0905.1694}}].

\bibitem{Cvetic:2010rq}
M.~Cvetic, I.~Garcia-Etxebarria and J.~Halverson, \emph{{Global F-theory
  Models: Instantons and Gauge Dynamics}},
  \href{http://dx.doi.org/10.1007/JHEP01(2011)073}{\emph{JHEP} {\bf 01} (2011)
  073}, [\href{http://arxiv.org/abs/1003.5337}{{\tt 1003.5337}}].

\bibitem{Cvetic:2012ts}
M.~Cvetic, R.~Donagi, J.~Halverson and J.~Marsano, \emph{{On Seven-Brane
  Dependent Instanton Prefactors in F-theory}},
  \href{http://dx.doi.org/10.1007/JHEP11(2012)004}{\emph{JHEP} {\bf 11} (2012)
  004}, [\href{http://arxiv.org/abs/1209.4906}{{\tt 1209.4906}}].

\bibitem{Cvetic:2011gp}
M.~Cvetic, I.~Garcia~Etxebarria and J.~Halverson, \emph{{Three Looks at
  Instantons in F-theory -- New Insights from Anomaly Inflow, String Junctions
  and Heterotic Duality}},
  \href{http://dx.doi.org/10.1007/JHEP11(2011)101}{\emph{JHEP} {\bf 11} (2011)
  101}, [\href{http://arxiv.org/abs/1107.2388}{{\tt 1107.2388}}].

\bibitem{Donagi:2010pd}
R.~Donagi and M.~Wijnholt, \emph{{MSW Instantons}},
  \href{http://dx.doi.org/10.1007/JHEP06(2013)050}{\emph{JHEP} {\bf 06} (2013)
  050}, [\href{http://arxiv.org/abs/1005.5391}{{\tt 1005.5391}}].

\bibitem{Grimm:2011sk}
T.~W. Grimm and R.~Savelli, \emph{{Gravitational Instantons and Fluxes from
  M/F-theory on Calabi-Yau fourfolds}},
  \href{http://dx.doi.org/10.1103/PhysRevD.85.026003}{\emph{Phys. Rev.} {\bf
  D85} (2012) 026003}, [\href{http://arxiv.org/abs/1109.3191}{{\tt
  1109.3191}}].

\bibitem{Grimm:2011dj}
T.~W. Grimm, M.~Kerstan, E.~Palti and T.~Weigand, \emph{{On Fluxed Instantons
  and Moduli Stabilisation in IIB Orientifolds and F-theory}},
  \href{http://dx.doi.org/10.1103/PhysRevD.84.066001}{\emph{Phys. Rev.} {\bf
  D84} (2011) 066001}, [\href{http://arxiv.org/abs/1105.3193}{{\tt
  1105.3193}}].

\bibitem{Heckman:2008es}
J.~J. Heckman, J.~Marsano, N.~Saulina, S.~Schafer-Nameki and C.~Vafa,
  \emph{{Instantons and SUSY breaking in F-theory}},
  \href{http://arxiv.org/abs/0808.1286}{{\tt 0808.1286}}.

\bibitem{Kerstan:2014dva}
M.~Kerstan, \emph{{Abelian gauge symmetries and fluxed instantons in
  compactifications of type IIB and F-theory}}.
\newblock PhD thesis, Heidelberg U., 2014.
\newblock \href{http://arxiv.org/abs/1402.3636}{{\tt 1402.3636}}.

\bibitem{Kerstan:2012cy}
M.~Kerstan and T.~Weigand, \emph{{Fluxed M5-instantons in F-theory}},
  \href{http://dx.doi.org/10.1016/j.nuclphysb.2012.07.008}{\emph{Nucl. Phys.}
  {\bf B864} (2012) 597--639}, [\href{http://arxiv.org/abs/1205.4720}{{\tt
  1205.4720}}].

\bibitem{Marsano:2008py}
J.~Marsano, N.~Saulina and S.~Schafer-Nameki, \emph{{An Instanton Toolbox for
  F-Theory Model Building}},
  \href{http://dx.doi.org/10.1007/JHEP01(2010)128}{\emph{JHEP} {\bf 01} (2010)
  128}, [\href{http://arxiv.org/abs/0808.2450}{{\tt 0808.2450}}].

\bibitem{Marsano:2011nn}
J.~Marsano, N.~Saulina and S.~Sch{\"a}fer-Nameki, \emph{{G-flux, M5 instantons,
  and U(1) symmetries in F-theory}},
  \href{http://dx.doi.org/10.1103/PhysRevD.87.066007}{\emph{Phys. Rev.} {\bf
  D87} (2013) 066007}, [\href{http://arxiv.org/abs/1107.1718}{{\tt
  1107.1718}}].

\bibitem{Martucci:2014ema}
L.~Martucci, \emph{{Topological duality twist and brane instantons in
  F-theory}}, \href{http://dx.doi.org/10.1007/JHEP06(2014)180}{\emph{JHEP} {\bf
  06} (2014) 180}, [\href{http://arxiv.org/abs/1403.2530}{{\tt 1403.2530}}].

\bibitem{Braun:2012nk}
A.~P. Braun, A.~Collinucci and R.~Valandro, \emph{{Algebraic description of
  G-flux in F-theory: new techniques for F-theory phenomenology}},
  \href{http://dx.doi.org/10.1002/prop.201200018}{\emph{Fortsch. Phys.} {\bf
  60} (2012) 934--940}, [\href{http://arxiv.org/abs/1202.5029}{{\tt
  1202.5029}}].

\bibitem{Braun:2011zm}
A.~P. Braun, A.~Collinucci and R.~Valandro, \emph{{G-flux in F-theory and
  algebraic cycles}},
  \href{http://dx.doi.org/10.1016/j.nuclphysb.2011.10.034}{\emph{Nucl. Phys.}
  {\bf B856} (2012) 129--179}, [\href{http://arxiv.org/abs/1107.5337}{{\tt
  1107.5337}}].

\bibitem{Braun:2014pva}
A.~P. Braun, A.~Collinucci and R.~Valandro, \emph{{Hypercharge flux in F-theory
  and the stable Sen limit}},
  \href{http://dx.doi.org/10.1007/JHEP07(2014)121}{\emph{JHEP} {\bf 07} (2014)
  121}, [\href{http://arxiv.org/abs/1402.4096}{{\tt 1402.4096}}].

\bibitem{Braun:2014xka}
A.~P. Braun and T.~Watari, \emph{{The Vertical, the Horizontal and the Rest:
  anatomy of the middle cohomology of Calabi-Yau fourfolds and F-theory
  applications}}, \href{http://dx.doi.org/10.1007/JHEP01(2015)047}{\emph{JHEP}
  {\bf 01} (2015) 047}, [\href{http://arxiv.org/abs/1408.6167}{{\tt
  1408.6167}}].

\bibitem{Collinucci:2012as}
A.~Collinucci and R.~Savelli, \emph{{On Flux Quantization in F-Theory II:
  Unitary and Symplectic Gauge Groups}},
  \href{http://dx.doi.org/10.1007/JHEP08(2012)094}{\emph{JHEP} {\bf 08} (2012)
  094}, [\href{http://arxiv.org/abs/1203.4542}{{\tt 1203.4542}}].

\bibitem{Grimm:2011fx}
T.~W. Grimm and H.~Hayashi, \emph{{F-theory fluxes, Chirality and Chern-Simons
  theories}}, \href{http://dx.doi.org/10.1007/JHEP03(2012)027}{\emph{JHEP} {\bf
  03} (2012) 027}, [\href{http://arxiv.org/abs/1111.1232}{{\tt 1111.1232}}].

\bibitem{Intriligator:2012ue}
K.~Intriligator, H.~Jockers, P.~Mayr, D.~R. Morrison and M.~R. Plesser,
  \emph{{Conifold Transitions in M-theory on Calabi-Yau Fourfolds with
  Background Fluxes}},
  \href{http://dx.doi.org/10.4310/ATMP.2013.v17.n3.a2}{\emph{Adv. Theor. Math.
  Phys.} {\bf 17} (2013) 601--699}, [\href{http://arxiv.org/abs/1203.6662}{{\tt
  1203.6662}}].

\bibitem{Jockers:2016bwi}
H.~Jockers, S.~Katz, D.~R. Morrison and M.~R. Plesser, \emph{{SU(N) transitions
  in M-theory on Calabi-Yau fourfolds and background fluxes}},
  \href{http://arxiv.org/abs/1602.07693}{{\tt 1602.07693}}.

\bibitem{Krause:2011xj}
S.~Krause, C.~Mayrhofer and T.~Weigand, \emph{{$G_4$ flux, chiral matter and
  singularity resolution in F-theory compactifications}},
  \href{http://dx.doi.org/10.1016/j.nuclphysb.2011.12.013}{\emph{Nucl. Phys.}
  {\bf B858} (2012) 1--47}, [\href{http://arxiv.org/abs/1109.3454}{{\tt
  1109.3454}}].

\bibitem{Krause:2012yh}
S.~Krause, C.~Mayrhofer and T.~Weigand, \emph{{Gauge Fluxes in F-theory and
  Type IIB Orientifolds}},
  \href{http://dx.doi.org/10.1007/JHEP08(2012)119}{\emph{JHEP} {\bf 08} (2012)
  119}, [\href{http://arxiv.org/abs/1202.3138}{{\tt 1202.3138}}].

\bibitem{Kuntzler:2012bu}
M.~Kuntzler and S.~Schafer-Nameki, \emph{{G-flux and Spectral Divisors}},
  \href{http://dx.doi.org/10.1007/JHEP11(2012)025}{\emph{JHEP} {\bf 11} (2012)
  025}, [\href{http://arxiv.org/abs/1205.5688}{{\tt 1205.5688}}].

\bibitem{Lin:2015qsa}
L.~Lin, C.~Mayrhofer, O.~Till and T.~Weigand, \emph{{Fluxes in F-theory
  Compactifications on Genus-One Fibrations}},
  \href{http://dx.doi.org/10.1007/JHEP01(2016)098}{\emph{JHEP} {\bf 01} (2016)
  098}, [\href{http://arxiv.org/abs/1508.00162}{{\tt 1508.00162}}].

\bibitem{Lin:2016vus}
L.~Lin and T.~Weigand, \emph{{G 4 -flux and standard model vacua in F-theory}},
  \href{http://dx.doi.org/10.1016/j.nuclphysb.2016.09.008}{\emph{Nucl. Phys.}
  {\bf B913} (2016) 209--247}, [\href{http://arxiv.org/abs/1604.04292}{{\tt
  1604.04292}}].

\bibitem{Lin:2016zha}
L.~Lin, \emph{{Gauge fluxes in F-theory compactifications}}.
\newblock PhD thesis, Inst. Appl. Math., Heidelberg, 2016.

\bibitem{Marsano:2010ix}
J.~Marsano, N.~Saulina and S.~Schafer-Nameki, \emph{{A Note on G-Fluxes for
  F-theory Model Building}},
  \href{http://dx.doi.org/10.1007/JHEP11(2010)088}{\emph{JHEP} {\bf 11} (2010)
  088}, [\href{http://arxiv.org/abs/1006.0483}{{\tt 1006.0483}}].

\bibitem{Marsano:2012bf}
J.~Marsano, N.~Saulina and S.~Sch{\"a}fer-Nameki, \emph{{Global Gluing and
  $G$-flux}}, \href{http://dx.doi.org/10.1007/JHEP08(2013)001}{\emph{JHEP} {\bf
  08} (2013) 001}, [\href{http://arxiv.org/abs/1211.1097}{{\tt 1211.1097}}].

\bibitem{Marsano:2011hv}
J.~Marsano and S.~Schafer-Nameki, \emph{{Yukawas, G-flux, and Spectral Covers
  from Resolved Calabi-Yau's}},
  \href{http://dx.doi.org/10.1007/JHEP11(2011)098}{\emph{JHEP} {\bf 11} (2011)
  098}, [\href{http://arxiv.org/abs/1108.1794}{{\tt 1108.1794}}].

\bibitem{Martucci:2015oaa}
L.~Martucci and T.~Weigand, \emph{{Hidden Selection Rules, M5-instantons and
  Fluxes in F-theory}},
  \href{http://dx.doi.org/10.1007/JHEP10(2015)131}{\emph{JHEP} {\bf 10} (2015)
  131}, [\href{http://arxiv.org/abs/1507.06999}{{\tt 1507.06999}}].

\bibitem{Martucci:2015dxa}
L.~Martucci and T.~Weigand, \emph{{Non-perturbative selection rules in
  F-theory}}, \href{http://dx.doi.org/10.1007/JHEP09(2015)198}{\emph{JHEP} {\bf
  09} (2015) 198}, [\href{http://arxiv.org/abs/1506.06764}{{\tt 1506.06764}}].

\bibitem{Mayrhofer:2013ara}
C.~Mayrhofer, E.~Palti and T.~Weigand, \emph{{Hypercharge Flux in IIB and
  F-theory: Anomalies and Gauge Coupling Unification}},
  \href{http://dx.doi.org/10.1007/JHEP09(2013)082}{\emph{JHEP} {\bf 09} (2013)
  082}, [\href{http://arxiv.org/abs/1303.3589}{{\tt 1303.3589}}].

\bibitem{Palti:2012dd}
E.~Palti, \emph{{A Note on Hypercharge Flux, Anomalies, and U(1)s in F-theory
  GUTs}}, \href{http://dx.doi.org/10.1103/PhysRevD.87.085036}{\emph{Phys. Rev.}
  {\bf D87} (2013) 085036}, [\href{http://arxiv.org/abs/1209.4421}{{\tt
  1209.4421}}].

\bibitem{Donagi:2009ra}
R.~Donagi and M.~Wijnholt, \emph{{Higgs Bundles and UV Completion in
  F-Theory}}, \href{http://dx.doi.org/10.1007/s00220-013-1878-8}{\emph{Commun.
  Math. Phys.} {\bf 326} (2014) 287--327},
  [\href{http://arxiv.org/abs/0904.1218}{{\tt 0904.1218}}].

\bibitem{Esole:2011sm}
M.~Esole and S.-T. Yau, \emph{{Small resolutions of SU(5)-models in F-theory}},
  \href{http://dx.doi.org/10.4310/ATMP.2013.v17.n6.a1}{\emph{Adv. Theor. Math.
  Phys.} {\bf 17} (2013) 1195--1253},
  [\href{http://arxiv.org/abs/1107.0733}{{\tt 1107.0733}}].

\bibitem{Cecotti:2010bp}
S.~Cecotti, C.~Cordova, J.~J. Heckman and C.~Vafa, \emph{{T-Branes and
  Monodromy}}, \href{http://dx.doi.org/10.1007/JHEP07(2011)030}{\emph{JHEP}
  {\bf 07} (2011) 030}, [\href{http://arxiv.org/abs/1010.5780}{{\tt
  1010.5780}}].

\bibitem{Hayashi:2009ge}
H.~Hayashi, T.~Kawano, R.~Tatar and T.~Watari, \emph{{Codimension-3
  Singularities and Yukawa Couplings in F-theory}},
  \href{http://dx.doi.org/10.1016/j.nuclphysb.2009.07.021}{\emph{Nucl. Phys.}
  {\bf B823} (2009) 47--115}, [\href{http://arxiv.org/abs/0901.4941}{{\tt
  0901.4941}}].

\bibitem{Braun:2013cb}
A.~P. Braun and T.~Watari, \emph{{On Singular Fibres in F-Theory}},
  \href{http://dx.doi.org/10.1007/JHEP07(2013)031}{\emph{JHEP} {\bf 07} (2013)
  031}, [\href{http://arxiv.org/abs/1301.5814}{{\tt 1301.5814}}].

\bibitem{Tate1975}
J.~Tate, \emph{Algorithm for determining the type of a singular fiber in an
  elliptic pencil}, pp.~33--52.
\newblock Springer Berlin Heidelberg, Berlin, Heidelberg, 1975.
\newblock 10.1007/BFb0097582.

\bibitem{Bershadsky:1996nh}
M.~Bershadsky, K.~A. Intriligator, S.~Kachru, D.~R. Morrison, V.~Sadov and
  C.~Vafa, \emph{{Geometric singularities and enhanced gauge symmetries}},
  \href{http://dx.doi.org/10.1016/S0550-3213(96)90131-5}{\emph{Nucl. Phys.}
  {\bf B481} (1996) 215--252}, [\href{http://arxiv.org/abs/hep-th/9605200}{{\tt
  hep-th/9605200}}].

\bibitem{Sen:1996vd}
A.~Sen, \emph{{F theory and orientifolds}},
  \href{http://dx.doi.org/10.1016/0550-3213(96)00347-1}{\emph{Nucl. Phys.} {\bf
  B475} (1996) 562--578}, [\href{http://arxiv.org/abs/hep-th/9605150}{{\tt
  hep-th/9605150}}].

\bibitem{Sen:1997gv}
A.~Sen, \emph{{Orientifold limit of F theory vacua}},
  \href{http://dx.doi.org/10.1103/PhysRevD.55.R7345}{\emph{Phys. Rev.} {\bf
  D55} (1997) R7345--R7349}, [\href{http://arxiv.org/abs/hep-th/9702165}{{\tt
  hep-th/9702165}}].

\bibitem{Aluffi:2009tm}
P.~Aluffi and M.~Esole, \emph{{New Orientifold Weak Coupling Limits in
  F-theory}}, \href{http://dx.doi.org/10.1007/JHEP02(2010)020}{\emph{JHEP} {\bf
  02} (2010) 020}, [\href{http://arxiv.org/abs/0908.1572}{{\tt 0908.1572}}].

\bibitem{Esole:2012tf}
M.~Esole and R.~Savelli, \emph{{Tate Form and Weak Coupling Limits in
  F-theory}}, \href{http://dx.doi.org/10.1007/JHEP06(2013)027}{\emph{JHEP} {\bf
  06} (2013) 027}, [\href{http://arxiv.org/abs/1209.1633}{{\tt 1209.1633}}].

\bibitem{Banks:1996nj}
T.~Banks, M.~R. Douglas and N.~Seiberg, \emph{{Probing F theory with branes}},
  \href{http://dx.doi.org/10.1016/0370-2693(96)00808-8}{\emph{Phys. Lett.} {\bf
  B387} (1996) 278--281}, [\href{http://arxiv.org/abs/hep-th/9605199}{{\tt
  hep-th/9605199}}].

\bibitem{Seiberg:1994rs}
N.~Seiberg and E.~Witten, \emph{{Electric - magnetic duality, monopole
  condensation, and confinement in N=2 supersymmetric Yang-Mills theory}},
  \href{http://dx.doi.org/10.1016/0550-3213(94)90124-4,
  10.1016/0550-3213(94)00449-8}{\emph{Nucl. Phys.} {\bf B426} (1994) 19--52},
  [\href{http://arxiv.org/abs/hep-th/9407087}{{\tt hep-th/9407087}}].

\bibitem{GarciaEtxebarria:2007zv}
I.~Garcia-Etxebarria and A.~M. Uranga, \emph{{Non-perturbative superpotentials
  across lines of marginal stability}},
  \href{http://dx.doi.org/10.1088/1126-6708/2008/01/033}{\emph{JHEP} {\bf 01}
  (2008) 033}, [\href{http://arxiv.org/abs/0711.1430}{{\tt 0711.1430}}].

\bibitem{GarciaEtxebarria:2008pi}
I.~Garcia-Etxebarria, F.~Marchesano and A.~M. Uranga, \emph{{Non-perturbative
  F-terms across lines of BPS stability}},
  \href{http://dx.doi.org/10.1088/1126-6708/2008/07/028}{\emph{JHEP} {\bf 07}
  (2008) 028}, [\href{http://arxiv.org/abs/0805.0713}{{\tt 0805.0713}}].

\bibitem{Beasley:2004ys}
C.~Beasley and E.~Witten, \emph{{New instanton effects in supersymmetric QCD}},
  \href{http://dx.doi.org/10.1088/1126-6708/2005/01/056}{\emph{JHEP} {\bf 01}
  (2005) 056}, [\href{http://arxiv.org/abs/hep-th/0409149}{{\tt
  hep-th/0409149}}].

\bibitem{Beasley:2005iu}
C.~Beasley and E.~Witten, \emph{{New instanton effects in string theory}},
  \href{http://dx.doi.org/10.1088/1126-6708/2006/02/060}{\emph{JHEP} {\bf 02}
  (2006) 060}, [\href{http://arxiv.org/abs/hep-th/0512039}{{\tt
  hep-th/0512039}}].

\bibitem{Hanany:2005ve}
A.~Hanany and K.~D. Kennaway, \emph{{Dimer models and toric diagrams}},
  \href{http://arxiv.org/abs/hep-th/0503149}{{\tt hep-th/0503149}}.

\bibitem{Franco:2005rj}
S.~Franco, A.~Hanany, K.~D. Kennaway, D.~Vegh and B.~Wecht, \emph{{Brane dimers
  and quiver gauge theories}},
  \href{http://dx.doi.org/10.1088/1126-6708/2006/01/096}{\emph{JHEP} {\bf 01}
  (2006) 096}, [\href{http://arxiv.org/abs/hep-th/0504110}{{\tt
  hep-th/0504110}}].

\bibitem{Feng:2005gw}
B.~Feng, Y.-H. He, K.~D. Kennaway and C.~Vafa, \emph{{Dimer models from mirror
  symmetry and quivering amoebae}},
  \href{http://dx.doi.org/10.4310/ATMP.2008.v12.n3.a2}{\emph{Adv. Theor. Math.
  Phys.} {\bf 12} (2008) 489--545},
  [\href{http://arxiv.org/abs/hep-th/0511287}{{\tt hep-th/0511287}}].

\bibitem{Franco:2006es}
S.~Franco and A.~M.~. Uranga, \emph{{Dynamical SUSY breaking at meta-stable
  minima from D-branes at obstructed geometries}},
  \href{http://dx.doi.org/10.1088/1126-6708/2006/06/031}{\emph{JHEP} {\bf 06}
  (2006) 031}, [\href{http://arxiv.org/abs/hep-th/0604136}{{\tt
  hep-th/0604136}}].

\bibitem{Franco:2007ii}
S.~Franco, A.~Hanany, D.~Krefl, J.~Park, A.~M. Uranga and D.~Vegh,
  \emph{{Dimers and orientifolds}},
  \href{http://dx.doi.org/10.1088/1126-6708/2007/09/075}{\emph{JHEP} {\bf 09}
  (2007) 075}, [\href{http://arxiv.org/abs/0707.0298}{{\tt 0707.0298}}].

\bibitem{Forcella:2008au}
D.~Forcella, I.~Garcia-Etxebarria and A.~Uranga, \emph{{E3-brane instantons and
  baryonic operators for D3-branes on toric singularities}},
  \href{http://dx.doi.org/10.1088/1126-6708/2009/03/041}{\emph{JHEP} {\bf 03}
  (2009) 041}, [\href{http://arxiv.org/abs/0806.2291}{{\tt 0806.2291}}].

\bibitem{Bergh:aa}
M.~V. den Bergh, \emph{Non-commutative crepant resolutions},
  \href{http://arxiv.org/abs/math/0211064}{{\tt math/0211064}}.

\bibitem{Berenstein:aa}
D.~Berenstein and R.~G. Leigh, \emph{Resolution of stringy singularities by
  non-commutative algebras},  \href{http://arxiv.org/abs/hep-th/0105229}{{\tt
  hep-th/0105229}}.

\bibitem{Aspinwall:2004jr}
P.~S. Aspinwall, \emph{{D-branes on Calabi-Yau manifolds}},  in \emph{{Progress
  in string theory. Proceedings, Summer School, TASI 2003, Boulder, USA, June
  2-27, 2003}}, pp.~1--152, 2004.
\newblock \href{http://arxiv.org/abs/hep-th/0403166}{{\tt hep-th/0403166}}.
\newblock \href{http://dx.doi.org/10.1142/9789812775108_0001}{DOI}.

\bibitem{Herbst:2008jq}
M.~Herbst, K.~Hori and D.~Page, \emph{{Phases Of N=2 Theories In 1+1 Dimensions
  With Boundary}},  \href{http://arxiv.org/abs/0803.2045}{{\tt 0803.2045}}.

\bibitem{Collinucci:2014qfa}
A.~Collinucci and R.~Savelli, \emph{{T-branes as branes within branes}},
  \href{http://dx.doi.org/10.1007/JHEP09(2015)161}{\emph{JHEP} {\bf 09} (2015)
  161}, [\href{http://arxiv.org/abs/1410.4178}{{\tt 1410.4178}}].

\bibitem{Klebanov:1998hh}
I.~R. Klebanov and E.~Witten, \emph{{Superconformal field theory on
  three-branes at a Calabi-Yau singularity}},
  \href{http://dx.doi.org/10.1016/S0550-3213(98)00654-3}{\emph{Nucl. Phys.}
  {\bf B536} (1998) 199--218}, [\href{http://arxiv.org/abs/hep-th/9807080}{{\tt
  hep-th/9807080}}].

\bibitem{Bershadsky:1995qy}
M.~Bershadsky, C.~Vafa and V.~Sadov, \emph{{D-branes and topological field
  theories}}, \href{http://dx.doi.org/10.1016/0550-3213(96)00026-0}{\emph{Nucl.
  Phys.} {\bf B463} (1996) 420--434},
  [\href{http://arxiv.org/abs/hep-th/9511222}{{\tt hep-th/9511222}}].

\bibitem{Garcia-Etxebarria:2015lif}
I.~Garc{\'\i}a-Etxebarria, F.~Quevedo and R.~Valandro, \emph{{Global String
  Embeddings for the Nilpotent Goldstino}},
  \href{http://dx.doi.org/10.1007/JHEP02(2016)148}{\emph{JHEP} {\bf 02} (2016)
  148}, [\href{http://arxiv.org/abs/1512.06926}{{\tt 1512.06926}}].

\bibitem{Aganagic:2003xq}
M.~Aganagic, K.~A. Intriligator, C.~Vafa and N.~P. Warner, \emph{{The Glueball
  superpotential}},
  \href{http://dx.doi.org/10.4310/ATMP.2003.v7.n6.a4}{\emph{Adv. Theor. Math.
  Phys.} {\bf 7} (2003) 1045--1101},
  [\href{http://arxiv.org/abs/hep-th/0304271}{{\tt hep-th/0304271}}].

\bibitem{Intriligator:2003xs}
K.~A. Intriligator, P.~Kraus, A.~V. Ryzhov, M.~Shigemori and C.~Vafa, \emph{{On
  low rank classical groups in string theory, gauge theory and matrix models}},
  \href{http://dx.doi.org/10.1016/j.nuclphysb.2003.12.030}{\emph{Nucl. Phys.}
  {\bf B682} (2004) 45--82}, [\href{http://arxiv.org/abs/hep-th/0311181}{{\tt
  hep-th/0311181}}].

\bibitem{Aganagic:2007py}
M.~Aganagic, C.~Beem and S.~Kachru, \emph{{Geometric transitions and dynamical
  SUSY breaking}},
  \href{http://dx.doi.org/10.1016/j.nuclphysb.2007.11.032}{\emph{Nucl. Phys.}
  {\bf B796} (2008) 1--24}, [\href{http://arxiv.org/abs/0709.4277}{{\tt
  0709.4277}}].

\bibitem{GarciaEtxebarria:2008iw}
I.~Garcia-Etxebarria, \emph{{D-brane instantons and matrix models}},
  \href{http://dx.doi.org/10.1088/1126-6708/2009/07/017}{\emph{JHEP} {\bf 07}
  (2009) 017}, [\href{http://arxiv.org/abs/0810.1482}{{\tt 0810.1482}}].

\bibitem{Argurio:2007qk}
R.~Argurio, M.~Bertolini, S.~Franco and S.~Kachru, \emph{{Meta-stable vacua and
  D-branes at the conifold}},
  \href{http://dx.doi.org/10.1088/1126-6708/2007/06/017}{\emph{JHEP} {\bf 06}
  (2007) 017}, [\href{http://arxiv.org/abs/hep-th/0703236}{{\tt
  hep-th/0703236}}].

\bibitem{Argurio:2007vqa}
R.~Argurio, M.~Bertolini, G.~Ferretti, A.~Lerda and C.~Petersson,
  \emph{{Stringy instantons at orbifold singularities}},
  \href{http://dx.doi.org/10.1088/1126-6708/2007/06/067}{\emph{JHEP} {\bf 06}
  (2007) 067}, [\href{http://arxiv.org/abs/0704.0262}{{\tt 0704.0262}}].

\bibitem{Bianchi:2007wy}
M.~Bianchi, F.~Fucito and J.~F. Morales, \emph{{D-brane instantons on the T**6
  / Z(3) orientifold}},
  \href{http://dx.doi.org/10.1088/1126-6708/2007/07/038}{\emph{JHEP} {\bf 07}
  (2007) 038}, [\href{http://arxiv.org/abs/0704.0784}{{\tt 0704.0784}}].

\bibitem{Ibanez:2007rs}
L.~E. Ibanez, A.~N. Schellekens and A.~M. Uranga, \emph{{Instanton Induced
  Neutrino Majorana Masses in CFT Orientifolds with MSSM-like spectra}},
  \href{http://dx.doi.org/10.1088/1126-6708/2007/06/011}{\emph{JHEP} {\bf 06}
  (2007) 011}, [\href{http://arxiv.org/abs/0704.1079}{{\tt 0704.1079}}].

\bibitem{Blumenhagen:2009qh}
R.~Blumenhagen, M.~Cvetic, S.~Kachru and T.~Weigand, \emph{{D-Brane Instantons
  in Type II Orientifolds}},
  \href{http://dx.doi.org/10.1146/annurev.nucl.010909.083113}{\emph{Ann. Rev.
  Nucl. Part. Sci.} {\bf 59} (2009) 269--296},
  [\href{http://arxiv.org/abs/0902.3251}{{\tt 0902.3251}}].

\bibitem{Imai:2001cq}
S.~Imai and T.~Yokono, \emph{{Comments on orientifold projection in the
  conifold and $SO\times Sp$ duality cascade}},
  \href{http://dx.doi.org/10.1103/PhysRevD.65.066007}{\emph{Phys. Rev.} {\bf
  D65} (2002) 066007}, [\href{http://arxiv.org/abs/hep-th/0110209}{{\tt
  hep-th/0110209}}].

\bibitem{Blumenhagen:2007zk}
R.~Blumenhagen, M.~Cvetic, D.~Lust, R.~Richter and T.~Weigand,
  \emph{{Non-perturbative Yukawa Couplings from String Instantons}},
  \href{http://dx.doi.org/10.1103/PhysRevLett.100.061602}{\emph{Phys. Rev.
  Lett.} {\bf 100} (2008) 061602}, [\href{http://arxiv.org/abs/0707.1871}{{\tt
  0707.1871}}].

\bibitem{Cecotti:2009zf}
S.~Cecotti, M.~C.~N. Cheng, J.~J. Heckman and C.~Vafa, \emph{{Yukawa Couplings
  in F-theory and Non-Commutative Geometry}},
  \href{http://arxiv.org/abs/0910.0477}{{\tt 0910.0477}}.

\bibitem{Hori:2000ck}
K.~Hori, A.~Iqbal and C.~Vafa, \emph{{D-branes and mirror symmetry}},
  \href{http://arxiv.org/abs/hep-th/0005247}{{\tt hep-th/0005247}}.

\bibitem{Bouchard:2011ya}
V.~Bouchard and P.~Sulkowski, \emph{{Topological recursion and mirror curves}},
  \href{http://dx.doi.org/10.4310/ATMP.2012.v16.n5.a3}{\emph{Adv. Theor. Math.
  Phys.} {\bf 16} (2012) 1443--1483},
  [\href{http://arxiv.org/abs/1105.2052}{{\tt 1105.2052}}].

\bibitem{Callan:1997kz}
C.~G. Callan and J.~M. Maldacena, \emph{{Brane death and dynamics from the
  Born-Infeld action}},
  \href{http://dx.doi.org/10.1016/S0550-3213(97)00700-1}{\emph{Nucl. Phys.}
  {\bf B513} (1998) 198--212}, [\href{http://arxiv.org/abs/hep-th/9708147}{{\tt
  hep-th/9708147}}].

\bibitem{Strominger:1995cz}
A.~Strominger, \emph{{Massless black holes and conifolds in string theory}},
  \href{http://dx.doi.org/10.1016/0550-3213(95)00287-3}{\emph{Nucl. Phys.} {\bf
  B451} (1995) 96--108}, [\href{http://arxiv.org/abs/hep-th/9504090}{{\tt
  hep-th/9504090}}].

\bibitem{Aspinwall:1995zi}
P.~S. Aspinwall, \emph{{Enhanced gauge symmetries and K3 surfaces}},
  \href{http://dx.doi.org/10.1016/0370-2693(95)00957-M}{\emph{Phys. Lett.} {\bf
  B357} (1995) 329--334}, [\href{http://arxiv.org/abs/hep-th/9507012}{{\tt
  hep-th/9507012}}].

\bibitem{Donagi:2012ts}
R.~Donagi, S.~Katz and M.~Wijnholt, \emph{{Weak Coupling, Degeneration and Log
  Calabi-Yau Spaces}},  \href{http://arxiv.org/abs/1212.0553}{{\tt 1212.0553}}.

\bibitem{Clingher:2012rg}
A.~Clingher, R.~Donagi and M.~Wijnholt, \emph{{The Sen Limit}},
  \href{http://dx.doi.org/10.4310/ATMP.2014.v18.n3.a2}{\emph{Adv. Theor. Math.
  Phys.} {\bf 18} (2014) 613--658}, [\href{http://arxiv.org/abs/1212.4505}{{\tt
  1212.4505}}].

\bibitem{Collinucci:2008pf}
A.~Collinucci, F.~Denef and M.~Esole, \emph{{D-brane Deconstructions in IIB
  Orientifolds}},
  \href{http://dx.doi.org/10.1088/1126-6708/2009/02/005}{\emph{JHEP} {\bf 02}
  (2009) 005}, [\href{http://arxiv.org/abs/0805.1573}{{\tt 0805.1573}}].

\bibitem{Blumenhagen:2008zz}
R.~Blumenhagen, V.~Braun, T.~W. Grimm and T.~Weigand, \emph{{GUTs in Type IIB
  Orientifold Compactifications}},
  \href{http://dx.doi.org/10.1016/j.nuclphysb.2009.02.011}{\emph{Nucl. Phys.}
  {\bf B815} (2009) 1--94}, [\href{http://arxiv.org/abs/0811.2936}{{\tt
  0811.2936}}].

\bibitem{Bondal:aa}
A.~Bondal and D.~Orlov, \emph{Derived categories of coherent sheaves},
  {\emph{Proceedings of the ICM, Vol. II (Beijing, 2002), 47--56} (2002) },
  [\href{http://arxiv.org/abs/math/0206295}{{\tt math/0206295}}].

\bibitem{Aspinwall:2009isa}
P.~S. Aspinwall, T.~Bridgeland, A.~Craw, M.~R. Douglas, A.~Kapustin, G.~W.
  Moore et~al., \emph{{Dirichlet branes and mirror symmetry}}, vol.~4 of
  \emph{Clay Mathematics Monographs}.
\newblock AMS, Providence, RI, 2009.

\end{thebibliography}\endgroup

\end{document}